\documentclass[twocolumn]{aastex631}

\usepackage{hyperref}
\usepackage{nameref}

\begin{document}

\title{The rise of the galactic empire: ultraviolet luminosity functions  at \boldmath{$z\sim17$} and \boldmath{$z\sim25$} estimated with the MIDIS$+$NGDEEP ultra-deep JWST/NIRCam dataset}
\shorttitle{The Rise of the Galactic Empire}
\shortauthors{P\'erez-Gonz\'alez et al.}

\correspondingauthor{Pablo G. P\'erez-Gonz\'alez}
\email{pgperez@cab.inta-csic.es}

\author[0000-0003-4528-5639]{Pablo G. P\'erez-Gonz\'alez}
\affiliation{Centro de Astrobiolog\'{\i}a (CAB), CSIC-INTA, Ctra. de Ajalvir km 4, Torrej\'on de Ardoz, E-28850, Madrid, Spain}

\author[0000-0002-3005-1349]{G{\"o}ran {\"O}stlin}
\affiliation{Department of Astronomy, Stockholm University, Oscar Klein Centre, AlbaNova University Centre, 106 91 Stockholm, Sweden}

\author[0000-0001-6820-0015]{Luca Costantin}
\affiliation{Centro de Astrobiolog\'{\i}a (CAB), CSIC-INTA, Ctra. de Ajalvir km 4, Torrej\'on de Ardoz, E-28850, Madrid, Spain}

\author[0000-0003-0470-8754]{Jens Melinder}
\affiliation{Department of Astronomy, Stockholm University, Oscar Klein Centre, AlbaNova University Centre, 106 91 Stockholm, Sweden}

\author[0000-0001-8519-1130]{Steven L. Finkelstein}
\affiliation{Department of Astronomy, The University of Texas at Austin, Austin, TX 78712, USA}

\author[0000-0002-6748-6821]{Rachel S. Somerville}
\affiliation{Center for Computational Astrophysics, Flatiron Institute, 162 5th Avenue, New York, NY, 10010, USA}

\author[0000-0002-8053-8040]{Marianna Annunziatella}
\affiliation{Centro de Astrobiolog\'{\i}a (CAB), CSIC-INTA, Ctra. de Ajalvir km 4, Torrej\'on de Ardoz, E-28850, Madrid, Spain}


\author[0000-0002-7093-1877]{Javier Álvarez-Márquez}
\affiliation{Centro de Astrobiolog\'{\i}a (CAB), CSIC-INTA, Ctra. de Ajalvir km 4, Torrej\'on de Ardoz, E-28850, Madrid, Spain}

\author[0000-0002-9090-4227]{Luis Colina}
\affiliation{Centro de Astrobiolog\'{\i}a (CAB), CSIC-INTA, Ctra. de Ajalvir km 4, Torrej\'on de Ardoz, E-28850, Madrid, Spain}

\author{Avishai Dekel}
\affiliation{Racah Institute of Physics, The Hebrew University of Jerusalem, Jerusalem 91904, Israel}

\author[0000-0001-7113-2738]{Henry C. Ferguson}
\affiliation{Space Telescope Science Institute, 3700 San Martin Drive, Baltimore, MD 21218, USA}

\author[0000-0001-7890-4964]{Zhaozhou Li}
\affiliation{Racah Institute of Physics, The Hebrew University of Jerusalem, Jerusalem 91904, Israel}

\author[0000-0003-3466-035X]{{L. Y. Aaron} {Yung}}
\affiliation{Space Telescope Science Institute, 3700 San Martin Drive, Baltimore, MD 21218, USA}


\author[0000-0002-9921-9218]{Micaela B. Bagley}
\affiliation{Department of Astronomy, The University of Texas at Austin, Austin, TX, USA}
\affiliation{Astrophysics Science Division, NASA Goddard Space Flight Center, 8800 Greenbelt Rd, Greenbelt, MD 20771, USA}

\author[0000-0002-3952-8588]{Leindert A. Boogaard}\affiliation{Leiden Observatory, Leiden University, PO~Box~9513, 2300~RA~Leiden, The Netherlands}

\author[0000-0002-4193-2539]{Denis Burgarella}
\affiliation{Aix Marseille Université, CNRS, CNES, LAM, Marseille, France}

\author[0000-0003-2536-1614]{Antonello Calabr\`o}
\affiliation{INAF Osservatorio Astronomico di Roma, Via Frascati 33, 00078 Monte Porzio Catone, Rome, Italy}

\author[0000-0001-8183-1460]{Karina I. Caputi}
\affiliation{Kapteyn Astronomical Institute, University of Groningen, P.O. Box 800, 9700AV Groningen, The Netherlands}
\affiliation{Cosmic Dawn Center (DAWN), Copenhagen, Denmark}

\author[0000-0001-8551-071X]{Yingjie Cheng}
\affiliation{University of Massachusetts Amherst, 710 North Pleasant Street, Amherst, MA 01003-9305, USA}

\author[0000-0001-5414-5131]{Mark Dickinson}
\affiliation{NSF's National Optical-IR Astronomy Research Laboratory, 950 N. Cherry Ave., Tucson, AZ 85719, USA}

\author[0000-0001-6049-3132]{Andreas Eckart}
\affiliation{I.Physikalisches Institut der Universit\"at zu K\"oln, Z\"ulpicher Str. 77, 50937 K\"oln, Germany}

\author[0000-0002-7831-8751]{Mauro Giavalisco}
\affiliation{University of Massachusetts Amherst, 710 North Pleasant Street, Amherst, MA 01003-9305, USA}

\author[0000-0001-9885-4589]{Steven Gillman}
\affiliation{Cosmic Dawn Center (DAWN), Denmark}
\affiliation{DTU Space, Technical University of Denmark, Elektrovej, Building 328, 2800, Kgs. Lyngby, Denmark}

\author[0000-0002-2554-1837]{Thomas R. Greve}
\affiliation{Cosmic Dawn Center (DAWN), Denmark}
\affiliation{DTU Space, Technical University of Denmark, Elektrovej, Building 328, 2800, Kgs. Lyngby, Denmark}
\affiliation{Dept.~of Physics and Astronomy, University College London, Gower Street, London WC1E 6BT, United Kingdom}

\author[0000-0001-9626-9642]{Mahmoud Hamed}
\affiliation{Centro de Astrobiolog\'{\i}a (CAB), CSIC-INTA, Ctra. de Ajalvir km 4, Torrej\'on de Ardoz, E-28850, Madrid, Spain}

\author[0000-0001-6145-5090]{Nimish P. Hathi}
\affiliation{Space Telescope Science Institute, 3700 San Martin Drive, Baltimore, MD 21218, USA}
 
\author[0000-0002-4571-2306]{Jens Hjorth}
\affiliation{DARK, Niels Bohr Institute, University of Copenhagen, Jagtvej 155A, 2200 Copenhagen, Denmark}

\author[0000-0002-1416-8483]{Marc Huertas-Company}
\affiliation{Instituto de Astrofísica de Canarias (IAC), La Laguna, E-38205,
Spain}
\affiliation{Universidad de La Laguna. Avda. Astrofísico Fco. Sanchez, La La-
guna, Tenerife, Spain}
\affiliation{Observatoire de Paris, LERMA, PSL University, 61 avenue de
l’Observatoire, F-75014 Paris, France}
\affiliation{Université Paris-Cité, 5 Rue Thomas Mann, 75014 Paris, France}

\author[0000-0001-9187-3605]{Jeyhan S. Kartaltepe}
\affiliation{Laboratory for Multiwavelength Astrophysics, School of Physics and Astronomy, Rochester Institute of Technology, 84 Lomb Memorial Drive, Rochester, NY 14623, USA}

\author[0000-0002-6610-2048]{Anton M. Koekemoer}
\affiliation{Space Telescope Science Institute, 3700 San Martin Drive, Baltimore, MD 21218, USA}

\author[0000-0002-5588-9156]{Vasily Kokorev}
\affiliation{Department of Astronomy, The University of Texas at Austin, Austin, TX 78712, USA}

\author[0000-0002-0690-8824]{\'Alvaro Labiano}
\affiliation{Telespazio UK for the European Space Agency, ESAC, Camino Bajo del Castillo s/n, 28692 Villanueva de la Ca\~nada, Spain}

\author[0000-0001-5710-8395]{Danial Langeroodi}
\affiliation{DARK, Niels Bohr Institute, University of Copenhagen, Jagtvej 155A, 2200 Copenhagen, Denmark}

\author[0000-0002-9393-6507]{Gene C. K. Leung}
\affiliation{MIT Kavli Institute for Astrophysics and Space Research, 77 Massachusetts Ave., Cambridge, MA 02139, USA}

\author[0000-0002-5554-8896]{Priyamvada Natarajan}
\affiliation{Department of Astronomy, Yale University, New Haven, CT 06511, USA; Department of Physics, Yale University, New Haven, CT 06520, USA; Black Hole Initiative, Harvard University, 20 Garden Street, Cambridge, MA 02138, USA}

\author[0000-0001-7503-8482]{Casey Papovich}
\affiliation{Department of Physics and Astronomy, Texas A\&M University, College Station, TX, 77843-4242 USA}
\affiliation{George P.\ and Cynthia Woods Mitchell Institute for Fundamental Physics and Astronomy, Texas A\&M University, College Station, TX, 77843-4242 USA}

\author[0000-0002-9850-2708]{Florian Peissker}
\affiliation{(I. Physikalisches Institut der Universit\"at zu K\"oln, Z\"ulpicher Str. 77, 50937 K\"oln, Germany}

\author[0000-0001-8940-6768]{Laura Pentericci}
\affiliation{INAF Osservatorio Astronomico di Roma, Via Frascati 33, 00078 Monte Porzio Catone, Rome, Italy}

\author[0000-0003-3382-5941]{Nor Pirzkal}
\affiliation{ESA/AURA Space Telescope Science Institute}

\author[0000-0002-5104-8245]{Pierluigi Rinaldi}
\affiliation{Steward Observatory, University of Arizona, 933 North Cherry Avenue, Tucson, AZ 85721, USA}

\author[0000-0001-5434-5942]{Paul van der Werf}
\affiliation{Leiden Observatory, Leiden University, PO~Box~9513, 2300~RA~Leiden, The Netherlands}

\author[0000-0003-4793-7880]{Fabian Walter}
\affiliation{{Max Planck Institut f\"ur Astronomie, K\"onigstuhl 17, D-69117, Heidelberg, Germany}}

\begin{abstract}

We present a sample of six F200W and three F277W dropout sources identified as $16<z<25$ galaxy candidates using the deepest JWST/NIRCam data to date (5$\sigma$ depths $\sim31.5$~mag at $\geq2$~$\mu$m), provided by the MIRI Deep Imaging Survey (MIDIS) and the Next Generation Deep Extragalactic Exploratory Public survey  (NGDEEP). We estimate ultraviolet (UV) luminosity functions and densities at $z\sim17$ and $z\sim25$. The number density of galaxies with absolute magnitudes $-19<M_\mathrm{UV}<-18$ at $z\sim17$ ($z\sim25$) 
is a factor of 4 (25) smaller than at $z\sim12$; the luminosity density presents a similar evolution. Compared to state-of-the-art galaxy simulations, we find the need for an enhanced UV-photon production at $z=17-25$ in $\mathrm{M}_\mathrm{DM}=10^{8.5-9.5}$~M$_\odot$ dark matter halos, provided by an increase in the star formation efficiency at early times and/or by intense compact starbursts with enhanced emissivity linked to strong burstiness, low or primordial gas metallicities, and/or a top-heavy initial mass function. 
There are few robust theoretical predictions for the evolution of galaxies above $z\sim20$ in the literature, however, the continuing rapid drop in the halo mass function would predict a more rapid evolution than we observe if photon production efficiencies remained constant. 
Our $z>16$ candidates present mass-weighted ages around 30~Myr, and attenuations $\mathrm{A(V)}<0.1$~mag.  Their average stellar mass is $\mathrm{M}_\bigstar\sim10^{7}\,\mathrm{M}_\odot$, implying a stellar-to-baryon mass fraction around 10\% if the emissivity increases with redshift, or significantly higher otherwise.  Three candidates present very blue UV spectral slopes ($\beta\sim-3$) compatible with Pop III young ($\lesssim10$~Myr) stars and/or high escape fractions of ionizing photons; the rest have $\beta\sim-2.5$ similar to $z=10-12$ samples.

\end{abstract}



\section{Introduction}
\label{sec:intro}

After almost three years of scientific operations of JWST, there is a consensus on the analysis of the $z>10$ Universe performed with a variety of datasets (see \citealt{2024arXiv240521054A} for a review) indicating that the Universe was more active in the formation of galaxies during the first half billion years ($z>9$) than anticipated based on empirical extrapolations of pre-JWST observations at lower redshifts, as well as the predictions of state-of-the-art physics-based simulations that were calibrated to match pre-JWST observations \citep{2024ApJ...965..169A,2022ApJ...938L..15C,2023ApJ...948L..14C,2024arXiv240714973C,2024MNRAS.533.3222D,2023MNRAS.518.6011D,2022ApJ...940L..55F,2023ApJ...946L..13F,2024ApJ...969L...2F,2024ApJ...964...71H,2023ApJS..265....5H,2023ApJ...954L..46L,2024MNRAS.527.5004M,2023ApJ...951L...1P,2024ApJ...970...31R,2023ApJ...942L...8Y,2024ApJ...966...74W}.

Virtually all surveys reaching different depths, exploring different sky regions with deep, small areas as well as shallow, large area strategies, and with different filter sets (e.g, CEERS: \citealt{2025arXiv250104085F}; PRIMER: \citealt{2024MNRAS.533.3222D}; NGDEEP: \citealt{2024ApJ...965L...6B}; COSMOS-Web: \citealt{2023ApJ...954...31C}, JADES: \citealt{2023arXiv230602465E,2023ApJS..269...16R}; MIDIS: \citealt{2024arXiv241119686O}, UNCOVER: \citealt{2024ApJ...974...92B}; GLASS: \citealt{2022ApJ...935..110T}), find numerous $z>9$ galaxy candidates, in excess of expectations. Many of these candidates have been validated spectroscopically \citep{2023Natur.622..707A,2023ApJ...951L..22A,2023NatAs...7..622C,2024ApJ...972..143C,2024ApJ...960...56H,2024Natur.633..318C,2025arXiv250511263N}. The cumulative number density of high redshift galaxies at the bright end of the luminosity function (apparent magnitudes brighter than F277W$\sim29$, typically probing an order of magnitude fainter than L$^*$ at $z\sim12$) exceed expectations from pre-JWST simulations by a factor of 10 at $z\sim9$ and nearly a factor of 100 up to $z\sim12$ (see, e.g., \citealt{2023ApJ...946L..13F,2024ApJ...969L...2F,2023ApJ...951L...1P}).

Probing redshifts higher than $z\sim12$ in both photometrically selected candidate samples and spectroscopic follow-up campaigns remains very challenging, although some efforts exist (apart from references above, see \citealt{2023ApJ...942L...9Y,2024arXiv241113640K,2025arXiv250405893C}). From the photometric point of view, hundreds of F150W dropouts have been selected and many confirmed with spectroscopy. But no F200W dropout galaxy has been confirmed, the closest being the two highest confirmed redshift source known, JADES-GS-z14-0 \citep{2024ApJ...970...31R,2024Natur.633..318C} and MoM-z14 \citep{2025arXiv250511263N}, which only disappear bluewards of $\sim1.8$~$\mu$m. Interestingly, several  $z>14$ relatively bright F200W dropouts (around magnitude 28 in F277W) selected as photometric candidates have also been confirmed to instead be low redshift interlopers, typically dusty galaxies at $z=4-5$ \citep{2022arXiv220802794N,2023ApJ...943L...9Z,2023ApJ...946L..16P,2023Natur.622..707A}. Overall, our statement about the consensus in the results obtained out to $z\sim12$ and their reliability, based on relatively large photometric samples and a statistically meaningful number of spectroscopic confirmations, cannot yet be extended to $z\gtrsim12$.

\begin{figure*}[ht!]
\centering
\includegraphics[clip, trim=1.4cm 0.8cm 2.0cm 2.0cm,scale=0.8]{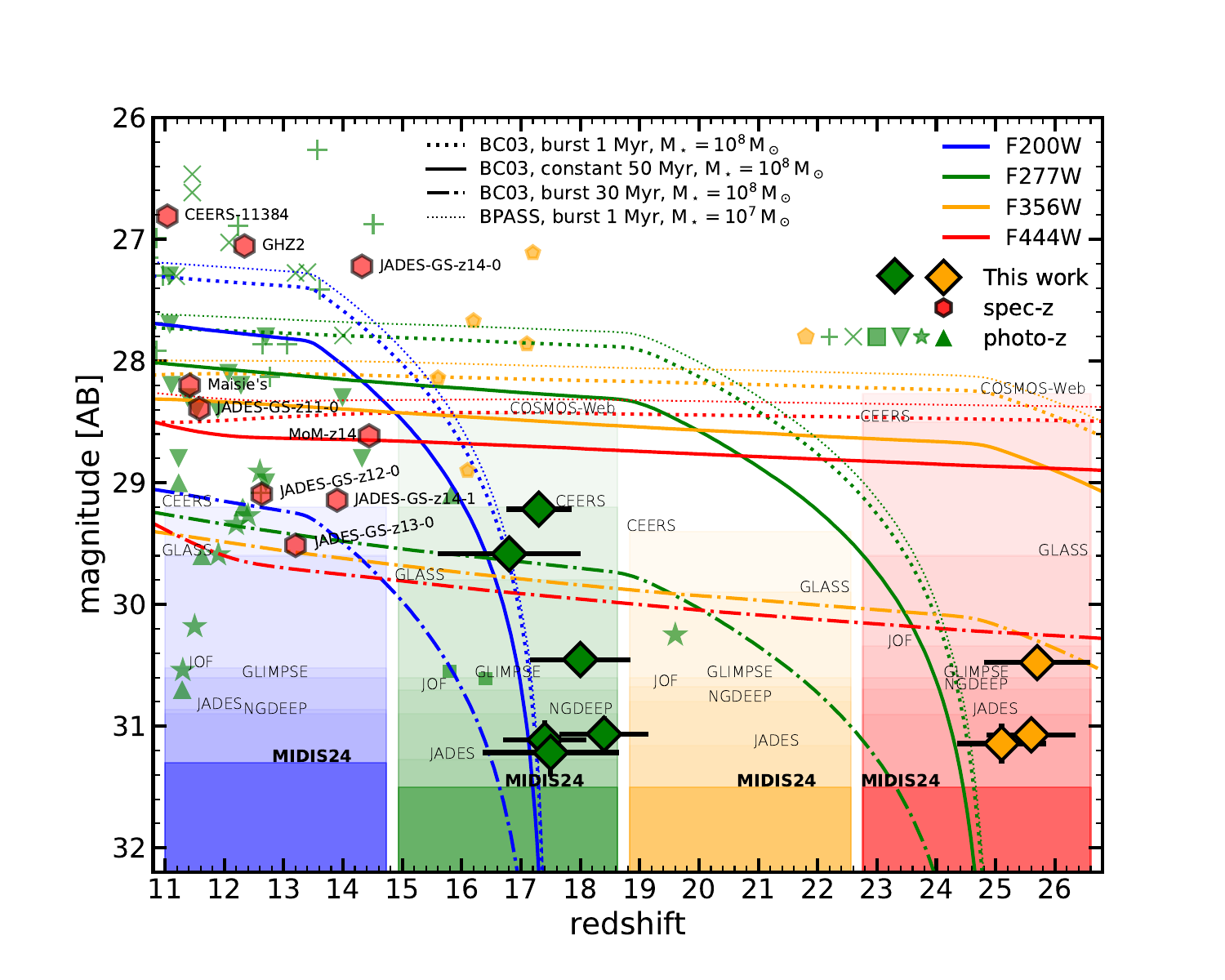}
\caption{\label{fig:depths}Compilation of spectroscopically confirmed galaxies (red hexagons, \citealt{2023Natur.622..707A,2023NatAs...7..622C,2024ApJ...972..143C,2024Natur.633..318C,2025arXiv250511263N}) and photometric samples (including \citealt{2023ApJ...954L..46L} -green pointing-up triangles-, \citealt{2024ApJ...969L...2F} -green pointing-down triangle-, \citealt{2023ApJ...951L...1P} -green stars-, \citealt{2024arXiv241113640K} -green squares-, \citealt{2024ApJ...965...98C} -green crosses-, \citealt{2025arXiv250405893C} -orange pentagons; one of their candidates lies outside our plot, at magnitude 25.5, three candidates were spectroscopically confirmed to be at low redshift and are not shown here-, and \citealt{2025arXiv250706292W} -green plus signs-) of galaxies at $z>11$. The $16<z<25$ galaxy candidates presented in this work are plotted with green and orange diamonds, the color representing the band used in the plot, the one closer to and redward of the Lyman break. We note that for most of the literature points, only the F444W flux is provided and shown in this plot. We also show their magnitude and photometric redshift uncertainties. Lines show the expected magnitudes as a function of redshift of a very young (1~Myr old) starburst with stellar mass $M_\star=10^8\,M_\odot$ according to the \citet{2003MNRAS.344.1000B} models (thick dotted line) and a less massive $M_\star=10^7\,M_\odot$ starburst according to the \citet{2017PASA...34...58E} BPASS models (thin dotted line). The expected magnitudes for a more extended star formation history (50~Myr constant star formation population) and a dormant galaxy (which experienced an instantaneous burst 30~Myr before the observation) are also shown (thick continuous and dashed-dotted lines, respectively). All models assume no dust. Shaded regions show the 5$\sigma$ depths of the major galaxy surveys used in the identification of $z>11$ galaxy candidates (see second paragraph in Introduction for references). The darkest regions refer to the dataset used in this paper (MIDIS24, which stands for the 2024 depth, including NGDEEP). The color for all data points, lines, and shaded regions indicates the NIRCam band considered. }
\end{figure*}

Part of the lack of consensus, leaving aside the very few spectroscopic confirmations beyond $z\sim12$ (only 5 galaxies, see references in Figure~\ref{fig:depths}), comes from the limited depth of NIRCam datasets in the relevant filters needed to detect galaxies in the first 350~Myr of the Universe. Two points have to be taken into account to understand the existence of the current redshift frontier at $z\sim12$. First, photometric samples typically reach magnitudes around F277W$\sim$29, consistent with the aim to detect Lyman breaks of at least 1.0-1.5~mag and considering that most surveys are limited to magnitude 30.5 in the filters longwards of 2~$\mu$m (even the deepest such as NGDEEP and JADES/JOF). Second, most of the spectroscopically confirmed $z>9$ galaxies (see caption of Figure~\ref{fig:depths} for references) have magnitudes around F277W$\sim28\pm1$, with a significant number of them having been assigned a redshift based on low spectral resolution NIRSpec prism data probing the continuum emission and the Lyman break rather than detecting (the typically quite faint) ultraviolet emission lines.

In this paper, we aim to construct a robust sample of $z>16$ galaxies by identifying F200W and F277W dropouts (probing roughly $16<z<21$ and $24<z<28$, respectively) in the deepest dataset obtained by JWST with its NIRCam instrument to date. This dataset results from combining the observations from: (1) the parallel field of the MIRI Deep Survey (MIDIS, \citealt{2024arXiv241119686O}), and its red extension (MIDIS-RED), whose prime observations were carried out with MIRI in the HUDF for nearly 100 hours; and (2) the Next Generation Deep Extragalactic Exploratory Public survey (NGDEEP), which observed the HUDF with NIRISS as prime for 60 hours \citep{2024ApJ...965L...6B}. The combination of these two surveys in all NIRCam broad bands from 2~$\mu$m redwards (i.e., F200W, F277W, F356W, and F444W) in the parallel field located to the South-East of the HUDF provides a unique dataset that reaches up to $\times2.5$ fainter fluxes than other small-area deep surveys such as JADES or GLIMPSE, and $\times10$ times fainter fluxes compared to large-area shallow surveys such as CEERS or PRIMER.

This paper is organized as follows. Section~\ref{sec:selection} presents our dataset, the NIRCam parallels of MIDIS and NGDEEP, and describes our method to select F200W and F277W dropout sources and identify $z\sim17$ and $z\sim25$ galaxy candidates. The sample is then used to construct UV luminosity functions and derive the early evolution of the luminosity density in the Universe, that are presented in Section~\ref{sec:results}. In this same section, we provide calculations of the efficiency of production of UV photons by dark matter halos, compare our results with galaxy formations simulations, and discuss the stellar masses, morphologies, and UV spectral slopes of our candidates. Section~\ref{sec:conclusions} summarizes our findings. 

Throughout this paper, we assume a flat cosmology with $\mathrm{\Omega_M\, =\, 0.3,\, \Omega_{\Lambda}\, =\, 0.7}$, and a Hubble constant $\mathrm{H_0\, =\, 70\, km\,s^{-1} Mpc^{-1}}$. We use AB magnitudes \citep{1983ApJ...266..713O}. All stellar mass estimates assume a universal Chabrier form \citep{2003PASP..115..763C} stellar initial mass function (IMF), unless stated otherwise.


\section{Selection of \texorpdfstring{$\MakeLowercase{z}>16$}{z>16}  galaxy candidates}
\label{sec:selection}

\subsection{Data}
\label{sec:data}

In this paper, we combine the NIRCam data coming from MIDIS (PID 1283, \citealt{2024arXiv241119686O}), extended to MIDIS-RED (PID 6511, \"Ostlin et al. 2025, in prep.) with the observations obtained by NGDEEP (PID 2079, \citealt{2024ApJ...965L...6B}). These surveys observed a region located to the SE of the HUDF with a single NIRCam pointing and different position angles, each program eventually employing slightly different central coordinates  and orientations. For this work, we cut the full observed field of view to the area in common between all NIRCam filters (namely, F115W, F150W, F200W, F277W, F356W, and F444W), combining the regions covered by MIDIS and NGDEEP, which adds up to 17.6~arcmin$^2$ (i.e., around 10\% smaller than 2 NIRCam pointings). 

The raw data were downloaded from the Mikulski Archive for Space Telescopes (MAST) at the Space Telescope Science Institute, and calibrated with the Rainbow pipeline, a bespoke version of the official \texttt{jwst} pipeline which includes additional steps to homogenize the background and remove artifacts to be able to achieve the deepest flux levels. The core of the Rainbow pipeline for this paper resides in the \texttt{jwst} pipeline, version v1.16.0, pmap 1298 for the reduction used in this paper. As an addition to that official reduction, we also applied a superbackground homogenization algorithm prior to obtaining the final mosaics, which takes care of background features such as wisps and $1/f$-noise removal. As explained by \citet{2024ApJ...968....4P} and \citet{2024arXiv241119686O} for MIRI data, for each stage 2 image that needs to be calibrated, we build a median background frame with all the other images taken with the same detector (after masking objects), properly scaled to the median background of the image being reduced.  In order to improve the overall background characterization,
 we create extended masks of every source in the field using a \texttt{NoiseChisel} \citep{2015ApJS..220....1A, 2024A&A...684A..99G} run on initial mosaics obtained directly from the official pipeline, and improve the masks with two extra mosaic-building iterations.
 We customize our setup in order to mask the faintest sources, the diffuse light from galaxy outskirts, and the spikes of the brightest stars.

%
To optimize the results, and given that wisps changed within our dataset, we only used the images taken within 2 months of each other to build the superbackground frame, also disregarding the  short exposures (noisier than the rest) taken for each filter by the NGDEEP program (corresponding to when the primary instrument NIRISS was taking imaging data instead of deep grism observations). We found that the superbackground procedure improves the detectability of faint sources in the final mosaics by 0.3-0.4~mag, compared to a more classical background homogenization algorithm based on row, column, and box filtering.

The whole dataset was registered to the same World Coordinate System (WCS) reference frame using the Hubble Legacy Field (HLF) catalog of \citet{2019ApJS..244...16W}, based on Gaia DR1.2 (\citealt{2016A&A...595A...1G}, \citealt{2016A&A...595A...2G}). For that purpose, we used the version of the {\it tweakreg} package published by the CEERS Team \citep{2023ApJ...946L..12B}, switching this task off in stage 3 of the pipeline. All images were drizzled to the same plate scale, namely, 0.03 arcsec/pixel. All images were PSF-matched to the reddest filter (F444W), using empirical PSF images constructed with $\sim10$ stars found in the MIDIS+NGDEEP data. The FWHM of the PSFs were checked to be consistent within 10\% with those determined for NIRCam based on dedicated commissioning data \citep{2023PASP..135d8001R}. Kernels to transform from one band to another were constructed with a Fourier transform method in \texttt{photutils} and using a cosine-bell filter to avoid high frequency noise. 


The depths of the NIRCam dataset used in this paper reaches maximum magnitude $5\sigma$ limits for point-like sources of 30.9 to 31.5 in all filters for the region with maximum overlap, and using 0.1\arcsec\, circular apertures corrected for PSF effects (i.e., to total magnitudes assuming point-like sources). The depth in the area covered only by MIDIS or NGDEEP alone is shallower by up to 0.5~mag in some filters. Special mention must be made of the three reddest bands, all reaching magnitudes beyond 31.1 in the full considered area, essential to select F200W dropouts (the two reddest to select F277W dropouts) down to $\sim31$~mag, significantly improving previous observational efforts (see Figure~\ref{fig:depths}). These selection limits are still adequate, with our limiting magnitudes in the dropout band, to detect the intense breaks (1.0-1.5~mag for the mentioned magnitude limit), most probably corresponding to the Lyman break. In Appendix~\ref{app:noise}, we give more details about the limiting magnitudes of our dataset compared to those from other surveys.

The coverage of the MIDIS$+$NGDEEP region with HST is very limited ($<30\%$ of the area in most bands), and much shallower ($>4$~mag) for the wavelengths in common (i.e., WFC3 vs NIRCam bands). Noticeably, very deep data in the F814W band (the deepest on the sky, see \citealt{2019ApJS..244...16W} and \citealt{2023ApJ...951L...1P}) are available for nearly half of our field of view, but its $5\sigma$ limiting magnitude is $\sim30$~mag and did not provide any extra constraint for our faint sources, which are undetected in the 1--2~$\mu$m NIRCam bands reaching  $\sim31$~mag (and beyond for stacked images).

The actual areas, exposure times, and depths found in different regions of the combined field of view are given in Table~\ref{tab:data}. We note that MIDIS-red is supposed to further increase the exposure time in short-wavelength (SW) and long-wavelength (LW) filters in the late Fall of 2025. 

The depths of the data used in this work are shown in Figure~\ref{fig:depths}, compared to other surveys (see also Appendix~\ref{app:noise}) and presenting several samples of $z>11$ galaxies selected photometrically and also those confirmed spectroscopically. In this figure, as well as in Figure~\ref{fig:colorcolor}, we also present the newly discovered $z>16$ galaxy candidates reported in this paper and a comparison with magnitudes and colors provided by representative stellar population models. 

\begin{deluxetable}{lccc}[ht]
\caption{Characteristics of the MIDIS$+$NGDEEP NIRCam datasets.}
\tabletypesize{\scriptsize}
\tablehead{\colhead{Filter} &  \colhead{$5\sigma$ depth (time)} &  \colhead{$5\sigma$ depth (time)} &  \colhead{$5\sigma$ depth (time)}\\
{}&MIDIS$+$NGDEEP & MIDIS & NGDEEP}
\startdata
F115W  & 31.1 (226~ks) & 30.6 (55~ks) & 31.0 (171~ks)\\
F150W  & 30.9 (138~ks) & 30.4 (58~ks) & 30.7 (80~ks)\\
F200W  & 31.3 (197~ks) & 31.3 (166~ks) & 30.6 (31~ks)\\
F277W  & 31.5 (135~ks) & 31.5 (55~ks) & 31.2 (80~ks)\\
F356W  & 31.5 (119~ks) & 31.3 (58~ks) & 31.1 (62~ks)\\
F444W  & 31.5 (321~ks) & 31.3 (166~ks) & 31.3 (155~ks)\\
  \enddata
\tablecomments{\label{tab:data}Table with information for all filters used in this work. We show the 5$\sigma$ depths corresponding to a point-like source measured in a 0.2\arcsec\, diameter circular aperture and corrected for the limited aperture using empirical PSFs. The three columns stand for the depth of the MIDIS+NGDEEP, MIDIS-only, and NGDEEP-only areas, extending for 2.8, 3.4, and 11.4 arcmin$^2$, respectively, for a total area considered in this paper of 17.6~arcmin$^2$.}
\end{deluxetable}

\subsection{Source detection and photometry}

Source detection was carried out with Source Extractor \citep{1996A&AS..117..393B}, executed on cold and hot modes to properly deal with the brightest extended galaxies as well as the faintest small sources \citep{2004ApJS..152..163R}. Detection was carried out in the F277W, F356W, and F444W filters, as well as in the stack of the three bands and stacks combining two of them, all obtained with constant weights. The different catalogs were cross-correlated with each other and a master list of sources was constructed by keeping all individual detections separated by more than 0.15\arcsec\, from any other source. The parent catalog is composed of 110,125 sources, detected in 17.6~arcmin$^2$.

Integrated photometry was measured for the parent catalog in \citet{1980ApJS...43..305K} apertures, assuming Kron factors of 2.5 for the cold SExtractor execution, and 1.1 for the hot one. Aperture photometry was also measured in circular apertures of diameter 0.2\arcsec, 0.3\arcsec, and 0.4\arcsec, applying aperture corrections for point-like sources, and scaling the final photometry with a constant factor obtained by averaging the flux ratios with respect to the the Kron aperture in a source by source basis. In this paper, given that we are interested in very high redshift faint sources, we used the 0.2\arcsec\, diameter aperture photometry for our fiducial spectral energy distributions, but evaluated the results with all four different photometric values (i.e., the direct Kron aperture and the three circular aperture photometries). 

Uncertainties in the flux measurements for each source and band were calculated by measuring the background noise locally with a procedure presented by \citet{2008ApJ...675..234P} and also used by \citet[][where a more quantitative characterization of the method is presented]{2023ApJ...951L...1P}, devised to take into account correlated noise. Briefly, we built artificial apertures composed of randomly-selected non-contiguous pixels (in fact, separated by more than 3 pixels from any other entering the statistics calculation) in the neighborhood of each source, adding up the number of pixels of the photometric aperture whose uncertainty we were trying to estimate. The whole procedure ensures that all the pixels included in the noise calculation are independent, and not affected by correlated noise. 

Given that a proper estimation of the noise and related error calculation is essential for the detection of $z>16$ galaxy candidates, we provide a thorough evaluation of the error calculation procedure in Appendix~\ref{app:noise}.

\subsection{Selection method}

\begin{deluxetable*}{llcccccccc}
\caption{Galaxy candidates at $16<z<25$ selected with the MIDIS$+$NGDEEP dataset.}
\centerwidetable 
\tabletypesize{\scriptsize}
\tablehead{\colhead{MIDIS ID} & \colhead{Galaxy name} & \colhead{RA (J2000)} & \colhead{DEC (J2000)} & \colhead{F115W} & \colhead{F150W} & \colhead{F200W} & \colhead{F277W} & \colhead{F356W} & \colhead{F444W}\\
& & [degrees] & [degrees] & [mag] & [mag] & [mag] & [mag] & [mag] & [mag]}
\startdata
\hline
\hline
\multicolumn{10}{c}{$z\sim17$ sample}\\
\hline
midisred0006800 & MIDIS-z17-1 & 53.28551340 & $-$27.87947000 & $>32.1$& $>32.2$& $>32.7$& $31.06\pm0.15$& $31.26\pm0.16$& $31.19\pm0.16$ \\
midisred0019207 & MIDIS-z17-2 & 53.27974630 & $-$27.86530520 & $>32.4$& $>31.7$& $>32.4$& $31.11\pm0.16$& $31.84\pm0.29$& $31.52\pm0.19$ \\
midisred0028033 & MIDIS-z17-3 & 53.30680240 & $-$27.85571070 & $>31.1$& $>31.4$& $>33.0$& $29.22\pm0.05$& $30.12\pm0.07$& $31.41\pm0.14$ \\
midisred0047123 & MIDIS-z17-4 & 53.23560300 & $-$27.81536380 & $>32.7$& $>32.3$& $>31.6$& $30.45\pm0.11$& $30.60\pm0.12$& $30.73\pm0.16$ \\
midisred0051552 & MIDIS-z17-5 & 53.25235830 & $-$27.80295390 & $>32.2$& $>31.6$& $>31.6$& $30.77\pm0.13$& $30.76\pm0.18$& $31.01\pm0.18$ \\
midisred0053786 & MIDIS-z17-6 & 53.24822550 & $-$27.79413320 & $>32.4$& $>30.3$& $>30.4$& $29.58\pm0.06$& $29.89\pm0.15$& $29.63\pm0.07$ \\
\hline
\hline
\multicolumn{10}{c}{$z\sim24$ sample}\\
\hline
midisred0019248 & MIDIS-z25-1 & 53.23226140 & $-$27.86526940 & $>32.6$& $>32.3$& $>33.0$& $>32.4$& $30.48\pm0.10$& $30.60\pm0.25$ \\
midisred0048457 & MIDIS-z25-2 & 53.23122790 & $-$27.81162440 & $>32.5$& $>31.9$& $>32.3$& $>33.0$& $31.14\pm0.16$& $31.63\pm0.29$ \\
midisred0049147 & MIDIS-z25-3 & 53.23929800 & $-$27.80989420 & $>33.0$& $>31.6$& $>32.3$& $>33.0$& $31.07\pm0.11$& $31.12\pm0.14$ \\
\enddata
\tablecomments{\label{tab:sample}We provide coordinates and magnitudes in all NIRCam bands, identifying each galaxy by its MIDIS parent catalog ID and the galaxy name chosen for this paper about $z>16$ galaxy candidates. Upper limits refer to 1-$\sigma$.}
\end{deluxetable*}

\begin{figure*}[ht!]
\centering
\includegraphics[clip, trim=3.1cm 1.0cm 3.2cm 0.0cm,scale=0.60]{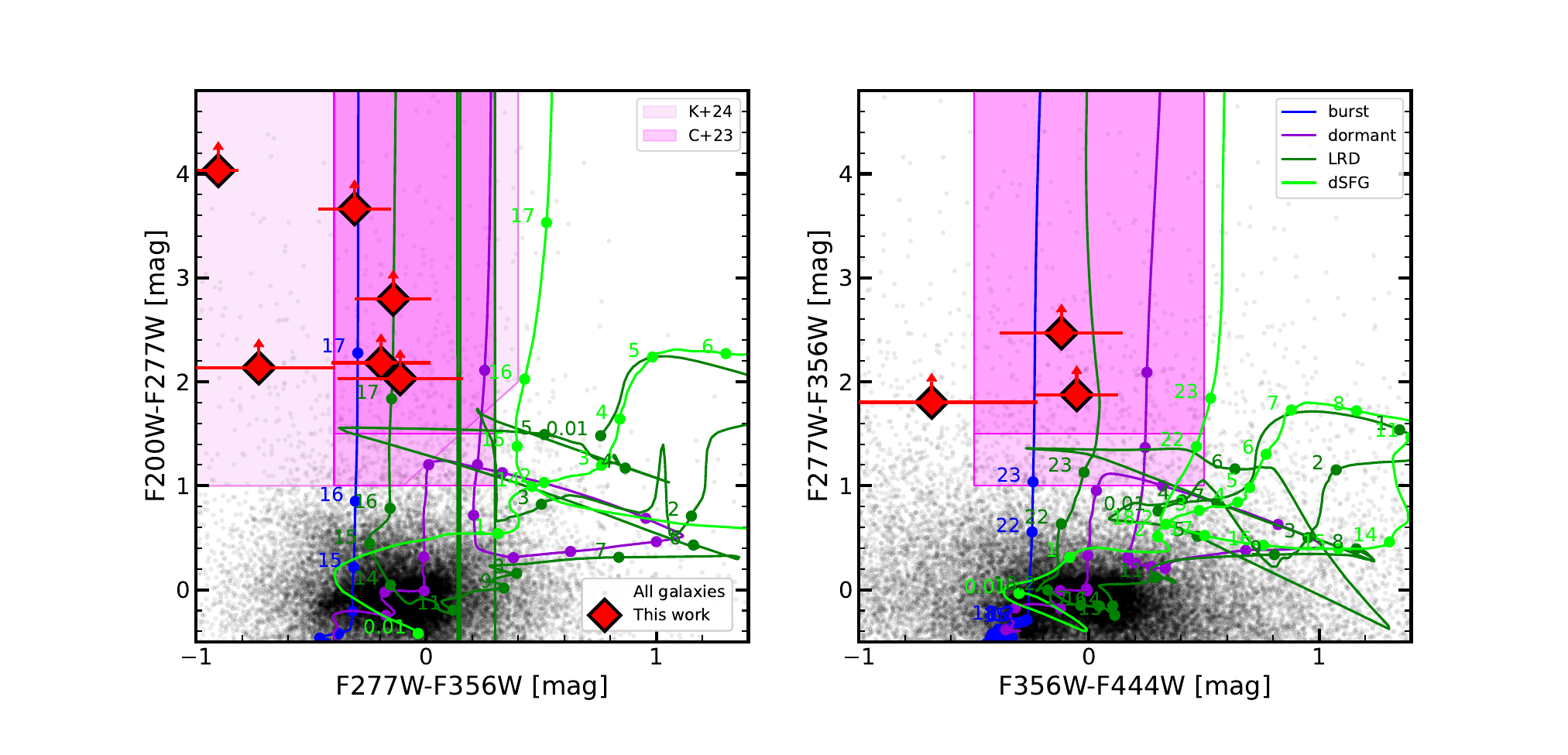}
\caption{\label{fig:colorcolor}Color-color diagrams characterizing the selection of F200W and F277W dropout galaxies. In each panel, our parent sample is shown with black dots. We remark that our sources have been selected by pre-identifying breaks, calculating photometric redshifts with prominent $z>10$ solutions, and visual vetting, i.e., our selection is not based directly on these color-color diagrams, but it is closely related to them. The $z\sim17$ and $z\sim25$ galaxy candidates are shown with red diamonds on the left and right panels, respectively. The color probing the Lyman break for these candidates (i.e., F200W-F277W and F277W-F356W) can only be measured as a lower limit (using 1-$\sigma$ detection thresholds for point-like sources), since we imposed a non-detection in the bluest band. In blue, we show the typical colors expected for a very young starburst as a function of redshift (representative values are marked with circles and labelled in the plots). In purple, we show the colors for a dormant galaxy, defined as a system that experienced a star-forming event 30~Myr ago. In light and dark green, we show the colors expected for typical interlopers in the selection of $z\gtrsim10$ galaxies, namely a little red dot template from \citet{2023arXiv231203065K} and a dusty star-forming galaxy model from \citet{2023ApJ...946L..16P}. These redshift tracks show that $z\gtrsim16$ galaxies would present F200W-F277W colors redder than 1~mag and F277W-F356W colors bluer than 0.4~mag, a locus that has been proposed by authors such as \citet[][also in Castellano et al. 2025, in prep, private communication]{2022ApJ...938L..15C,2023ApJ...948L..14C}  and \citet{2024arXiv241113640K} to select galaxies at these redshifts (regions shown in panels). The right panel shows similar behavior for $z\gtrsim23$ galaxies whose Lyman break is probed by the F277W-F356W color, while the F356W-F444W color probes the rest-frame UV spectral range.}
\end{figure*}

\begin{figure*}[ht!]
\centering
\includegraphics[clip, trim=1.5cm 5.0cm 2.0cm 1.5cm,scale=0.59]{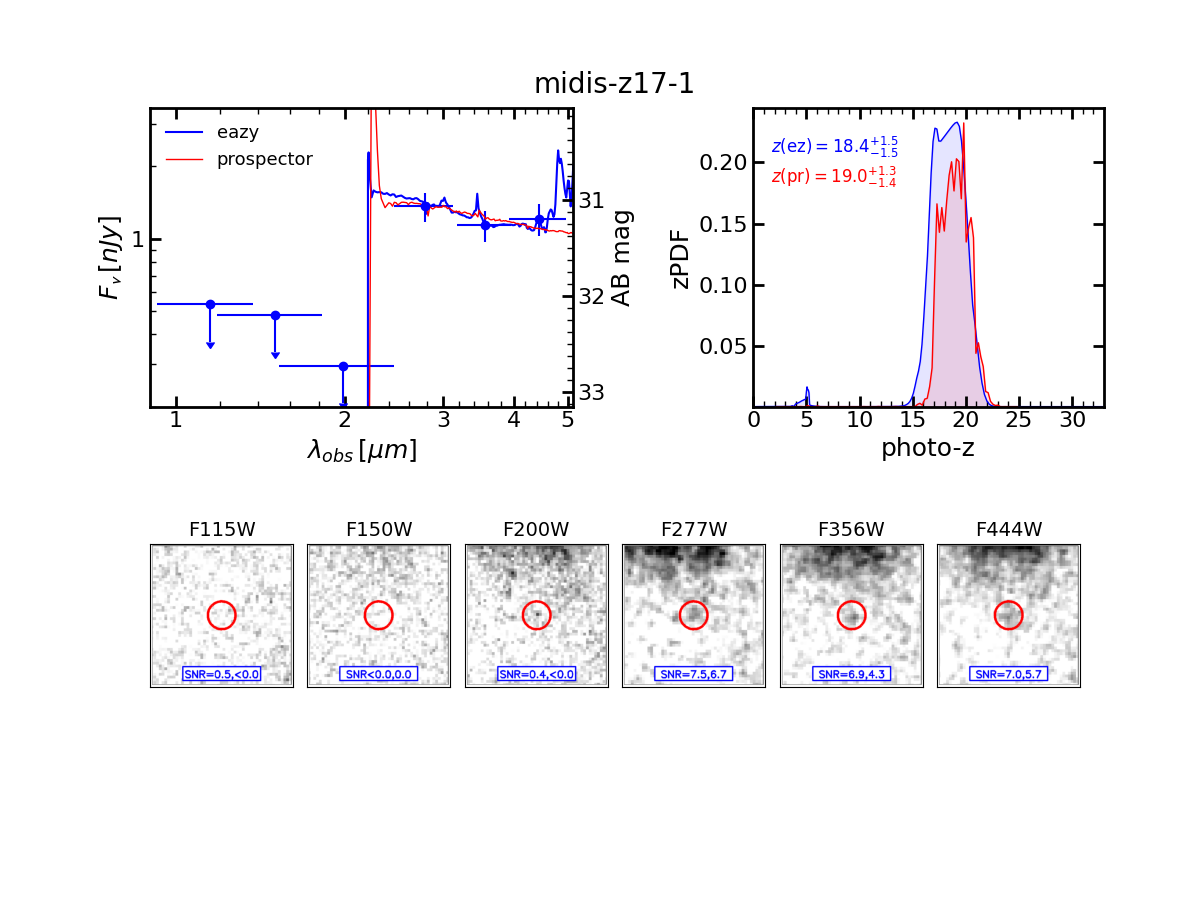}
\includegraphics[clip, trim=1.5cm 5.0cm 2.0cm 1.5cm,scale=0.59]{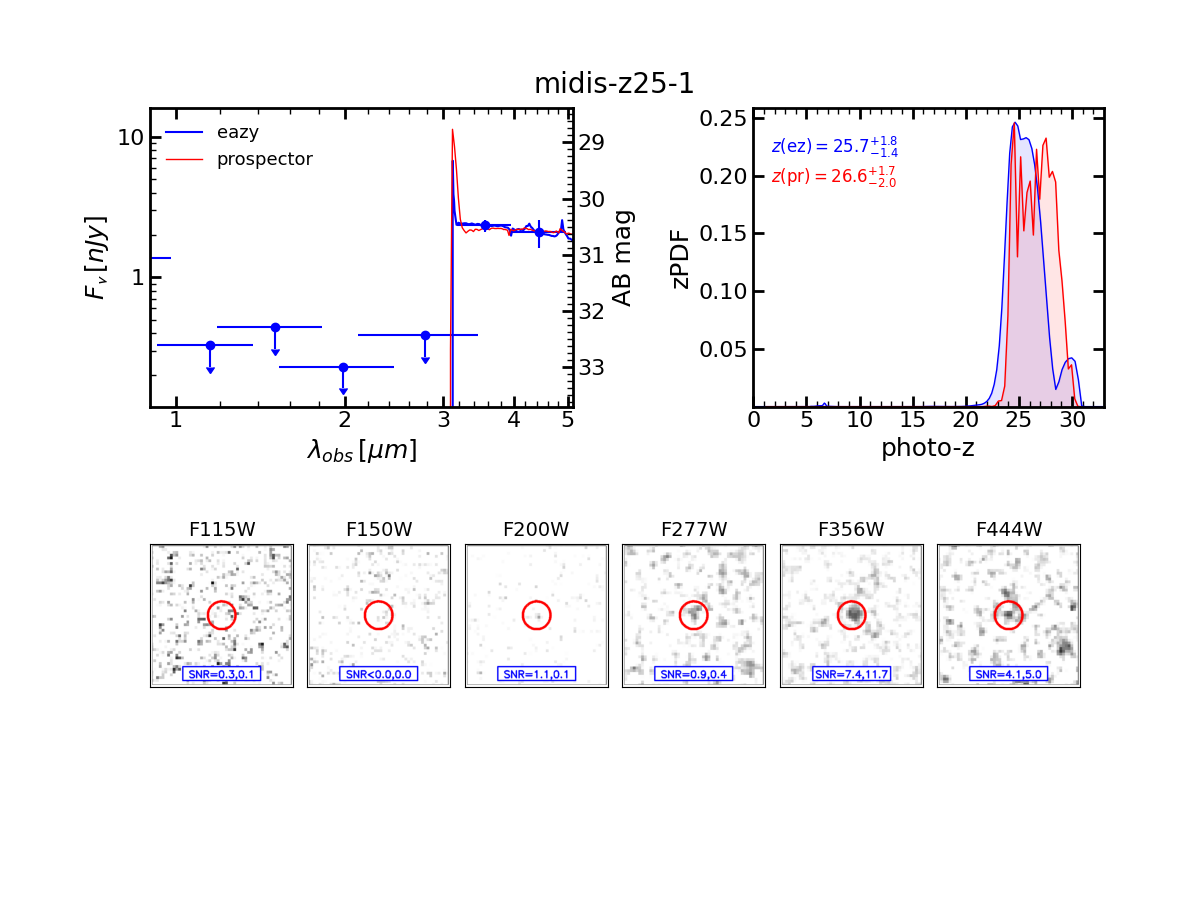}
\caption{\label{fig:sed_examples}Examples of the galaxies selected at $z\sim17$ and $z\sim25$ in our MIDIS$+$NGDEEP survey. For each galaxy, we provide the SED measured in a circular 0.2\arcsec\, diameter aperture (blue circles), and the template fits provided by the photometric redshift codes \texttt{eazy} and \texttt{prospector}. Arrows indicate $1\sigma$ upper limits. The top right panel shows the photometric redshift probability distribution function for the two codes. The bottom of the figure for each source presents $1.5\arcsec\times1.5\arcsec$ postage stamps in all NIRCam bands, with the source marked with a 0.30\arcsec\, diameter red circle. Intensity scales have been tweaked for each band to enhance the faintest signals, using a gray scale from white corresponding to the median background around the source to black for $3\sigma$. For each band, we quote the signal-to-noise ratio in 0.2\arcsec\, and 0.3\arcsec\, diameter apertures (negative values corresponding to a negative flux measurement).}
\end{figure*}

The selection of high redshift galaxy candidates follows the same method used by \citet{2023ApJ...946L..16P}, adopted from other works such as \citet{2023ApJ...946L..13F} and  \citet{2023ApJS..265....5H}. Taking into account  the faint nature expected for the $z>16$ galaxy candidates, as well as random effects that might be introduced by the arbitrary selection of a given photometric aperture, our selection method uses several spectral energy distributions (SEDs) for each source, constructed with the photometry  derived from the Kron aperture (whose average diameter is 0.32\arcsec\, for our candidates) and the three circular apertures mentioned in the previous section (with diameters $d=0.2,0.3,0.4\arcsec$). We note that all selection criteria must be met by all those four SEDs, which is a conservative approach since random variations might affect some of the apertures. We note that, typically, photometric apertures used in this kind of studies searching for very high redshift galaxy candidates have sizes of $d=0.2\arcsec$ (e.g., \citealt{2024ApJ...970...31R},  \citealt{2024arXiv241113640K}, and \citealt{2023ApJS..265....5H} -who also use $d=0.2\arcsec$ apertures-), $\sim0.3\arcsec$ (e.g., \citealt{2023ApJ...952L...7A}, \citealt{2024arXiv240714973C}, \citealt{2025ApJ...978...89H}, \citealt{2024MNRAS.527.5004M}, or \citealt{2023MNRAS.518.6011D} -who, in fact, uses different apertures for the SW and LW bands-), or Kron (with small Kron factors, e.g., \citealt{2023ApJ...946L..13F}, \citealt{2023ApJ...954L..46L}).

First, we selected all galaxies whose median flux across all four spectral energy distributions (SEDs), i.e., the SEDs derived from the Kron aperture and the three circular apertures, and using the F277W, F356W, and F444W filters, had a signal-to-noise ratio $\mathrm{SNR}>5$. This sample has an median and quartiles F444W magnitudes of $30.3\pm1.4$. Then, in order to select dropouts and to make them compatible with $z>16$, which would mean zero flux blue-wards of the Lyman break, we restricted the catalog to all sources with a median $\mathrm{SNR}<3$ in F115W and F150W, as well as the actual dropout band (F200W or F277W), resulting in a sample  with average F444W magnitudes $31.3\pm0.6$ and $31.4\pm0.7$ for the F200W and F277W dropouts (5633 and 386 sources, respectively), respectively. 

After the observational cuts, we analyzed the SEDs of the dropout galaxies to obtain photometric redshifts and other physical properties. Two independent codes were used for this purpose. First, we used the \texttt{eazy} code \citep{2008ApJ...686.1503B} with the default FSPS templates \citep{2010ascl.soft10043C}, adding a dusty galaxy template \citep{2013ApJS..206....8M} and the templates presented by \citet{2022arXiv221110035L} to optimize the analysis of high-redshift galaxies, which include emission-line galaxies with high equivalent widths (known to contaminate {\em JWST}\, high-redshift samples, see \citealt{2023ApJ...943L...9Z} and \citealt{2022arXiv220802794N}). We also used two templates with extreme equivalent width emission lines, one presented by \citet{2010ApJ...719.1168E}, the other based on the most extreme galaxy identified by \citet{2023ApJ...945...35T}. Finally, new obscured Active Galactic Nucleus (AGN) and stellar$+$AGN little red dot (LRD) templates from \citet{2023arXiv231203065K} and \citet{2024ApJ...968....4P} were also added. We did not use any magnitude prior nor a template error, allowing the redshift to take values between $z=0$ and $z=35$.

Photometric redshifts were also estimated with \texttt{prospector} \citep{2021ApJS..254...22J}, using FSPS templates with python bindings \citep{2010ascl.soft10043C,2014zndo.....12157F} and the \texttt{dynesty} nested sampler \citep{2020MNRAS.493.3132S}. For this fitting, we fitted for four free parameters: redshift $z$, stellar mass M$_{\star}$, dust attenuation E(B-V), and stellar population age. We assumed a parametric star formation history (delayed $\tau$ model), and intergalactic medium (IGM) absorption turned on. The stellar (and gas) metallicity was fixed at 0.03 $Z_{\odot}$ and all other parameters fixed at their FSPS default values (e.g., the SFH timescale was set to an e-folding time of $\tau = 1$ Gyr). We used uniform priors on $z\in[2.0,31.0]$, $\log{\mathrm{M}_{\star}}\in[6,10]$, stellar age between 1 Myr and age of the Universe at the given $z$, and a prior on $E(B-V)$ in the shape of a normal distribution centered and truncated at 0.0 with a standard deviation of 0.3.

We obtained  photometric redshift probability distribution functions (zPDFs) with both codes using the SEDs for the four different apertures mentioned in Section~\ref{sec:data}. The zPDFs were then analyzed to build a sample of $z\gtrsim16$ galaxy candidates.

Based on photometric redshifts, we further cut the original sample to those galaxies presenting a most probable (as found by integrating the zPDF) and the peak (that providing the smallest $\chi^2$ value) redshift above $z>10$, keeping only those galaxies with a $>70$\% cumulative probability of lying at $z>15$. The final cut was on $\chi^2$ values, ensuring that the difference between the minimum value of the zPDF at $z<10$ and $z>10$, $\Delta\log\chi^2$, was larger than 0.4~dex. In the final sample that we used to estimate luminosity functions (see the following section), we only kept candidates for which all apertures and the two codes provided photometric results compatible with $z>10$. The number of selected sources after these cuts was 733.



We note that most of the templates and models used by our two photometric redshift fitting codes present Ly$\alpha$ emission (see, e.g., Figure~\ref{fig:sed_examples}), which is unexpected at the high redshifts we are probing, well beyond the epoch of reionization. We remark, however, that Ly$\alpha$ emission has been observed  up to $z\sim13$ \citep[see][and references therein]{2025Natur.639..897W}. In any case, we tested the effect by building a new template set for \texttt{eazy} with no Ly$\alpha$ emission. Small differences (well within uncertainties) were found with respect to our fiducial estimation of photometric redshifts given the low EW of the line and the very small effect on the broad-band filter photometry in the 2--3~$\mu$m spectral range. On average for all galaxies with a peak at $15<z<30$, we found a difference of $\Delta z=0.2$ (to be compared with the typical uncertainties: $\delta z\sim\pm1.5$) between the estimates with models including and excluding Ly$\alpha$ emission. The redshift for the former were found to be lower, a fact that we interpret as a consequence of needing a higher redshift to reproduce the large observed flux ratio between the bands enclosing the Ly$\alpha$ jump in the absence of line emission.

This sample of 733 sources selected with observational and physical criteria  were also inspected visually. We vetted galaxies unaffected by artifacts such as star spikes and/or contamination by nearby objects (695 were removed due to these effects), as well as sources for which the individual bands and the stacks of the SW data (or also adding F277W for the $z\sim25$ candidates) provided $\mathrm{SNR}>2$ measurements in any of the photometric apertures (29 sources were removed due to this check).


Our final sample is composed of six F200W dropouts and three F277W dropouts, whose typical photometric redshifts are $z\sim17$ and $z\sim25$, respectively. Four of the nine candidates are located in the MIDIS area, and the others in the NGDEEP region. The sample is presented in Table~\ref{tab:sample}. 

\subsection{Candidates}

In this section, we present the candidates identified with our selection method that survived the vetting process, both procedures described in the previous subsection. 

Figure~\ref{fig:depths} presented this sample (green and orange diamonds). The bulk of our sample has F444W magnitudes between 30 and 31.5 mag, a brightness level that lies more than 1 magnitude beyond the depths achieved by prior medium area shallow surveys such as CEERS. The magnitudes are comparable to the 5$\sigma$ depths obtained by the deepest surveys carried out in 2022-2024, such as JADES or NGDEEP. The two brightest objects in our sample are comparable (but still slightly fainter) to the $z\sim17$ candidates presented by \citet{2024arXiv241113640K} using GLIMPSE data, the $15<z<20$ candidates in \citet{2025arXiv250405893C}, and with several spectroscopically confirmed $z\gtrsim13$ galaxies (see Fig.~1~caption for references). 

Comparing the magnitudes with the models depicted in Figure~\ref{fig:depths}, we infer that the stellar masses for our $z>16$ galaxy candidates are nearly an order of magnitude smaller than the average of the spectroscopically confirmed galaxies at $11<z<14.5$ ($\sim10^9$\,M$_\odot$), i.e., their stellar masses should be around $10^{7-8}$\,M$_\odot$ (where the BPASS simple stellar population models suggest an order of magnitude smaller stellar masses than BC03). We expect 0.3-0.4~dex variations depending on the star formation history. We will discuss the stellar mass estimates of our selected galaxies further in Section~\ref{sec:physical}.

Figure \ref{fig:colorcolor} shows color-color plots for the parent sample and our $z\sim17$ and $z\sim25$ galaxy candidates. We remark that our selection is based on break detection and photometric redshifts, not directly colors, but we compare with color-color criteria in this figure. We plot the color tracks as a function of redshift for various galaxy types previously detected by JWST at high redshift, including a template for a galaxy with a very recent starburst (1~Myr old); a dormant galaxy with a 30~Myr old star formation event; an LRD; and a dusty star-forming galaxy, which is the typical interloper in the search for $z>8$ galaxy candidates (see \citealt{2023ApJ...943L...9Z},  \citealt{2022arXiv220802794N}, or \citealt{2025arXiv250202637G}). All our $z\gtrsim16$ galaxy candidates have colors redder than 1.5~mag, implying the presence of a strong break, most probably identified with the Lyman break rather than the Balmer break, as shown by the selection locus presented by \citet{2023ApJ...948L..14C}  and \citet{2024arXiv241113640K} and marked in the figure. Remarkably, we detect two F200W dropout galaxies and one F277W with very blue colors in the bands probing the far-UV spectral region. We further discuss UV slopes in Section~\ref{sec:physical}. No interlopers of Galactic origin such as brown dwarfs are expected  in our sample, since typically those sources have red F356W-F444W colors (see, e.g., \citealt{2023ApJ...951L..48B} and \citealt{2024ApJ...964...66H}). We remark that there is a large number of sources (around 1500) whose colors are within the $z>16$ galaxy locus. Around 70\% of them do not comply with the $SNR>5$ cut in the bands redward of the break. Out of the rest, half of the sources are located near bright objects or directly in star spikes that contaminate their photometry in one or several bands. All the rest have been rejected from our final sample based mainly on the photometric redshift quality and the visual vetting. 

Figure~\ref{fig:sed_examples} presents the SEDs and postage stamps of one of the $z\sim17$ and one of the $z\sim25$ galaxy candidates in our sample. Similar figures for the rest of the candidates are provided in Appendix~\ref{appA}. Most of the sources show a secondary redshift probability peak at $5\lesssim z\lesssim7$. This peak is always obtained with models corresponding to a red source with prominent emission lines, a combination of the extreme templates which we added to the \texttt{eazy} run (see description of templates above). We remark that the \texttt{prospector} execution is biased against this type of distinct templates, which include AGN features not considered by \texttt{prospector}. This translates to less pronounced low-redshift solutions. On the other hand, the \texttt{prospector} photometric redshift estimation covers a wider range of stellar population properties (including, e.g., ages, timescales, or dust attenuation properties) compared to the more limited set of star-dominated templates used by \texttt{eazy}.

The observational properties of our main sample are given in Table~\ref{tab:sample}. In Appendix~\ref{appB}, we provide a table with a few sources that lie on the edge of our identification limits for all selection criteria except one;  they are dubious cases, and were not used in the calculations of the luminosity function presented in the following section.

\section{Results}
\label{sec:results}

\begin{figure*}[ht!]
\centering
\includegraphics[clip, trim=2.0cm 1.0cm 2.0cm 0.0cm,scale=0.68]{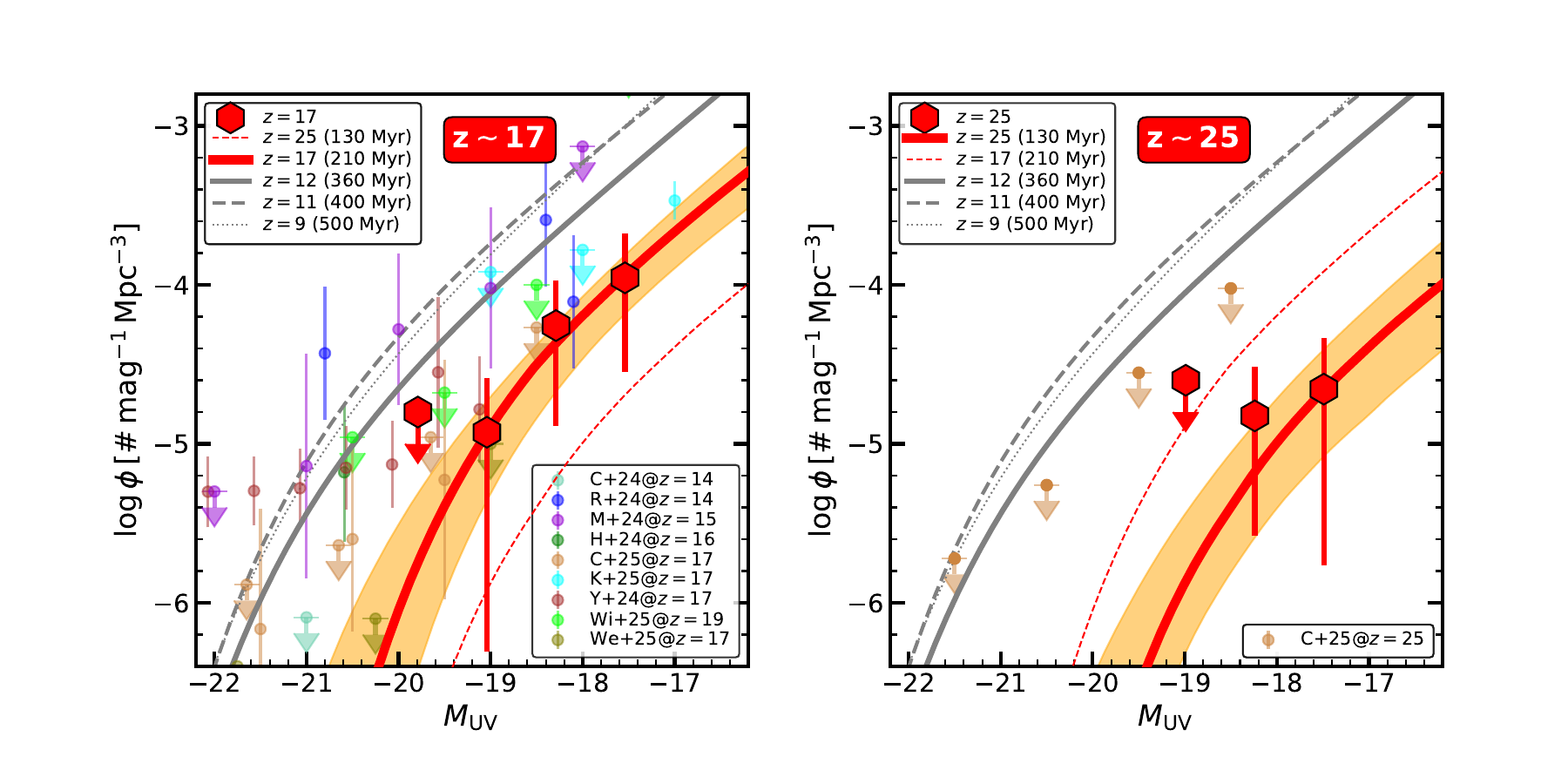}
\caption{\label{fig:lf}Luminosity functions at $z\sim17$ (left panel)  and $z\sim25$ (right panel) estimated in this paper (red diamonds) and \citet{1976ApJ...203..297S} fits (red line and $1-\sigma$ uncertainties in orange). On the left, symbols show literature estimates from \citet[mediumaquamarine]{2024ApJ...965...98C}, \citet[blue]{2024ApJ...970...31R}, \citet[violet]{2024arXiv241204211M}, \citet[green]{2024ApJ...960...56H}, \citet[peru]{2025arXiv250405893C}, \citet[cyan]{2024arXiv241113640K}, \citet[brown]{2023arXiv231115121Y},  \citet[lime]{2025arXiv250100984W}, and \citet[olive]{2025arXiv250706292W}. On the right panel, we show the upper limits at $20<z<30$ in \citet{2025arXiv250405893C}. The plots also show in gray the Schechter fits to the luminosity functions presented in \citet{2023ApJ...951L...1P} for $z=9,11,12$ (dashed, dotted, and continuous line, respectively), as well as the results from this paper for the other redshift bin (red dashed line, for $z\sim25$ on the left panel, for $z\sim17$ on the right one). The age of the Universe for the redshifts of all these luminosity functions are given in the legend.}
\end{figure*}

\subsection{UV luminosity functions}
\label{sec:lf}

\begin{figure*}[ht!]
\centering
\includegraphics[clip, trim=0.7cm 1.0cm 0.0cm 0.0cm,scale=0.5]{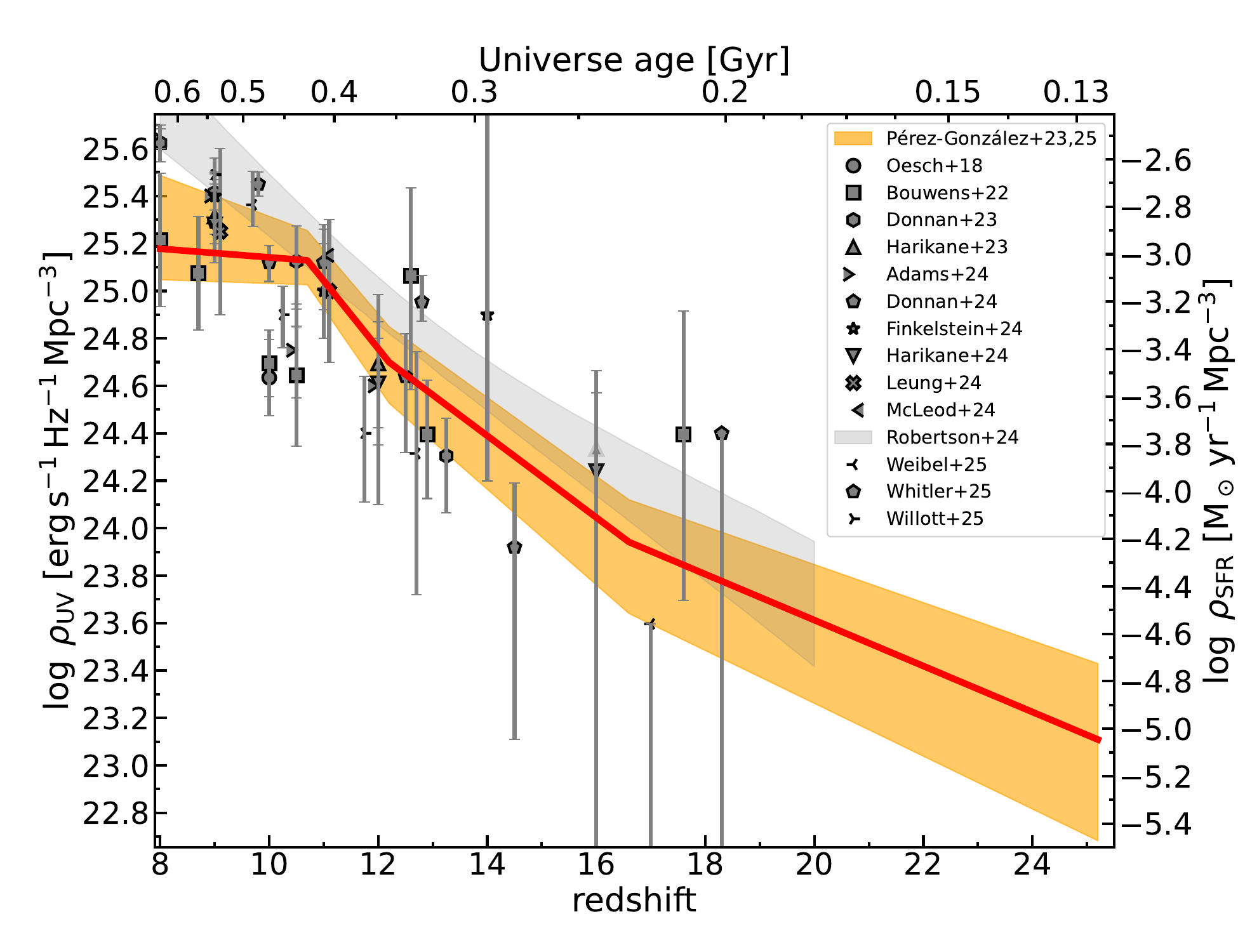}
\caption{\label{fig:sfrd}Evolution with redshift (age of the Universe given on top axis) of the UV luminosity density, transformed into SFR density on the right vertical axis (using the factor provided by \citealt{2014ARA&A..52..415M}, converted to a \citealt{2003PASP..115..763C} IMF; see main text for details). Our results are plotted with a red line joining the most probable values and orange shaded region for $1-\sigma$ errors. Estimates from the literature are plotted with gray symbols \citep{2018ApJ...855..105O,2022arXiv221206683B,2022arXiv221102607B,2023MNRAS.518.6011D,2024MNRAS.533.3222D,2023ApJS..265....5H,2024ApJ...965..169A,2024ApJ...969L...2F,2024ApJ...960...56H,2023ApJ...954L..46L,2024MNRAS.527.5004M,2024ApJ...970...31R,2025arXiv250706292W,2025arXiv250100984W,2024ApJ...966...74W}.}
\end{figure*}

\begin{deluxetable*}{crccrc}
\caption{\label{tab:lf}Luminosity function data points.}
\centerwidetable 
\tablehead{\colhead{$\mathrm{M}_\mathrm{UV}$} & \colhead{$\log\phi$} & Volume & \colhead{$\mathrm{M}_\mathrm{UV}$} & \colhead{$\log\phi$} & Volume\\
\colhead{[AB mag]} & \colhead{[Mpc$^{-3}$~mag$^{-1}$]} & \colhead{[$10^4$~Mpc$^3$]}  & \colhead{[AB mag]} & \colhead{[Mpc$^{-3}$~mag$^{-1}$]} & \colhead{[$10^4$~Mpc$^3$]} }
\startdata
\multicolumn{3}{c}{$z\sim17$} & \multicolumn{3}{c}{$z\sim25$} \\
$-19.79$ & $<-4.8$ & 7.8 & $-18.99$ & $<-4.6$ & 5.5\\
$-19.04$ & $-4.93^{+0.34}_{-1.38}$ & 7.8 & $-18.24$ & $-4.82^{+0.31}_{-0.75}$ & 4.7\\
$-18.29$ & $-4.26^{+0.28}_{-0.63}$ & 3.1 & $-17.49$ & $-4.66^{+0.33}_{-1.11}$ & 1.8\\
$-17.54$ & $-3.96^{+0.28}_{-0.58}$ & 1.3 & & & \\ 
\enddata
\end{deluxetable*}

\begin{deluxetable*}{lcc}
\caption{\label{tab:schechter}Luminosity function parameters.}
\tablehead{\colhead{Parameter} & \colhead{$z\sim17$} & \colhead{$z\sim25$}}
\startdata
$\alpha$         & $-2.12^{+0.15}_{-0.09}$  & $-2.12^{+0.13}_{-0.22}$ \\
$M^*$ [AB mag]   & $-19.0 ^{+1.0}_{-0.5}$ & $-18.5 ^{+0.8}_{-0.4}$\\
$\phi^*$ [$10^{-5}$~Mpc$^{-3}$~mag$^{-1}$] & $3.55 ^{+0.49}_{-0.15}$      & $1.22 ^{+0.58}_{-0.28}$ \\
$\log\rho_\mathrm{UV}$ [10$^{24}$~erg\,s$^{-1}$\,Hz$^{-1}$\,Mpc$^{-3}$] & $0.89^{+0.45}_{-0.49}$ & $0.13^{+0.14}_{-0.08}$ \\
\enddata
\tablecomments{Results for the \citet{1976ApJ...203..297S} parametrization fits to the luminosity functions at $z\sim17$ and $z\sim24$ presented in Fig.~\ref{fig:lf}. The last row shows the integrated luminosity for absolute magnitudes $M_\mathrm{UV}<-17$~mag.}
\end{deluxetable*}

To estimate luminosity functions, we calculate the absolute UV magnitudes ($M_\mathrm{UV}$, averaged in a 0.01~$\mu$m window around 0.15~$\mu$m) and their errors from SED fitting, considering both the photometric and redshift uncertainties. No dust correction has been applied. The luminosity functions were constructed using the $V_\mathrm{max}$ method \citep{1988MNRAS.232..431E}. Completeness was estimated in two different ways. The first one consisted in inserting artificial sources covering the magnitude range between 29 and 33~mag, and rerunning our detection algorithm. We find that our catalog is 80\% complete at $\mathrm{F356W}=31.0$~mag, 50\% at 31.4~mag, and 10\% at 31.6~mag. The second way is simulating multi-band photometry for sources of different magnitudes, with SEDs following the best-fitting templates (for low and high redshift solutions) provided by our \texttt{eazy} runs, and adding the characteristic noise of our dataset. Those SEDs were passed through the photometric redshift selection procedure presented in Section~\ref{sec:selection}. With this method, we found that the high redshift galaxy selection is $>80\%$ efficient at $16<z<20$ for  the F200W dropout sample, and $24<z<26$ for the F277W dropout sample. This method was also used to check the estimated survey effective volumes calculated with the $V_\mathrm{max}$ method (and provided in Table~\ref{tab:lf}).

In addition to errors calculated with a Monte Carlo method using the absolute magnitude and photometric redshift uncertainties, we took into account cosmic variance effects by adding in quadrature values calculated by \citet[][see also \citealt{2008ApJ...676..767T,2020MNRAS.496..754B,2020MNRAS.499.2401T}]{2021MNRAS.506..202U}. We estimate a 35--45\% (30--40\%) uncertainty in our luminosity function estimates at the bright (faint) ends due to field-to-field variations.

We show our luminosity functions at $z\sim17$ and $z\sim25$ in Figure~\ref{fig:lf}. Our results are given in Table~\ref{tab:lf}. We chose 0.75~mag luminosity bins to maximize the SNR of each number density point, typically including two objects. Central values are governed by the magnitudes of our candidates and the Monte Carlo method used to estimate the number densities. We fit the data points with \citet{1976ApJ...203..297S} functions without any prior using an MCMC method. The results are given in Table~\ref{tab:schechter}. 

Our estimate of the luminosity function is compared with others in the literature for the same absolute magnitude range $-19<M_{UV}<-18$ and similar redshifts from $z\sim14$ to $z\sim19$. Our number densities are 0.1-0.5~dex smaller than direct estimates taken from those  works. They are also consistent with the upper limits quoted by \citet{2024arXiv241204211M} and \citet{2025arXiv250100984W}. Our calculations have also been compared with those obtained by \citet{2025arXiv250405893C} for the same absolute magnitude range. The number densities estimated in our work are consistent with their upper limits, obtained by analyzing a number of deep fields adding up an area nearly 40 times larger than ours, but characterized by shallower depths.

Comparing with luminosity functions estimated at $z=9-12$, exemplified in Figure~\ref{fig:lf} with those presented by \citet{2023ApJ...951L...1P}, we find a $\times4$ number density evolution from $z\sim17$ to $z\sim12$ (corresponding to a 150~Myr Universe age difference), and $\times6$ from $z\sim25$ to $\sim17$ (in 80~Myr), implying rapid build-up of stars starting some time after the first 100~Myr of cosmic time. The dispersion in the luminosity function estimates at $z\sim12$ is 0.2~dex (see \citealt{2024MNRAS.533.3222D} for a recent compilation), so the factors quoted above for our luminosity function evolution have relative errors of $\sim50$\%. The dispersion in estimates at $z\sim17$ in the absolute magnitude range covered by our survey is larger, 0.4~dex, with some works obtaining much smaller evolution from $z\sim12$ to $z\sim17$.

\begin{figure*}[ht!]
\centering
\includegraphics[clip, trim=0.4cm 0.3cm 0.0cm 0.0cm,scale=0.4]{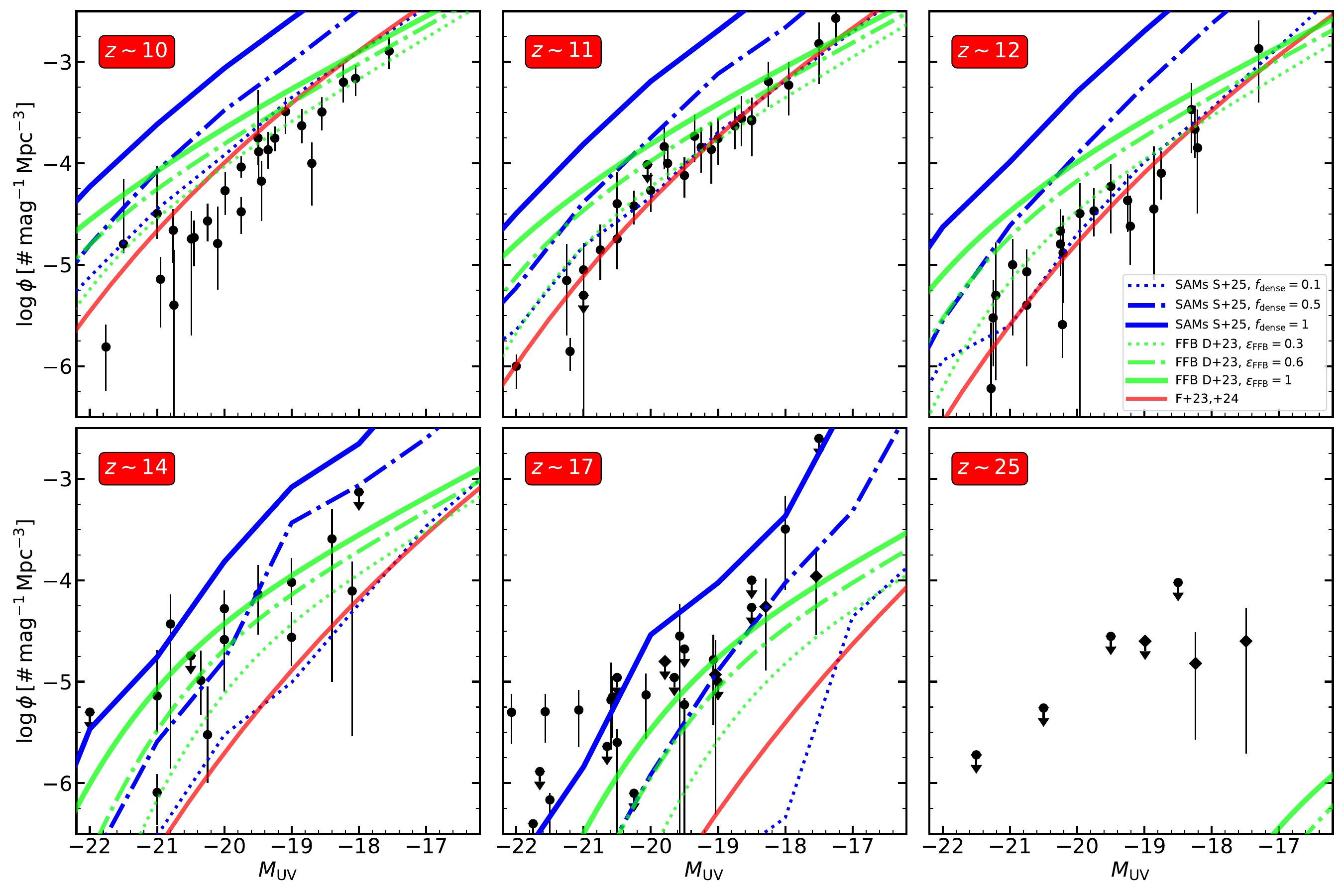}
\caption{\label{fig:sims1}Comparison of the observations and model predictions for the evolution of the UV luminosity function at $z>10$. The literature compilation includes the results in this paper as well as those in \citet{2024ApJ...965..169A,2022arXiv221206683B,2024ApJ...965...98C,2022ApJ...938L..15C,2025arXiv250405893C,2022ApJ...940L..55F,2024ApJ...969L...2F,2022ApJS..259...20H,2024ApJ...960...56H,2024arXiv241113640K,2023ApJ...954L..46L,2024arXiv241204211M,2022arXiv220802794N,2024ApJ...970...31R,2025arXiv250706292W,2025arXiv250100984W,2024ApJ...966...74W,2023arXiv231115121Y}
The models are: (1) the feedback-free burst models in \citet[FFB, see also \citealt{2024A&A...690A.108L}]{2023MNRAS.523.3201D}, including several star formation efficiencies, $\epsilon_{\rm FFB}=0.3,0.6,1.0$, and for their disc dust recipe (lime color); (2)  the models including the effects and blowout of dust from \citet[][see also \citealt{2024A&A...689A.310F}; red]{2023MNRAS.522.3986F}; and (3) the semi-analytic models with density-dependent star formation efficiency, with dense gas fraction $f_{\rm dense} =0.1,0.5,1$, from \citet{2025arXiv250505442S}, only up to $z\sim17$, as no predictions are available at $z\gtrsim 20$ (blue).}
\end{figure*}

\begin{figure*}[ht!]
\centering
\includegraphics[clip, trim=0.99cm 1.0cm 0.0cm 0.0cm,scale=0.41]{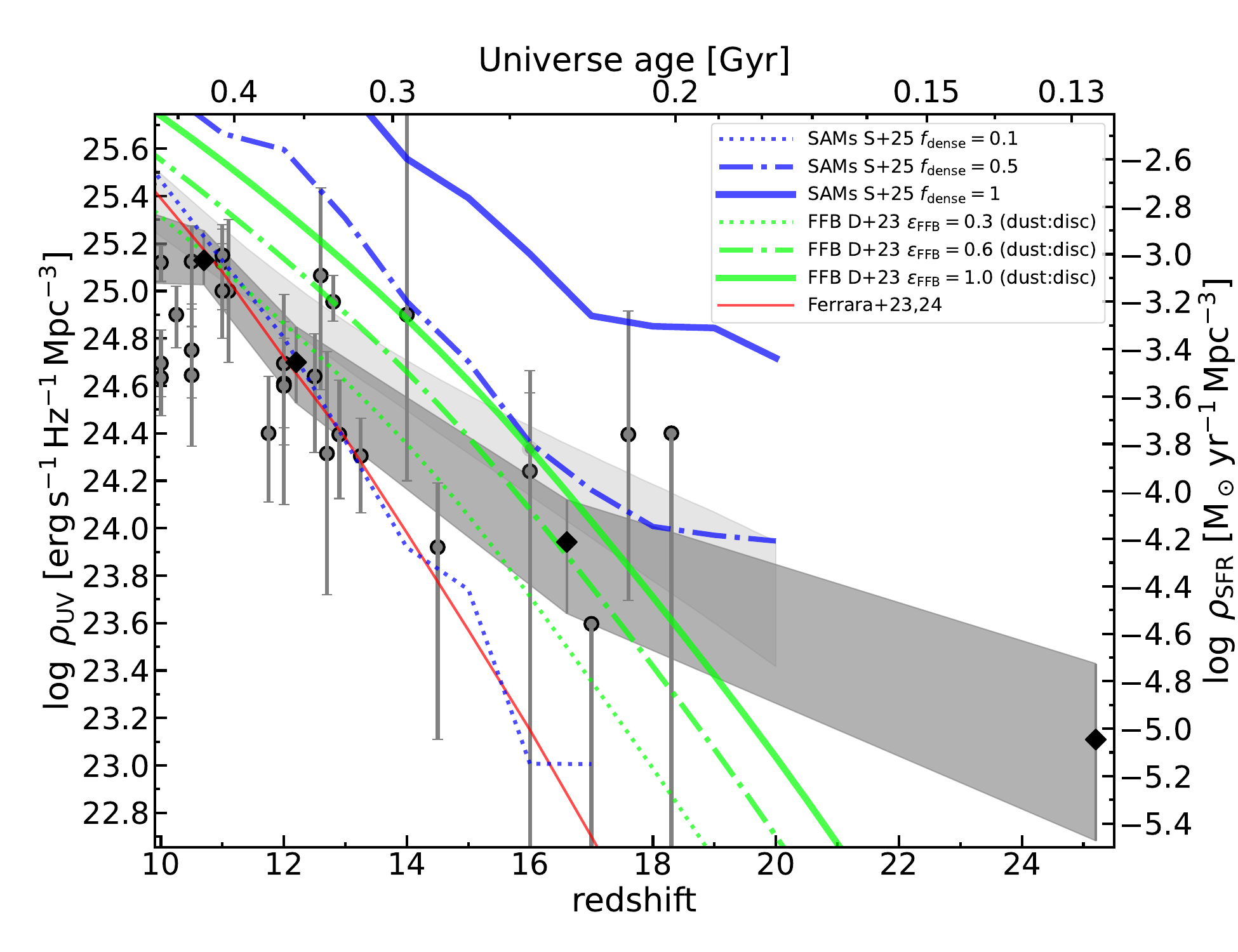}
\caption{\label{fig:sims2}Comparison of the observations and model predictions for the evolution of the UV luminosity density at $z>10$. Results from this paper and the literature (see Figure~\ref{fig:sfrd} for references) are compared with predictions from the same models depicted in Figure~\ref{fig:sims1}.}
\end{figure*}

\subsection{Cosmic UV luminosity density evolution}
\label{sec:csfrd}

Figure~\ref{fig:sfrd} presents the cosmic UV luminosity density obtained from integrating the luminosity functions shown in Figure~\ref{fig:lf} down to the same absolute magnitude, i.e., for $M_\mathrm{UV}\sim-17$~mag. We note that no dust correction has been applied. The plot also includes an axis with the translation from UV luminosity density to SFR density, assuming the conversion factor $0.7\times10^{-28}$~M$_\odot$~yr$^{-1}$~erg$^{-1}$~s~Hz, taken from \citet[][see also \citealt{1998ARA&A..36..189K}]{2014ARA&A..52..415M} and converted to a \citet{2003PASP..115..763C} IMF. As discussed in \citet{2014ARA&A..52..415M}, this conversion corresponds to a constant SFH burst with roughly 0.3~Z$_\odot$ metallicity (lower metallicities providing $\sim10$\% smaller factors), and for ages older than 300~Myr. Consistently, \citet{1998ARA&A..36..189K} mentions that this type of conversion is valid for constant SFH and timescales longer than 100~Myr, and the factor gets significantly larger for younger bursts ($\sim60$\% for 10~Myr old bursts; see also \citealt{2013seg..book.....F,2019ApJ...871..128T}). We will discuss more on the implications of using these factors in Section~\ref{sec:models}.

The density (in logarithmic scale) evolves roughly linearly with respect to redshift from $z\sim10$ to $z\sim$25, increasing by a factor of $\sim$2.5 in the 100 Myr from $z\sim14$ to $z\sim10$ based on values from the literature, and, taking into account the results in this work, it continues decreasing faster, by a factor of $\sim3$ in the previous 100 Myr ($z\sim14$ to $z\sim$18), and another factor of $\sim3$ in the previous
$\sim80$~Myr (back to $z\sim25$).

Using the UV luminosity density data points at $z=9,11,12$ from \citet{2023ApJ...951L...1P} and the results at $z\sim17$ and $z\sim25$ from this work, we fit the redshift dependence of the  evolution of the UV luminosity density in the $10<z<25$ interval, obtaining that it evolves as:

\begin{equation}
\rho_\mathrm{UV}\propto(1+z)^{-5.3\pm1.6}.
\end{equation}

\subsection{Comparison with cosmological galaxy formation models}
\label{sec:models}

Figures~\ref{fig:sims1} and \ref{fig:sims2} show comparisons of the evolution at $z>10$ of the observationally derived luminosity function and UV luminosity density with several recent galaxy evolution models. 
These include an updated version of the Santa Cruz Semi-Analytic Model \citep{Somerville2008,2024MNRAS.527.5929Y}, presented by \citet[][hereafter S25; see also \citealt{2025arXiv250418618Y}]{2025arXiv250505442S}, as well as the models of \citet{2023MNRAS.523.3201D} and \citet{2023MNRAS.522.3986F}. 

The updated Santa Cruz SAMs shown here are based on results from molecular cloud-scale simulations with radiative transfer by \citet{Menon2024}, which indicate that higher surface density clouds have higher star formation efficiencies, converting up to 85\% of their available gas into stars within a cloud lifetime, in contrast to a few percent in lower density, local Universe molecular clouds. The S25 models assume that the surface density of clouds is proportional to the average surface density of the overall ISM in galaxies. Because the ISM is expected to be denser at higher redshift, these models predict higher star formation efficiencies at early times. The main uncertain parameter in these models is the fraction of the ISM in dense star forming clouds, which is represented by the parameter $f_{\rm dense}$.

In the "feedback free starburst'' (FFB) model of \citet[][see also \citealt{2024A&A...690A.108L}]{2023MNRAS.523.3201D}, it is posited that when the galaxy free-fall time is shorter than the timescale for the first supernovae to begin to explode (a few Myr), star formation is more efficient because of the absence of strong supernovae driven winds. These conditions become more common in the most massive halos above $z\sim 10$ \citep{2023MNRAS.523.3201D}, again because of the higher expected density of the ISM. These models are also parameterized by the unknown efficiency of star formation, here denoted as $\epsilon_{\rm FFB}$. The efficiency depends on a redshift-dependent critical mass above which feedback is negligible and, therefore, star formation is enhanced.

In the \citet[][hereafter F23]{2023MNRAS.522.3986F} model, the shallower than expected evolution of the luminosity function and luminosity density at $z\gtrsim 10$ is explained as being due to ejection of dust via radiation pressure above a critical specific star formation rate, which again is expected to be more frequently attained in higher redshift galaxies. 

First considering the UV luminosity function comparison, the F23 model reproduces the observations at $z\sim 10$--12, but underproduces the galaxy number density at higher redshifts. The FFB models reproduce the observations at $z\sim 10$--12 well for the lower values of SFE $\epsilon_{\rm FFB} = 0.3$, but require higher values $\epsilon_{\rm FFB} \sim 1$ (above 0.5 if we consider uncertainties in the LF at $z\sim17$ presented in this paper) to reproduce the UVLF at $z\sim 14$--17. Even the $\epsilon_{\rm FFB} \sim 1$ FFB model lies below the observational estimate at $z\sim 25$. The S25 Santa Cruz SAM with density modulated star formation efficiency, similarly, matches the lower redshift UV luminosity functions well with the lower values of the dense gas fraction ($f_{\rm dense} = 0.1$), but higher values of $f_{\rm dense}$, approaching unity and at least around $f_{\rm dense}=0.5$, appear to be required to reproduce the observed UV LF at $z\sim 17$. Predictions above $z\sim 20$ for this model are not currently available.  

The integrated UV luminosity density tells a similar story. None of the theoretical models predicts as shallow a decline in $\rho_{\rm UV}$ from $z\sim 12$--17 as that implied by our observations. The discrepancy between the predicted very steep drop in $\rho_{\rm UV}$ from $z\sim 17$--24 with our results is even more dramatic. The $\log\rho_{\rm UV}$ evolution for models follows a $(1+z)^\gamma$ law with $\gamma\sim-6$ at $10\lesssim z\lesssim12$, very similar to observations, but that exponent is closer to $\gamma\sim-4$ at $z>12$. If we assume that the halo mass functions derived from the $\Lambda$CDM model are correct, then this means that the amount of (UV) light created in the halos considered in the simulations is significantly smaller than what is observed and it also implies that it evolves with redshift.

We note that the halo mass function at $z\gtrsim 20$ has not been well-characterized with modern N-body simulations with up-to-date cosmological parameters. The predictions of \citet{2023MNRAS.523.3201D}, shown here, adopt an extrapolation of a fit to a halo mass function based on N-body simulations analyzed at much lower redshifts and over a very different halo mass range. The semi-analytic models of S25 are based on halo merger trees extracted from the N-body simulations presented by \citet{2024MNRAS.530.4868Y}, but due to resolution limitations of those simulations, the merger trees do not extend beyond $z \sim 20$. The predictions at the highest redshifts shown here should therefore be taken with some caution. 

To further understand the implications of our observed luminosity functions, and assuming that the $\Lambda$CDM cosmology holds, we calculate the mass-to-light ratios needed to reproduce these current observations. For this calculation, we start with the dark matter halo mass functions obtained by \citet{2024MNRAS.530.4868Y}\footnote{Similar results were obtained when using the functions provided by the {\sc hmf} code for \citet{2008ApJ...688..709T} halos, extended to high redshift by \citet{2013ApJ...770...57B}}. Then,  we calculate the ratio for a given number density to convert that mass function to the luminosity functions (both in per dex units) estimated for $z\sim9$ to $z\sim25$ by \citet{2023ApJ...951L...1P} and this paper. This operation is analogous to the abundance matching technique used to compare stellar and dark matter mass functions \citep[e.g.,][]{2010ApJ...717..379B,2011ApJ...742...16T}. However, we remark that we rank halos according to dark matter halo mass and UV luminosity, and not stellar mass. The stellar mass calculations  include larger uncertainties than the directly observed luminosities, but the caveat is that bursts can make a galaxy jump to a higher luminosity state than the average it would get for its halo abundance \citep[see, e.g.][]{2023MNRAS.521..497M}. The final result of the abundance matching exercise is a mass-to-light ratio (M/L) that depends on the mass of the halo and redshift. The M/L obtained with this method refers to dark matter halo masses. We translate these into baryon mass by multiplying by the relative baryon abundance with respect to dark matter, and then we apply a star formation efficiency $\epsilon$, i.e., the fraction of mass transformed to stars relative to the total amount of baryons in the halo.

\begin{figure*}[ht!]
\centering
\includegraphics[clip, trim=2.7cm 0.0cm 19.0cm 0.0cm,scale=0.7]{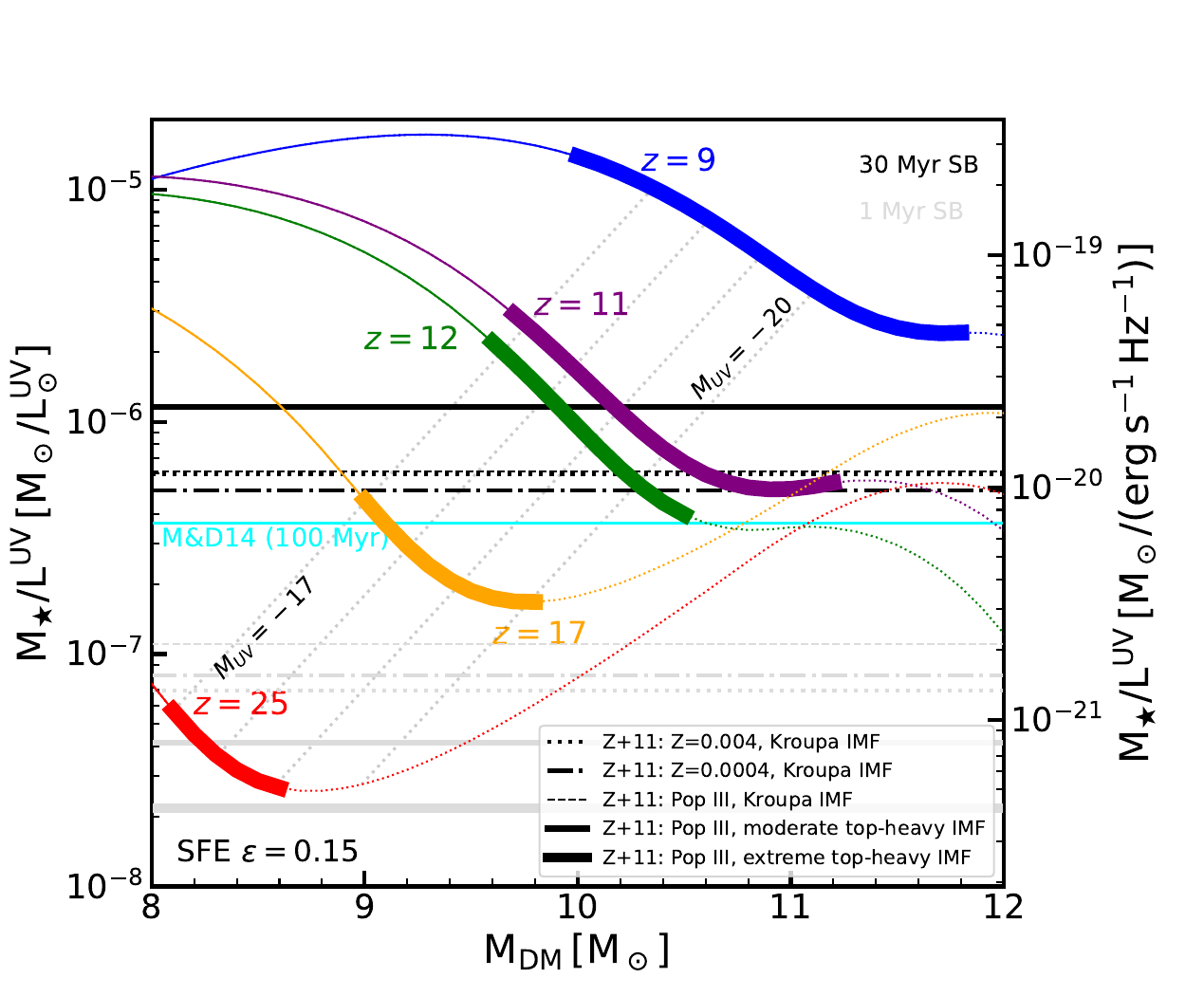}
\caption{\label{fig:m/l}Stellar mass-to-light ratios as a function of halo mass for redshifts $z\sim9$ to $z\sim25$. The left y-axis give UV luminosity densities in solar units (assuming an AB absolute magnitude for the Sun $M^{UV}_\odot=17.30$, \citealt{2018ApJS..236...47W}), the right y-axis in CGS units. The vertical axes could also be read in terms of dark matter mass-to-light ratios considering the baryon-to-dark matter density ratio ($\sim16$\%) and the star formation efficiency, defined as stellar mass divided by baryon mass, assumed to be 15\% in the plot (i.e., multiplying by 41.7). The right y-axis can also be translated to mass-to-light ratios in terms of $\nu\mathrm{L}_\nu$ and the bolometric solar luminosity, in $\mathrm{M}_\odot/\mathrm{L}_\odot$ units, \textcolor{blue}{by multiplying by $1.9\times10^{18}$~erg~s$^{-1}$~Hz$^{-1}$~L$_\odot^{-1}$.} The curved lines provide results based on an abundance matching comparison between luminosity functions from this work and \citet{2023ApJ...951L...1P} and dark matter mass functions from \citet{2024MNRAS.530.4868Y}. Thick segments show the calculations based on the range of absolute magnitudes actually probed by those works, while the thin lines show results based on extrapolated Schechter fits. We differentiate between the faint/light ends of the luminosity and mass functions, which might be well characterized by the current data with a faint-end slope (where we use a continuous line) and the bright/massive ends, whose behavior is more complex and unconstrained for most redshifts above $z=10$ (where we used a dotted line). Oblique lines show the UV absolute magnitudes corresponding to the dark matter masses according to the abundance matching exercise, running from -17 to -20 every one magnitude (the extreme values are marked).  Horizontal lines show mass-to-light ratios for a compilation of relevant stellar population models from \citet{2011ApJ...740...13Z}, with metallicities spanning expected values from 1/5~Z$_\odot$ to Pop-III stars, and different IMFs including \citet{2001MNRAS.322..231K} and top-heavy parametrizations. Lines in black and light gray show predictions for 30 and 1~Myr old instantaneous bursts (instantaneous meaning $\delta$ functions). We note that the models with the most extreme top-heavy IMF do not extend to 30~Myr, those massive stars die earlier. This effect also affects the relative mass-to-light ratios between the Kroupa and top-heavy IMFs for 30~Myr (reversed compared to 1~Myr predictions). The cyan line marks the mass-to-light ratio for the UV luminosity to SFR calibration used in \citet{2014ARA&A..52..415M}, which assume a constant SFH and an age of 100~Myr.}
\end{figure*}

Figure~\ref{fig:m/l} shows our results for the derived mass-to-light ratios for $\epsilon=0.15$ \citep[similar to the values adopted by \citealt{2025ApJ...980..249L};][]{2022ApJS..259...20H}. We compare the mass-to-light ratios with those obtained from a comprehensive list of stellar population models for different stellar ages (typical of $z>10$ galaxies; see, for example, \citealt{2024Natur.633..318C,2024arXiv240518462H,2023NatAs...7..611R}) spanning the whole range of expected metallicities as well as a variety of possible IMFs
 \citep{2011ApJ...740...13Z}.  

Figure~\ref{fig:m/l} shows that, based on this abundance matching exercise, the typical host halo mass for the faintest galaxies that are currently observable by JWST shifts from $10^{11}$~M$_\odot$ at $z\sim 9$ down to $\sim 10^{9.4}$~M$_\odot$ at $z\sim17$ and even further down to $\sim 10^{8.4}$~M$_\odot$ at $z\sim25$. In particular, the intrinsic UV magnitude of the populations we detect at $z\sim 17$ and $z\sim 25$ are similar ($-18 < M_{\rm UV} < -17$), yet the host halo mass required to match the detected abundance is an order of magnitude smaller at the highest redshift. Leaving aside the possibility of non-stellar contributions to the UV luminosity, this implies that either halos are more efficient at converting baryons into stars at earlier epochs (i.e., the star formation efficiency increases with redshift), or that the stellar populations are brighter at higher redshifts for a given amount of stars formed, i.e., stars are more efficient producing photons or, said in an alternative way, their emissivity is higher.


To investigate how our calculations compare with the expected stellar mass-to-light ratios from stellar population models, in Figure~\ref{fig:m/l} we also show $M_\bigstar/L^{\rm UV}$ from a variety of stellar population models with a mean age of 1 Myr (i.e., a burst) or 30 Myr (age found for spectroscopically confirmed $z>10$ galaxies, see, e.g., \citealt{2025NatAs...9..155Z}) and low ($<0.1$~Z$_\odot$) to primordial metallicities. The results for a 30 Myr old stellar population are consistent with the required $M_\bigstar/L^{\rm UV}$ at $z=17$ for our adopted (constant) value of $\epsilon=0.15$. The factor of approximately three lower values of $M_\bigstar/L^{\rm UV}$ at $z\sim 25$ imply that these $z\sim25$ galaxies, if the $\epsilon$ is constant (15\% assumed in the plot, a scenario considered in galaxy simulations such as those in \citealt{2025MNRAS.536..988F}), should host much younger stellar populations (i.e., catching the galaxies in a very young burst), and/or a top-heavy IMF \citep[see,e.g.,][]{2025MNRAS.536.1018L}, so their emissivity is higher and their mass-to-light ratios smaller. Alternatively, if at $z\sim25$ the star formation efficiency changes but the typical stellar properties at $z\sim17$ and $z\sim25$ are similar, the relative efficiency between these two epochs (and halos of different masses) must change. For example, if the efficiency at $z\sim17$ stays at 15\%, the $z\sim25$ galaxies must present higher values to go up in the plot to the same $M_\bigstar/L^{\rm UV}$ values; or if the $z\sim25$ galaxies remain in the $\epsilon=0.15$ efficiency and mass-to-light ratios typical of young low-metallicity starburst, $z\sim17$ galaxies should have smaller star formation efficiencies to match those stellar ages. Given that the probability of catching a galaxy in a 1~Myr old starburst should be smaller than catching it with older ages, the first scenario seems more representative. However, the detectability of the highest redshift galaxies might bias the selection to the youngest and brightest evolutionary stages, which, would favor the second scenario.


Directly related to the previous point, we remark that Figure~\ref{fig:m/l} is constructed with an abundance matching method applied to dark matter mass functions and UV luminosity functions. Bursty star formation histories might enhance the UV photon production of halos with smaller masses than those corresponding to the same number density of the luminosity function. The small number of galaxies in the sample discussed in this paper would worsen these effects, as we might be biased preferentially towards galaxies that are experiencing a very recent or ongoing burst \citep[see, e.g., the effect of star formation stochasticity in][]{2023MNRAS.521..497M,2024ApJ...975..192G,2023MNRAS.525.3254S}. 

We remark, as exemplified in the two scenarios described above, that the effect of the emissivity and the star formation efficiency discussed in Figure~\ref{fig:m/l} are degenerate. Concerning the emissivity, in that figure we demonstrate that the metallicity (within the low-Z regime) can account for up to $\sim40$\% differences in the mass-to-light ratios. The IMF would have a much more significant impact, it can make galaxies up to a factor of $\times3.5$ brighter for the same mass, according to the \citet{2011ApJ...740...13Z} Pop III models shown in the plot. In addition, differences in stellar ages affect the brightness of galaxies by more than an order of magnitude in just 30~Myr (an age used as reference in Figure~\ref{fig:m/l}, and taken from papers such as \citealt{2025NatAs...9..155Z} and our own analysis described in the next section). Therefore, burstiness is another factor that affects the results shown in Figure~\ref{fig:m/l} (see \citealt{2023MNRAS.525.3254S} and \citealt{2024ApJ...975..192G}).

Taking into account these considerations, the 15\% star formation efficiency used in Figure~\ref{fig:m/l} for reference can change significantly if we consider different emissivities. For example, at $z\sim17$, if the typical age of the stellar populations in those galaxy candidates is $\sim1$~Myr (i.e., if they are experiencing a burst), we would need smaller efficiencies, $\epsilon\sim0.01$.

It is also noteworthy to consider further the meaning and implications of the star formation efficiencies introduced in Figures~\ref{fig:sims1} to \ref{fig:m/l}, given that the definitions in simulations and in our calculations present important but subtle differences. For Figure~\ref{fig:m/l}, the efficiency $\epsilon$ is just a constant factor used to compare dark matter mass-to-light ratios (directly obtained from the abundance matching exercise) to stellar population models (apart from the baryon to dark matter fraction). For galaxy formation simulations, typically the star formation efficiency is more complex, it depends on the dark matter halo mass (see, e.g., equation 7 in \citealt{2025arXiv250418618Y}) or the density in the star-forming knots (see, e.g., \citealt{2023MNRAS.523.3201D}).

In addition, the predictions for the UV luminosity functions (and, in general, for the luminosity of galaxies) then comes in the simulations from the assumption of some SFR to UV luminosity factor, which include certain stellar population properties (e.g., metallicity or IMF). For example, \citet{2025arXiv250418618Y} assume the same SFR-to-UV-luminosity ratio we used for Figure~\ref{fig:sfrd}, which comes from \citet{2014ARA&A..52..415M} after correcting to a \citet{2003PASP..115..763C} IMF. For reference, this factor is shown in Figure~\ref{fig:m/l} for a 100~Myr burst (cyan line). Applying that factor to all the dark matter halos probed at $z\sim17$ (orange curve), we would obtain efficiencies of $\sim10\%$ for M$_\mathrm{DM}=10^{9.0}$~M$_\odot$ halos (the orange curve should come down by using a smaller than 15\% efficiency), and $\sim40$\% for M$_\mathrm{DM}=10^{9.8}$~M$_\odot$ halos (we need a higher efficiency to scale the orange curve up), consistent with the findings in \citet{2025arXiv250418618Y} (cf. their Figure~5). 

In the case of the FFB models, the predictions  for the luminosity function at $z=17$ and especially at $z=25$, as displayed in Figure~\ref{fig:lf}, seem to underestimate the observational calculations even when very high maximum star-formation efficiency is adopted at the FFB phase. However, the current FFB predictions at these very high redshifts should be taken with a grain of salt. The main reason is, indeed, that the UV emissivity assumed in this current FFB modeling is rather arbitrary, adopting the value assumed to be valid at $z=10$, which is likely an underestimate by a factor of a few. It assumes $\mathrm{M}_\bigstar/\mathrm{L}^\mathrm{UV}=6-9 \times10^{-7}$~M$_\odot$/L$^\mathrm{UV}_\odot$ (very similar to the 30~Myr models in Figure~\ref{fig:m/l}) for $\mathrm{M}_\bigstar=10^{8-10}$~M$_\odot$. However, we learn from Figure~\ref{fig:m/l} that $\mathrm{M}_\bigstar/\mathrm{L}^\mathrm{UV}\sim4\times 10^{-8}$~M$_\odot$/L$^\mathrm{UV}_\odot$ is required at $z=25$ based on abundance matching with $\epsilon=0.15$.  Other reasons for uncertainty concerning the comparison to the current FFB predictions include the facts that the high efficiency is applied only to the FFB galaxies, the predictions for non-FFB galaxies are based on ad-hoc extrapolations of the UniverseMachine results to $z\sim20$ \citep{2019MNRAS.488.3143B}, and the definitions of efficiency are somewhat different than in the current work. More realistic FFB predictions will be performed soon along the lines above, to be applied at $17<z<25$.

Given the effects of the properties discussed above, the 60-100\% efficiencies discussed in the simulation papers and the 15\% efficiency used in Figure~\ref{fig:m/l} should be interpreted with caution, in combination with emissivities, and are certainly not inconsistent.

We make two final remarks about this comparison with models. First, we have left apart the possible non-stellar origin of the luminosity of high redshift galaxies \citep[see, e.g., the discussion about primordial black holes in][see also \citealt{2024ApJ...961L..39S}]{2025arXiv250318850M}. And finally, even our results at $z\sim25$, which are not reproduced by the simulations (falling short by an order of magnitude in the figures discussed in this section), have been demonstrated to not violate the $\Lambda$CDM paradigm \citep{2025arXiv250418618Y}. 

We conclude that Figure~\ref{fig:m/l} provides robust constraints about the amount of UV photons that must be created from  halos of different masses (and redshift), but how those photons are created by stars (or AGN) depends on a variety of factors whose detailed study is beyond the scope of this paper. Our analysis points out to emissivity as the main ruling factor, related to stellar properties such as metallicity, and especially age (also introduced as burstiness, stochasticity or variability in the literature, as given in references mentioned in this section) and the IMF.


\subsection{Stellar populations of the \texorpdfstring{$\MakeLowercase{z}\sim17$}{z~17} 
  and \texorpdfstring{$\MakeLowercase{z}\sim25$}{z~25} galaxy candidates}
\label{sec:physical}

\begin{deluxetable*}{llcccccccc}
\caption{Physical properties of the galaxy candidates at $14<z<25$.}
\centerwidetable 
\tabletypesize{\scriptsize}
\tablehead{
\colhead{Galaxy name} &
\colhead{code}&
\colhead{redshift} & 
\colhead{$\beta$} &
\colhead{$\log \mathrm{M}_\bigstar$} & \colhead{$\mathrm{A(V)}$} & \colhead{$\mathrm{t}_\mathrm{m-w}$} &
\colhead{$<\mathrm{SFR}>$} &
\colhead{$r^c_\mathrm{eff}$}&
\colhead{$M_\mathrm{UV}$}\\
\colhead{} &
\colhead{} &
\colhead{} &
\colhead{} &
\colhead{[M$_\odot$]} &
\colhead{[mag]} &
\colhead{[Myr]} &
\colhead{[M$_\odot$/yr]} &
\colhead{[pc]} &
\colhead{[mag]} 
}
\startdata
\hline
\hline
\multicolumn{8}{c}{$z\sim17$ sample}\\
\hline
midis-z17-1  & sy-BC03  & $18.4_{-1.5}^{+1.5}$  & $-2.7^{+0.6}_{-0.2}$ &  $7.5^{+0.5}_{-0.5}$ &  $0.00^{+0.21}_{-0.00}$ &  $24\pm1$ & $1.1\pm0.5$ & $45_{-10}^{+17}$ & $-17.3\pm0.5$\\ 
             & sy-Z11   & $18.4_{-1.5}^{+1.5}$  & $-3.0^{+0.4}_{-0.1}$ &  $7.3^{+0.6}_{-0.4}$ &  $0.00^{+0.19}_{-0.00}$ &  $13\pm4$ & $1.0\pm0.4$  \\ 
             & sy-BPASS & $18.4_{-1.5}^{+1.5}$  & $-2.6^{+0.4}_{-0.3}$ &  $5.9^{+0.4}_{-0.1}$ &  $0.00^{+0.10}_{-0.00}$ &  $1\pm1$ & $0.4\pm0.1$  \\ 
             & pr-FSPS  & $19.0_{-1.4}^{+1.2}$  & $-2.3^{+0.1}_{-0.1}$ &  $7.4^{+7.6}_{-7.1}$ &  $0.27^{+0.46}_{-0.10}$ &  $154\pm62$ & $0.1\pm0.1$  \\ 
midis-z17-2  & sy-BC03  & $17.4_{-1.1}^{+1.4}$  & $-3.0^{+0.6}_{-0.1}$ &  $6.5^{+0.6}_{-0.1}$ &  $0.00^{+0.10}_{-0.00}$ &  $3\pm1$ & $0.6\pm0.1$ & $165_{-28}^{+39}$ & $-17.1\pm0.5$\\ 
             & sy-Z11   & $17.4_{-1.1}^{+1.4}$  & $-3.0^{+0.7}_{-0.1}$ &  $6.6^{+0.8}_{-0.1}$ &  $0.00^{+0.10}_{-0.00}$ &  $1\pm1$ & $4.2\pm0.8$  \\ 
             & sy-BPASS & $17.4_{-1.1}^{+1.4}$  & $-2.9^{+0.3}_{-0.2}$ &  $5.9^{+0.4}_{-0.1}$ &  $0.00^{+0.10}_{-0.00}$ &  $1\pm1$ & $0.4\pm0.1$  \\ 
             & pr-FSPS  & $18.9_{-1.9}^{+0.8}$  & $-2.1^{+0.1}_{-0.1}$ &  $7.2^{+7.4}_{-6.9}$ &  $0.19^{+0.39}_{-0.07}$ &  $132\pm56$ & $0.1\pm0.1$  \\ 
midis-z17-3  & sy-BC03  & $17.3_{-0.2}^{+1.1}$  & $-3.0^{+0.2}_{-0.1}$ &  $7.3^{+0.2}_{-0.2}$ &  $0.00^{+0.10}_{-0.00}$ &  $2\pm1$ & $6.8\pm1.2$ & $30_{-2}^{+4}$ & $-19.0\pm0.3$\\ 
             & sy-Z11   & $17.3_{-0.2}^{+1.1}$  & $-3.0^{+0.2}_{-0.1}$ &  $7.3^{+0.2}_{-0.2}$ &  $0.00^{+0.10}_{-0.00}$ &  $1\pm1$ & $6.8\pm1.3$  \\ 
             & sy-BPASS & $17.3_{-0.2}^{+1.1}$  & $-3.0^{+0.2}_{-0.1}$ &  $5.6^{+0.2}_{-0.2}$ &  $0.00^{+0.10}_{-0.00}$ &  $1\pm1$ & $0.2\pm0.0$  \\ 
             & pr-FSPS  & $19.1_{-0.1}^{+0.1}$  & $-2.2^{+0.1}_{-0.1}$ &  $6.7^{+6.8}_{-6.7}$ &  $0.01^{+0.10}_{-0.00}$ &  $4\pm2$ & $0.6\pm0.6$  \\ 
midis-z17-4  & sy-BC03  & $18.0_{-1.5}^{+1.7}$  & $-2.6^{+0.2}_{-0.4}$ &  $7.7^{+0.6}_{-0.3}$ &  $0.11^{+0.24}_{-0.01}$ &  $25\pm1$ & $1.6\pm0.8$ & $97_{-12}^{+13}$ & $-17.9\pm0.4$\\ 
             & sy-Z11   & $18.0_{-1.5}^{+1.7}$  & $-2.7^{+0.2}_{-0.1}$ &  $7.0^{+0.8}_{-0.2}$ &  $0.03^{+0.16}_{-0.02}$ &  $42\pm1$ & $1.4\pm0.5$  \\ 
             & sy-BPASS & $18.0_{-1.5}^{+1.7}$  & $-2.5^{+0.4}_{-0.3}$ &  $6.8^{+0.6}_{-0.1}$ &  $0.10^{+0.10}_{-0.01}$ &  $43\pm1$ & $4.7\pm0.9$  \\ 
             & pr-FSPS  & $19.1_{-0.5}^{+0.5}$  & $-1.1^{+0.1}_{-0.1}$ &  $7.6^{+7.8}_{-7.3}$ &  $0.45^{+0.60}_{-0.27}$ &  $113\pm64$ & $0.2\pm0.2$  \\ 
midis-z17-5  & sy-BC03  & $17.5_{-2.4}^{+2.3}$  & $-2.8^{+0.6}_{-0.6}$ &  $6.7^{+0.6}_{-0.5}$ &  $0.00^{+0.10}_{-0.00}$ &  $20\pm1$ & $1.7\pm0.6$ & $155_{-19}^{+26}$ & $-17.0\pm0.4$\\ 
             & sy-Z11   & $17.5_{-2.4}^{+2.3}$  & $-3.0^{+0.8}_{-0.1}$ &  $7.2^{+0.6}_{-0.3}$ &  $0.00^{+0.23}_{-0.00}$ &  $12\pm2$ & $1.2\pm0.4$  \\ 
             & sy-BPASS & $17.5_{-2.4}^{+2.3}$  & $-2.5^{+0.4}_{-0.3}$ &  $6.2^{+0.3}_{-0.1}$ &  $0.00^{+0.10}_{-0.00}$ &  $1\pm1$ & $1.3\pm0.2$  \\ 
             & pr-FSPS  & $18.3_{-2.5}^{+2.1}$  & $-2.2^{+0.1}_{-0.1}$ &  $7.4^{+7.6}_{-7.1}$ &  $0.33^{+0.57}_{-0.13}$ &  $120\pm44$ & $0.1\pm0.1$  \\ 
midis-z17-6  & sy-BC03  & $16.8_{-0.8}^{+2.4}$  & $-2.4^{+0.2}_{-0.2}$ &  $8.3^{+0.2}_{-0.2}$ &  $0.01^{+0.17}_{-0.01}$ &  $59\pm1$ & $3.3\pm0.8$ & $213_{-17}^{+23}$ & $-18.6\pm0.3$\\ 
             & sy-Z11   & $16.8_{-0.8}^{+2.4}$  & $-2.6^{+0.4}_{-0.3}$ &  $8.2^{+0.2}_{-0.7}$ &  $0.00^{+0.20}_{-0.00}$ &  $36\pm4$ & $3.8\pm0.9$  \\ 
             & sy-BPASS & $16.8_{-0.8}^{+2.4}$  & $-2.7^{+0.1}_{-0.1}$ &  $6.8^{+0.1}_{-0.1}$ &  $0.00^{+0.10}_{-0.00}$ &  $1\pm1$ & $4.7\pm0.9$  \\ 
             & pr-FSPS  & $16.5_{-0.9}^{+2.1}$  & $-2.2^{+0.1}_{-0.1}$ &  $7.7^{+7.9}_{-7.5}$ &  $0.22^{+0.37}_{-0.10}$ &  $165\pm56$ & $0.2\pm0.2$  \\ 
midis-z25-1  & sy-BC03  & $25.7_{-1.4}^{+1.8}$  & $-2.6^{+0.6}_{-0.2}$ &  $7.1^{+0.6}_{-0.5}$ &  $0.19^{+0.10}_{-0.01}$ &  $14\pm1$ & $4.1\pm0.8$ & $33_{-7}^{+10}$ & $-18.2\pm0.4$\\ 
             & sy-Z11   & $25.7_{-1.4}^{+1.8}$  & $-2.7^{+0.6}_{-0.1}$ &  $7.1^{+1.5}_{-0.1}$ &  $0.10^{+0.10}_{-0.01}$ &  $31\pm1$ & $13.7\pm2.8$  \\ 
             & sy-BPASS & $25.7_{-1.4}^{+1.8}$  & $-2.6^{+0.6}_{-0.5}$ &  $5.6^{+0.4}_{-0.2}$ &  $0.11^{+0.10}_{-0.01}$ &  $9\pm1$ & $0.2\pm0.0$  \\ 
             & pr-FSPS  & $26.6_{-2.0}^{+1.7}$  & $-2.0^{+0.1}_{-0.1}$ &  $7.5^{+7.7}_{-7.2}$ &  $0.33^{+0.55}_{-0.13}$ &  $90\pm38$ & $0.2\pm0.2$  \\ 
midis-z25-2  & sy-BC03  & $25.1_{-1.0}^{+1.5}$  & $-3.0^{+0.7}_{-0.1}$ &  $6.8^{+0.1}_{-0.1}$ &  $0.00^{+0.10}_{-0.00}$ &  $2\pm1$ & $2.0\pm0.4$ & $59_{-18}^{+17}$ & $-17.6\pm0.5$\\ 
             & sy-Z11   & $25.1_{-1.0}^{+1.5}$  & $-3.0^{+0.1}_{-0.1}$ &  $6.9^{+0.1}_{-0.1}$ &  $0.00^{+0.10}_{-0.00}$ &  $1\pm1$ & $7.1\pm1.3$  \\ 
             & sy-BPASS & $25.1_{-1.0}^{+1.5}$  & $-3.0^{+0.1}_{-0.1}$ &  $6.5^{+0.1}_{-0.1}$ &  $0.00^{+0.10}_{-0.00}$ &  $1\pm1$ & $2.4\pm0.7$  \\ 
             & pr-FSPS  & $24.9_{-1.5}^{+2.3}$  & $-2.1^{+0.1}_{-0.1}$ &  $7.2^{+7.5}_{-7.0}$ &  $0.29^{+0.52}_{-0.10}$ &  $89\pm36$ & $0.1\pm0.1$  \\ 
midis-z25-3  & sy-BC03  & $25.6_{-1.6}^{+1.5}$  & $-2.4^{+0.8}_{-0.4}$ &  $7.9^{+0.6}_{-0.6}$ &  $0.07^{+0.23}_{-0.07}$ &  $37\pm1$ & $2.7\pm0.7$ & $36_{-11}^{+18}$ & $-17.7\pm0.4$\\ 
             & sy-Z11   & $25.6_{-1.6}^{+1.5}$  & $-2.7^{+0.6}_{-0.1}$ &  $7.4^{+0.8}_{-0.5}$ &  $0.07^{+0.24}_{-0.07}$ &  $32\pm14$ & $3.0\pm1.1$  \\ 
             & sy-BPASS & $25.6_{-1.6}^{+1.5}$  & $-2.4^{+0.3}_{-0.4}$ &  $6.3^{+0.7}_{-0.3}$ &  $0.14^{+0.10}_{-0.01}$ &  $58\pm1$ & $2.4\pm0.7$  \\ 
             & pr-FSPS  & $26.6_{-1.3}^{+1.6}$  & $-2.3^{+0.1}_{-0.1}$ &  $7.4^{+7.7}_{-7.2}$ &  $0.32^{+0.55}_{-0.13}$ &  $76\pm33$ & $0.2\pm0.2$  \\ 
\enddata
\tablecomments{\label{tab:props}Physical properties of the sample of $z>16$ galaxy candidates presented in this paper. We provide: (1) photometric redshift, more specifically, the most probable according to the integral of the zPDF, including errors (\texttt{eazy} value for \textsc{synthesizer-AGN}, free parameter in \texttt{prospector}); (2) UV slope measured from SED-fitting; (3) stellar masses, attenuations, mass-weighted ages, and average SFRs, provided by the SED modeling using \textsc{synthesizer-AGN} (with three different stellar emission libraries: BC03, Z11, and BPASS), and \texttt{prospector} (using the FSPS library). (4) Circularized effective radii. (5) Absolute magnitudes including errors accounting for the photometric uncertainties and zPDFs.}
\end{deluxetable*}

We estimated stellar masses for our $z>16$ galaxy candidates using \texttt{prospector} (see Section~\ref{sec:selection}) as well as the \textsc{synthesizer-AGN} code \citep{2003MNRAS.338..508P,2008ApJ...675..234P}. For the latter, we used three different stellar emission libraries, namely, \citet[][BC03 hereafter]{2003MNRAS.344.1000B}, \citet[][Z11]{2011ApJ...740...13Z}, and \citet[][BPASS]{2017PASA...34...58E}. In all cases, we only considered models with metallicities lower than 0.2~Z$_\odot$, expected and observed at these redshifts \citep[e.g.,][]{2023MNRAS.518..425C,2023ApJS..269...33N,2024ApJ...962...24S}. As for the IMF, we used \citet{2003PASP..115..763C}
for BC03 and BPASS, and \citet{2001MNRAS.322..231K} for Z11. We assumed delayed-exponentials to model the star formation history, with timescales ranging from 1~Myr to 1~Gyr. Dust was considered using the \citet{2000ApJ...533..682C} attenuation law (differently affecting the stellar and gas emission, with a constant factor in this modeling) and assuming $V$-band attenuations smaller than 1~mag. The redshift was fixed to the most probable value provided by \texttt{eazy}. Jointly with stellar masses, we also estimated UV spectral slopes $\beta$ for each code. Uncertainties in all the SED fitting results were calculated using a Monte Carlo method by varying the flux data points according to their photometric errors and refitting. Results for each galaxy are given in Table~\ref{tab:props}.

\begin{figure}
\centering
\includegraphics[clip, trim=1.0cm 0.2cm 2.0cm 1.5cm,scale=0.45]{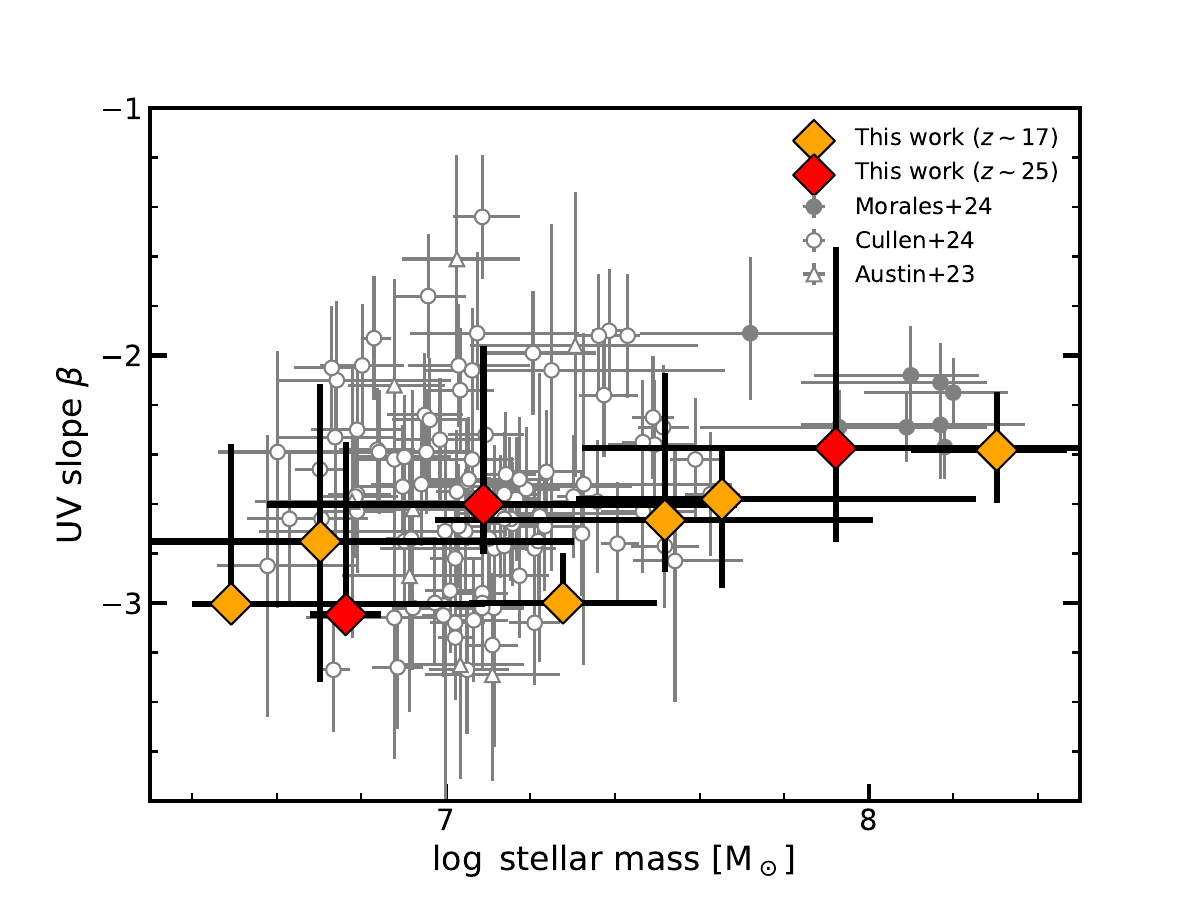}
\caption{\label{fig:massbeta}UV spectral slope $\beta$ (obtained from SED fitting, with errors accounting for photometric uncertainties; for the plot, we use the estimations based on \textsc{synthesizer-AGN} and BC03 models) $vs.$ stellar mass. The results for the $z\sim17$ and $z\sim25$ samples presented in this work are compared with values for $z>10$ galaxies in the compilations discussed in  \citet{2023MNRAS.520...14C}, \citet{2024arXiv240410751A}, and \citet{2024ApJ...964L..24M}.
}
\end{figure}

Figure~\ref{fig:massbeta} shows our $z>16$ galaxy candidates in a $\beta$ $vs.$ M$_\bigstar$ plot, compared to other lower redshift samples in the literature including $z>10$ galaxies (and typically $z<12)$. For this plot, we use the estimates based on \textsc{synthesizer-AGN} and BC03 models.

Typical stellar masses (median and quartiles\footnote{We note that the provided figures should be interpreted as real scatter in our sample, and therefore we use a different sub/super-script format compared to how we provide uncertainties.}) for our sample are $\log \mathrm{M}_\bigstar/\mathrm{M}_\odot=7.4_{6.8}^{7.9}$, for  BC03, $7.3_{7.1}^{7.4}$ for Z11, and $6.2_{5.9}^{6.4}$ for the BPASS models. The nearly constant star formation assumed by \texttt{prospector} yields $\log \mathrm{M}_\bigstar/\mathrm{M}_\odot=7.4_{7.3}^{7.5}$. The difference between the median stellar mass for the $z\sim17$ and $z\sim25$ samples is 0.2-0.4~dex. Comparing these estimates with the dark matter halo masses obtained by using rank-preserving abundance matching shown in Figure~\ref{fig:m/l}, we obtain stellar-to-dark matter mass ratios around a few percent at $z\sim25$, and 10 times smaller at $z\sim17$.

Most of our galaxies show UV slopes and masses similar to the comparison samples (at $10\lesssim z\lesssim12$) plotted in the figure (from \citealt{2023MNRAS.520...14C}, \citealt{2024arXiv240410751A}, and \citealt{2024ApJ...964L..24M}), which imply the presence of very young (1--10~Myr) stellar populations with significant nebular emission or older stars, resulting in slopes around $\beta\sim-2.5$. Notably, we detect three galaxies with extreme UV colors ($\beta\sim-3$), two at $z\sim17$ and one at $z\sim25$. In fact, their SED is extremely blue and can only be reproduced with very young and metal-poor models (as anticipated in our discussion of Figure~\ref{fig:m/l}), as well as very small or negligible dust obscuration (e.g. \citealt{2025ApJ...982....7N}), and also including significant HeII$\lambda$1640 emission.

In fact, the SED of these extremely blue galaxies could only be fitted by switching off the nebular continuum emission, i.e., by assuming that the escape fraction of ionizing photons is low and the UV is dominated by stellar emission from very young and metal-poor stars with low dust content. This could be explained by the presence of strong bursts of star formation (as in the feedback-free burst models of \citealt{2023MNRAS.523.3201D}, or analogous to the possible bursty nature of high redshift galaxies discussed in \citealt{2023MNRAS.525.3254S} or \citealt{2024ApJ...975..192G}) that quickly clear the inter-stellar medium, as suggested for the dust by \citet{2023MNRAS.522.3986F, 2024A&A...689A.310F}.

Finally, we note that, except for the bluest galaxies, the UV slope of the $z>17$ candidates are generally consistent with those expected from the PopIII stellar systems observed at various stages of their initial bursts, as discussed by \citet{2024arXiv240803189K}. For age $t<10$ Myr, the predicted far UV emissivity and the slope of the continuum emission of these systems is the result of different proportions of stellar and nebular continuum, which is expected around $\beta\approx -2.5$, assuming stellar populations synthesis models and IMF as made here. Galaxies with the bluest UV continuum, i.e., $\beta\approx-3$, would be consistent with this scenario only for IMF that include very massive stars or, as mentioned before, in presence of mechanisms capable to suppress the nebular continuum.

\begin{figure*}[ht!]
\centering
\includegraphics[clip, trim=3.0cm 0.2cm 2.0cm 1.5cm,scale=0.57]{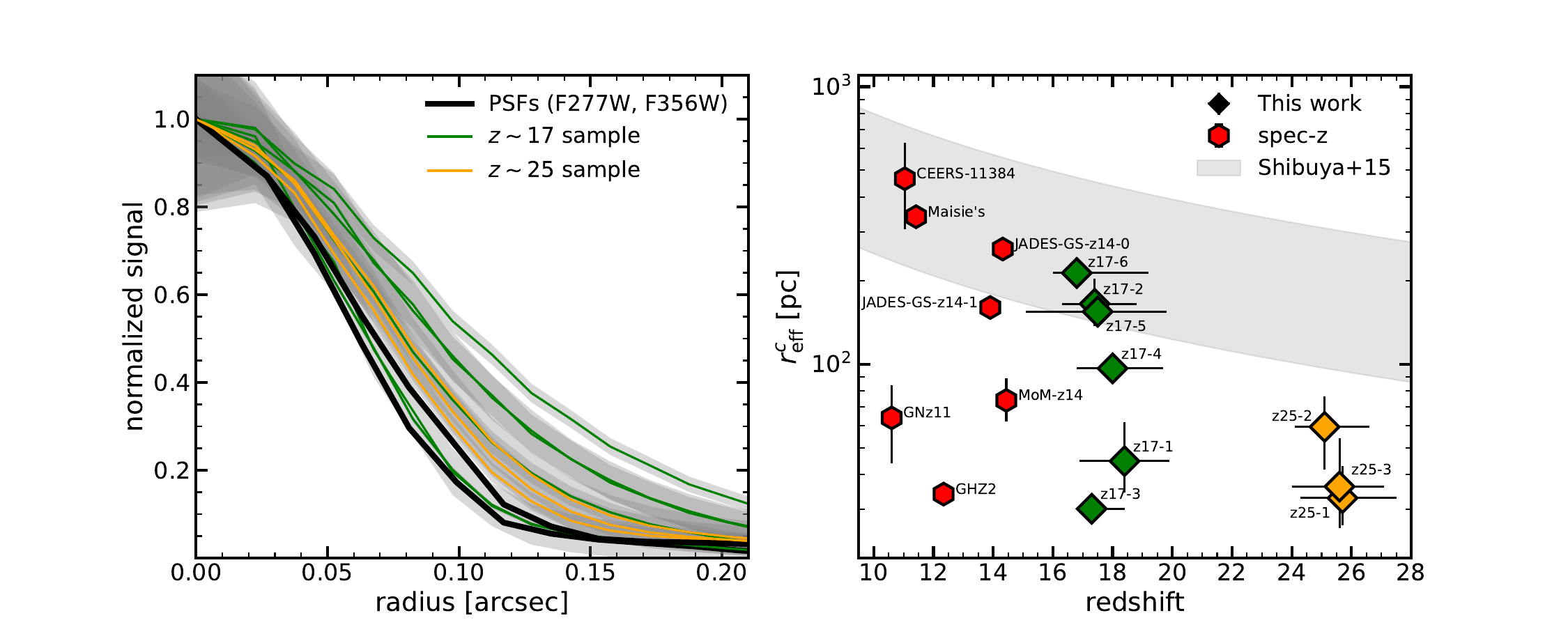}
\caption{\label{fig:morph} On the left, light profiles of the sources in our sample of $z>16$ galaxy candidates. The $z\sim17$ sources are shown in green (profiles built with the F277W data) and the $z\sim25$ sources in orange (using F356W data), with uncertainties depicted as a gray shade. For reference, we also plot in black the profiles for the F277W and F356W empirical PSF (built with stars in the MIDIS+NGDEEP dataset). On the right, we show circularized effective radii $vs.$ redshift for spectroscopically confirmed galaxies at $z>10$ (red hexagons, see Figure~\ref{fig:depths} for references), and our $z>16$ galaxy candidates, using the same colors as on the left panel. We label the points with the identification of each source. We remark that midis-z17-1, midis-z17-3, and all $z\sim25$ candidates are consistent with a point-like source, and therefore their sizes must be interpreted as upper limits. The  gray shaded region shows the evolution of sizes presented in \citet{2015ApJS..219...15S}.
}
\end{figure*}

The average for the mass-weighted age according to the BC03 modeling is 44~Myr for the whole sample, 55~Myr for the $z\sim17$ subsample and 17~Myr for the $z\sim25$ sources, suggesting that the onset of the star formation in the $z\sim17$ ($z\sim25$) sample would be $z_0\sim20$  ($z_0\sim$28). These ages are similar to those obtained for spectroscopically confirmed galaxies at $z=10-14$ \citep{2024Natur.633..318C,2024arXiv240518462H,2025NatAs...9..155Z}. With the Z11 models, we obtain mass-weighted ages 37, 44, and 21~Myr for the whole sample and the $z\sim17$ and  $z\sim25$ selections respectively. For BPASS, the ages are significantly younger, all below 10~Myr. \texttt{prospector} obtains 60, 66 and 42~Myr for the whole sample and the $z\sim17$ and  $z\sim25$ selections respectively. The stellar ages we obtain, combined with the information provided in Figure~\ref{fig:m/l}, imply that a star formation efficiency (i.e., mass in stars divided by mass in baryons) around 15\% (that is assumed in Figure~\ref{fig:m/l}) matches well the observations at $z\sim17$. This is consistent with the results reported in, for example, \citet{2022ApJS..259...20H} for lower redshifts (but still $z\gtrsim10$). For $z\sim25$, since the ages are still not as young as the 1~Myr shown in Figure~\ref{fig:m/l}, we need $\times 2-3$ times higher star formation efficiencies (around 40\%). The SED fitting results also imply that emissivities are changing with redshift (at least, due to age), as noted in the previous subsection.

The average attenuation is $\mathrm{A(V)}=0.04$~mag for the whole sample and sub-samples for all \textsc{synthesizer-AGN} models. \texttt{prospector}, however, gives an average of $\mathrm{A(V)}=0.3$~mag, as the constant SFH models need higher attenuations to fit the SEDs.

\subsection{Morphology of the \texorpdfstring{$\MakeLowercase{z}\sim17$}{z~17} 
  and \texorpdfstring{$\MakeLowercase{z}\sim25$}{z~25} galaxy candidates}
\label{sec:morphology}

We performed a morphological analysis of our $z>16$ galaxy candidates using the \texttt{pysersic} code \citep{2023JOSS....8.5703P}. We fitted the images corresponding to the bluest band redwards of the break, i.e., the one presenting a better spatial resolution and enough signal to carry out the analysis. More specifically, we used the F277W and F356W images for the $z\sim17$ and $z\sim25$ samples, respectively. We used single-component fits following the \citet{1968adga.book.....S} profiles,  allowing the S\'ersic index to vary within $n\in[0.5,2.0]$ and ellipticity smaller than $\epsilon<0.5$. We provide circularized effective radii in Table~\ref{tab:props}.

Figure~\ref{fig:morph} shows our results of the morphological analysis of our sample. On the left, we show that two of the $z\sim17$ galaxy candidates are nearly point-like, the rest shows a measurable extension. For $z\sim25$, all candidates are consistent, within errors, with the profile for a point-like source (or marginally resolved).

On the right panel of Figure~\ref{fig:morph}, we compare our sample with spectroscopically confirmed galaxies at $z>10$ (see Figure~\ref{fig:depths} for references). We remark that some sources are consistent with being point-like, so their sizes must be interpreted as upper limits. Around one third of our sample follows the redshift evolution of galaxy sizes presented in \citet{2015ApJS..219...15S}. The sizes for the rest of sources are smaller than 100~pc, consistent with galaxies such as GNz11 or GHz2 which have been claimed to host AGN \citep[][see also \citealt{2024ApJ...961L..39S} and \citealt{2025arXiv250318850M} for a discussion about the appearance of supermassive black holes in the very early Universe]{2024Natur.627...59M,2024ApJ...972..143C}, or MoM14 \citep{2025arXiv250511263N}, which presents a high nitrogen abundance similar to the two previously mentioned galaxies \citep{2023MNRAS.523.3516C,2023A&A...677A..88B,2025ApJ...981..136S}. 
 
\section{Summary and conclusions}
\label{sec:conclusions}

With exposure times in broad-band NIRCam filters between 33 and 90 hours in the common area, and 15 to 45~hours in non-overlapping regions, the combination of MIDIS and NGDEEP data provides the deepest JWST dataset to date, reaching $5\sigma$ depths fainter than 31~mag. We utilize these data to search for $z>16$ galaxies identified as F200W and F277W dropouts.

We find six $z\sim17$ and three $z\sim25$ galaxy candidates, after a selection based on photometry, photometric redshift probability distributions, and visual inspection. The estimated average stellar masses for these sources are  $\log\mathrm{M}_\bigstar/\mathrm{M}_\odot\sim7.4$, with mass-weighted stellar ages around 30~Myr, and very low dust content $\mathrm{A(V)}\sim0.05$~mag. Typical UV spectral slopes are $\beta\sim-2.5$, with three galaxies presenting extremely blue values compatible with very metal-poor models, significant HeII$\lambda$1640 emission, and/or a low to negligible contribution to the UV emission from the nebular continuum that could be explained with high escape fractions of ionizing photons. The candidates are compact, with two out of the six $z\sim17$ galaxies and all $z\sim25$ candidates exhibiting point-like morphologies translating to circularized effective radii around 50~pc, and four sources at $z\sim17$ being more extended, radii around  100-250~pc.

We construct UV luminosity functions and integrate them to get the cosmic luminosity density. Compared with results in the literature at $z\leq12$,  we find that the number density of galaxies around M$_\mathrm{UV}\sim-18.5$~mag decreases by a factor of $\sim4$ from $z\sim12$ to $z\sim17$ and a factor of $\sim25$ to $z\sim25$. A similar evolution is observed for the luminosity density, which evolves proportional to $(1+z)^{-5.3}$ at $z\gtrsim10$. Compared to a compilation of recent galaxy evolution models (\citealt{2023MNRAS.523.3201D}, \citealt{2024A&A...690A.108L}, \citealt{2023MNRAS.522.3986F}, \citealt{2025arXiv250505442S}, \citealt{2025arXiv250418618Y}), we find relatively good agreement up to $z\sim17$ with models in which the star formation efficiency was higher at early times, with a marked evolution at $z\gtrsim12$ (but see also \citealt{2025MNRAS.536..988F}). 
There are currently few robust predictions of the number density of galaxies at $z\gtrsim 20$ in the literature, however, the continuing rapid drop in the comoving number density of halos that are above the atomic cooling limit ($\mathrm{M}_h \sim 10^7 \mathrm{M}_\odot$ at $z=20$) suggests that an additional increase in the production efficiency of UV photons per unit halo mass may be required at $z \gtrsim 17$. This could be due to a higher conversion efficiency of gas into stars, more efficient production of UV light per unit stellar mass (e.g. from stellar populations with a top heavy IMF), or a contribution to the UV light from accreting black holes. 

Using an abundance matching technique, we infer that the galaxies we have selected as $z\sim17$ and $z\sim25$ candidates reside in  M$_\mathrm{DM}=10^{9.5}$~M$_\odot$ and  $10^{8.5}$~M$_\odot$ halos, respectively. Given that these galaxy populations have similar intrinsic rest-UV luminosities, this again suggests that the typical conversion efficiency of baryons into stars, or the photon production efficiency per unit stellar mass formed, or both, must have evolved to higher values over this time interval. 
Our analysis suggests typical mass-to-light ratios around $10^{-6}$~M$_\odot$/L$_{\odot}^\mathrm{UV}$ for stellar masses, and $10^{-4.5}$~M$_\odot$/L$_{\odot}^{UV}$ for dark matter masses. Assuming that 15\% of the halo baryon budget has been turned into stars, the reported mass-to-light values are consistent with stellar population models with young ages (1-30 Myr), low metallicities ($<0.1$~Z$_\odot$), and a top-heavy IMF. A constant star formation efficiency  would imply that an evolution on the emissivity of halos must exist due stronger burstiness, lower metallicities, a top-heavy IMF and/or non-stellar emission (i.e., from AGN) at higher redshifts. If the emissivity of galaxies does not evolve with redshift, i.e., galaxies are characterized by the same mass-to-light ratios independent of their redshift, and if that emissivity is dominated by stars, then the star formation efficiency must evolve, increasing with redshift, from a few percent at $z\sim10$ to $\sim20$\% at $z\sim17$, and at least $\sim60$\% at $z\sim25$.

Overall, our results indicate very rapid evolution of the formation of galaxies from $z\sim25$ to $z\sim12$. This points to the epoch between 100 and 350~Myr after the Big Bang as the rise of the galaxy formation empire, when galaxies started to be ubiquitous in the Universe. From the observational point of view, our conclusion in this paper is that selecting robust galaxy candidates at $z\sim17$ and $z\sim25$ requires, given the strong evolution, reaching magnitudes in the selection bands (F277W and redder filters) of at least 31~mag, $\sim$1.5~mag deeper in the dropout filter (F200W and F277W), achieavable with $\geq$100~hours integrations per band, expecting up to ten candidates per NIRCam pointing.

\clearpage

\begin{acknowledgements}
We thank the two anonymous referees for their useful comments that have helped to strengthen the robustness of the results presented in this paper.
We thank Andrea Ferrara and Brant Robertson  for providing their latest simulation predictions and observational results for this paper, and Marco Castellano for useful discussions. P.G.P.-G. and L.C. acknowledge support from grant PID2022-139567NB-I00 funded by Spanish Ministerio de Ciencia e Innovaci\'on MCIN/AEI/10.13039/501100011033, FEDER {\it Una manera de hacer Europa}.  This work has made use of the Rainbow Cosmological Surveys Database, which is operated by the Centro de Astrobiología (CAB), CSIC-INTA. The project that gave rise to these results received the support of a fellowship from the “la Caixa” Foundation (ID 100010434). The fellowship code is LCF/BQ/PR24/12050015. G.\"O. and J.M. acknowledges support from the Swedish National Space Agency (SNSA) and the Swedish Research Council (VR). L.C. and L.C. acknowledge support by grant PIB2021-127718NB-100 from the Spanish Ministry of Science and Innovation/State Agency of Research MCIN/AEI/10.13039/501100011033 and by “ERDF A way of making Europe”. 
This work was supported by research grants (VIL16599, VIL54489) from VILLUM FONDEN. P.N. acknowledges support from the Gordon and Betty Moore Foundation and the John Templeton Foundation that fund the Black Hole Initiative (BHI) at Harvard University. K.I.C. acknowledges funding from the Dutch Research Council (NWO) through the award of the Vici Grant VI.C.212.036 and funding from the Netherlands Research School for Astronomy (NOVA).  The Cosmic Dawn Center (DAWN) is funded by the Danish National Research Foundation (DNRF) under grant No. 140. A.E. and F.P. acknowledge support through the German Space Agency DLR 50OS1501 and DLR 50OS2001 from 2015 to 2023. RSS is supported by the Flatiron Institute, which is operated by the Simons Foundation. 

This work is based on observations made with the NASA/ESA/CSA James Webb Space Telescope. The data were obtained from the Mikulski Archive for Space Telescopes (MAST) at the Space Telescope Science Institute, which is operated by the Association of Universities for Research in Astronomy, Inc., under NASA contract NAS 5-03127 for JWST. These observations are associated with programs \#1283, \#2079, and \#6511.
The specific observations analyzed can be accessed for the MIRI Deep Survey, the NGDEEP Epochs 1 and 2 NIRCam imaging data, and the MIDIS-red epoch 1 dataset, which can be obtained from the following MAST repositories: \dataset[doi: 10.17909/je9x-d314]{https://doi.org/10.17909/je9x-d314}, \dataset[doi: 10.17909/d3rq-qg92]{https://doi.org/10.17909/d3rq-qg92}, and \dataset[doi: 10.17909/jwes-ed69]{https://doi.org/10.17909/jwes-ed69}.

\end{acknowledgements}

\facilities{HST (ACS), JWST (NIRCam)}

\bibliography{midisngdeep_zgt16}{}

\begin{thebibliography}{}
\expandafter\ifx\csname natexlab\endcsname\relax\def\natexlab#1{#1}\fi
\providecommand{\url}[1]{\href{#1}{#1}}
\providecommand{\dodoi}[1]{doi:~\href{http://doi.org/#1}{\nolinkurl{#1}}}
\providecommand{\doeprint}[1]{\href{http://ascl.net/#1}{\nolinkurl{http://ascl.net/#1}}}
\providecommand{\doarXiv}[1]{\href{https://arxiv.org/abs/#1}{\nolinkurl{https://arxiv.org/abs/#1}}}

\bibitem[{{Adamo} {et~al.}(2024){Adamo}, {Atek}, {Bagley}, {Ba{\~n}ados}, {Barrow}, {Berg}, {Bezanson}, {Brada{\v{c}}}, {Brammer}, {Carnall}, {Chisholm}, {Coe}, {Dayal}, {Eisenstein}, {Eldridge}, {Ferrara}, {Fujimoto}, {de Graaff}, {Habouzit}, {Hutchison}, {Kartaltepe}, {Kassin}, {Kriek}, {Labb{\'e}}, {Maiolino}, {Marques-Chaves}, {Maseda}, {Mason}, {Matthee}, {McQuinn}, {Meynet}, {Naidu}, {Oesch}, {Pentericci}, {P{\'e}rez-Gonz{\'a}lez}, {Rigby}, {Roberts-Borsani}, {Schaerer}, {Shapley}, {Stark}, {Stiavelli}, {Strom}, {Vanzella}, {Wang}, {Wilkins}, {Williams}, {Willott}, {Wylezalek}, \& {Nota}}]{2024arXiv240521054A}
{Adamo}, A., {Atek}, H., {Bagley}, M.~B., {et~al.} 2024, arXiv e-prints, arXiv:2405.21054, \dodoi{10.48550/arXiv.2405.21054}

\bibitem[{{Adams} {et~al.}(2024){Adams}, {Conselice}, {Austin}, {Harvey}, {Ferreira}, {Trussler}, {Juod{\v{z}}balis}, {Li}, {Windhorst}, {Cohen}, {Jansen}, {Summers}, {Tompkins}, {Driver}, {Robotham}, {D'Silva}, {Yan}, {Coe}, {Frye}, {Grogin}, {Koekemoer}, {Marshall}, {Pirzkal}, {Ryan}, {Maksym}, {Rutkowski}, {Willmer}, {Hammel}, {Nonino}, {Bhatawdekar}, {Wilkins}, {Bradley}, {Broadhurst}, {Cheng}, {Dole}, {Hathi}, \& {Zitrin}}]{2024ApJ...965..169A}
{Adams}, N.~J., {Conselice}, C.~J., {Austin}, D., {et~al.} 2024, \apj, 965, 169, \dodoi{10.3847/1538-4357/ad2a7b}

\bibitem[{{Akhlaghi} \& {Ichikawa}(2015)}]{2015ApJS..220....1A}
{Akhlaghi}, M., \& {Ichikawa}, T. 2015, \apjs, 220, 1, \dodoi{10.1088/0067-0049/220/1/1}

\bibitem[{{Arrabal Haro} {et~al.}(2023{\natexlab{a}}){Arrabal Haro}, {Dickinson}, {Finkelstein}, {Kartaltepe}, {Donnan}, {Burgarella}, {Carnall}, {Cullen}, {Dunlop}, {Fern{\'a}ndez}, {Fujimoto}, {Jung}, {Krips}, {Larson}, {Papovich}, {P{\'e}rez-Gonz{\'a}lez}, {Amor{\'\i}n}, {Bagley}, {Buat}, {Casey}, {Chworowsky}, {Cohen}, {Ferguson}, {Giavalisco}, {Huertas-Company}, {Hutchison}, {Kocevski}, {Koekemoer}, {Lucas}, {McLeod}, {McLure}, {Pirzkal}, {Seill{\'e}}, {Trump}, {Weiner}, {Wilkins}, \& {Zavala}}]{2023Natur.622..707A}
{Arrabal Haro}, P., {Dickinson}, M., {Finkelstein}, S.~L., {et~al.} 2023{\natexlab{a}}, \nat, 622, 707, \dodoi{10.1038/s41586-023-06521-7}

\bibitem[{{Arrabal Haro} {et~al.}(2023{\natexlab{b}}){Arrabal Haro}, {Dickinson}, {Finkelstein}, {Fujimoto}, {Fern{\'a}ndez}, {Kartaltepe}, {Jung}, {Cole}, {Burgarella}, {Chworowsky}, {Hutchison}, {Morales}, {Papovich}, {Simons}, {Amor{\'\i}n}, {Backhaus}, {Bagley}, {Bisigello}, {Calabr{\`o}}, {Castellano}, {Cleri}, {Dav{\'e}}, {Dekel}, {Ferguson}, {Fontana}, {Gawiser}, {Giavalisco}, {Harish}, {Hathi}, {Hirschmann}, {Holwerda}, {Huertas-Company}, {Koekemoer}, {Larson}, {Lucas}, {Mobasher}, {P{\'e}rez-Gonz{\'a}lez}, {Pirzkal}, {Rose}, {Santini}, {Trump}, {de la Vega}, {Wang}, {Weiner}, {Wilkins}, {Yang}, {Yung}, \& {Zavala}}]{2023ApJ...951L..22A}
---. 2023{\natexlab{b}}, \apjl, 951, L22, \dodoi{10.3847/2041-8213/acdd54}

\bibitem[{{Austin} {et~al.}(2023){Austin}, {Adams}, {Conselice}, {Harvey}, {Ormerod}, {Trussler}, {Li}, {Ferreira}, {Dayal}, \& {Juod{\v{z}}balis}}]{2023ApJ...952L...7A}
{Austin}, D., {Adams}, N., {Conselice}, C.~J., {et~al.} 2023, \apjl, 952, L7, \dodoi{10.3847/2041-8213/ace18d}

\bibitem[{{Austin} {et~al.}(2024){Austin}, {Conselice}, {Adams}, {Harvey}, {Duan}, {Trussler}, {Li}, {Juodzbalis}, {Ormerod}, {Ferreira}, {Westcott}, {Harris}, {Wilkins}, {Bhatawdekar}, {Caruana}, {Coe}, {Cohen}, {Driver}, {D'Silva}, {Frye}, {Furtak}, {Grogin}, {Hathi}, {Holwerda}, {Jansen}, {Koekemoer}, {Marshall}, {Nonino}, {Ortiz}, {Pirzkal}, {Robotham}, {Ryan}, {Summers}, {Willmer}, {Windhorst}, {Yan}, \& {Zackrisson}}]{2024arXiv240410751A}
{Austin}, D., {Conselice}, C.~J., {Adams}, N.~J., {et~al.} 2024, arXiv e-prints, arXiv:2404.10751, \dodoi{10.48550/arXiv.2404.10751}

\bibitem[{{Bagley} {et~al.}(2023){Bagley}, {Finkelstein}, {Koekemoer}, {Ferguson}, {Arrabal Haro}, {Dickinson}, {Kartaltepe}, {Papovich}, {P{\'e}rez-Gonz{\'a}lez}, {Pirzkal}, {Somerville}, {Willmer}, {Yang}, {Yung}, {Fontana}, {Grazian}, {Grogin}, {Hirschmann}, {Kewley}, {Kirkpatrick}, {Kocevski}, {Lotz}, {Medrano}, {Morales}, {Pentericci}, {Ravindranath}, {Trump}, {Wilkins}, {Calabr{\`o}}, {Cooper}, {Costantin}, {de la Vega}, {Hilbert}, {Hutchison}, {Larson}, {Lucas}, {McGrath}, {Ryan}, {Wang}, \& {Wuyts}}]{2023ApJ...946L..12B}
{Bagley}, M.~B., {Finkelstein}, S.~L., {Koekemoer}, A.~M., {et~al.} 2023, \apjl, 946, L12, \dodoi{10.3847/2041-8213/acbb08}

\bibitem[{{Bagley} {et~al.}(2024){Bagley}, {Pirzkal}, {Finkelstein}, {Papovich}, {Berg}, {Lotz}, {Leung}, {Ferguson}, {Koekemoer}, {Dickinson}, {Kartaltepe}, {Kocevski}, {Somerville}, {Yung}, {Backhaus}, {Casey}, {Castellano}, {Ch{\'a}vez Ortiz}, {Chworowsky}, {Cox}, {Dav{\'e}}, {Davis}, {Estrada-Carpenter}, {Fontana}, {Fujimoto}, {Gardner}, {Giavalisco}, {Grazian}, {Grogin}, {Hathi}, {Hutchison}, {Jaskot}, {Jung}, {Kewley}, {Kirkpatrick}, {Larson}, {Matharu}, {Natarajan}, {Pentericci}, {P{\'e}rez-Gonz{\'a}lez}, {Ravindranath}, {Rothberg}, {Ryan}, {Shen}, {Simons}, {Snyder}, {Trump}, \& {Wilkins}}]{2024ApJ...965L...6B}
{Bagley}, M.~B., {Pirzkal}, N., {Finkelstein}, S.~L., {et~al.} 2024, \apjl, 965, L6, \dodoi{10.3847/2041-8213/ad2f31}

\bibitem[{{Behroozi} {et~al.}(2019){Behroozi}, {Wechsler}, {Hearin}, \& {Conroy}}]{2019MNRAS.488.3143B}
{Behroozi}, P., {Wechsler}, R.~H., {Hearin}, A.~P., \& {Conroy}, C. 2019, \mnras, 488, 3143, \dodoi{10.1093/mnras/stz1182}

\bibitem[{{Behroozi} {et~al.}(2010){Behroozi}, {Conroy}, \& {Wechsler}}]{2010ApJ...717..379B}
{Behroozi}, P.~S., {Conroy}, C., \& {Wechsler}, R.~H. 2010, \apj, 717, 379, \dodoi{10.1088/0004-637X/717/1/379}

\bibitem[{{Behroozi} {et~al.}(2013){Behroozi}, {Wechsler}, \& {Conroy}}]{2013ApJ...770...57B}
{Behroozi}, P.~S., {Wechsler}, R.~H., \& {Conroy}, C. 2013, \apj, 770, 57, \dodoi{10.1088/0004-637X/770/1/57}

\bibitem[{{Beiler} {et~al.}(2023){Beiler}, {Cushing}, {Kirkpatrick}, {Schneider}, {Mukherjee}, \& {Marley}}]{2023ApJ...951L..48B}
{Beiler}, S.~A., {Cushing}, M.~C., {Kirkpatrick}, J.~D., {et~al.} 2023, \apjl, 951, L48, \dodoi{10.3847/2041-8213/ace32c}

\bibitem[{{Bertin} \& {Arnouts}(1996)}]{1996A&AS..117..393B}
{Bertin}, E., \& {Arnouts}, S. 1996, \aaps, 117, 393, \dodoi{10.1051/aas:1996164}

\bibitem[{{Bezanson} {et~al.}(2024){Bezanson}, {Labbe}, {Whitaker}, {Leja}, {Price}, {Franx}, {Brammer}, {Marchesini}, {Zitrin}, {Wang}, {Weaver}, {Furtak}, {Atek}, {Coe}, {Cutler}, {Dayal}, {van Dokkum}, {Feldmann}, {F{\"o}rster Schreiber}, {Fujimoto}, {Geha}, {Glazebrook}, {de Graaff}, {Greene}, {Juneau}, {Kassin}, {Kriek}, {Khullar}, {Maseda}, {Mowla}, {Muzzin}, {Nanayakkara}, {Nelson}, {Oesch}, {Pacifici}, {Pan}, {Papovich}, {Setton}, {Shapley}, {Smit}, {Stefanon}, {Taylor}, \& {Williams}}]{2024ApJ...974...92B}
{Bezanson}, R., {Labbe}, I., {Whitaker}, K.~E., {et~al.} 2024, \apj, 974, 92, \dodoi{10.3847/1538-4357/ad66cf}

\bibitem[{{Bhowmick} {et~al.}(2020){Bhowmick}, {Somerville}, {Di Matteo}, {Wilkins}, {Feng}, \& {Tenneti}}]{2020MNRAS.496..754B}
{Bhowmick}, A.~K., {Somerville}, R.~S., {Di Matteo}, T., {et~al.} 2020, \mnras, 496, 754, \dodoi{10.1093/mnras/staa1605}

\bibitem[{{Bouwens} {et~al.}(2022{\natexlab{a}}){Bouwens}, {Illingworth}, {Oesch}, {Stefanon}, {Naidu}, {van Leeuwen}, \& {Magee}}]{2022arXiv221206683B}
{Bouwens}, R., {Illingworth}, G., {Oesch}, P., {et~al.} 2022{\natexlab{a}}, arXiv e-prints, arXiv:2212.06683.
\newblock \doarXiv{2212.06683}

\bibitem[{{Bouwens} {et~al.}(2022{\natexlab{b}}){Bouwens}, {Stefanon}, {Brammer}, {Oesch}, {Herard-Demanche}, {Illingworth}, {Matthee}, {Naidu}, {van Dokkum}, \& {van Leeuwen}}]{2022arXiv221102607B}
{Bouwens}, R.~J., {Stefanon}, M., {Brammer}, G., {et~al.} 2022{\natexlab{b}}, arXiv e-prints, arXiv:2211.02607, \dodoi{10.48550/arXiv.2211.02607}

\bibitem[{{Brammer} {et~al.}(2008){Brammer}, {van Dokkum}, \& {Coppi}}]{2008ApJ...686.1503B}
{Brammer}, G.~B., {van Dokkum}, P.~G., \& {Coppi}, P. 2008, \apj, 686, 1503, \dodoi{10.1086/591786}

\bibitem[{{Bruzual} \& {Charlot}(2003)}]{2003MNRAS.344.1000B}
{Bruzual}, G., \& {Charlot}, S. 2003, \mnras, 344, 1000, \dodoi{10.1046/j.1365-8711.2003.06897.x}

\bibitem[{{Bunker} {et~al.}(2023){Bunker}, {Saxena}, {Cameron}, {Willott}, {Curtis-Lake}, {Jakobsen}, {Carniani}, {Smit}, {Maiolino}, {Witstok}, {Curti}, {D'Eugenio}, {Jones}, {Ferruit}, {Arribas}, {Charlot}, {Chevallard}, {Giardino}, {de Graaff}, {Looser}, {L{\"u}tzgendorf}, {Maseda}, {Rawle}, {Rix}, {Del Pino}, {Alberts}, {Egami}, {Eisenstein}, {Endsley}, {Hainline}, {Hausen}, {Johnson}, {Rieke}, {Rieke}, {Robertson}, {Shivaei}, {Stark}, {Sun}, {Tacchella}, {Tang}, {Williams}, {Willmer}, {Baker}, {Baum}, {Bhatawdekar}, {Bowler}, {Boyett}, {Chen}, {Circosta}, {Helton}, {Ji}, {Kumari}, {Lyu}, {Nelson}, {Parlanti}, {Perna}, {Sandles}, {Scholtz}, {Suess}, {Topping}, {{\"U}bler}, {Wallace}, \& {Whitler}}]{2023A&A...677A..88B}
{Bunker}, A.~J., {Saxena}, A., {Cameron}, A.~J., {et~al.} 2023, \aap, 677, A88, \dodoi{10.1051/0004-6361/202346159}

\bibitem[{{Calzetti} {et~al.}(2000){Calzetti}, {Armus}, {Bohlin}, {Kinney}, {Koornneef}, \& {Storchi-Bergmann}}]{2000ApJ...533..682C}
{Calzetti}, D., {Armus}, L., {Bohlin}, R.~C., {et~al.} 2000, \apj, 533, 682, \dodoi{10.1086/308692}

\bibitem[{{Cameron} {et~al.}(2023){Cameron}, {Katz}, {Rey}, \& {Saxena}}]{2023MNRAS.523.3516C}
{Cameron}, A.~J., {Katz}, H., {Rey}, M.~P., \& {Saxena}, A. 2023, \mnras, 523, 3516, \dodoi{10.1093/mnras/stad1579}

\bibitem[{{Carniani} {et~al.}(2024){Carniani}, {Hainline}, {D'Eugenio}, {Eisenstein}, {Jakobsen}, {Witstok}, {Johnson}, {Chevallard}, {Maiolino}, {Helton}, {Willott}, {Robertson}, {Alberts}, {Arribas}, {Baker}, {Bhatawdekar}, {Boyett}, {Bunker}, {Cameron}, {Cargile}, {Charlot}, {Curti}, {Curtis-Lake}, {Egami}, {Giardino}, {Isaak}, {Ji}, {Jones}, {Kumari}, {Maseda}, {Parlanti}, {P{\'e}rez-Gonz{\'a}lez}, {Rawle}, {Rieke}, {Rieke}, {Del Pino}, {Saxena}, {Scholtz}, {Smit}, {Sun}, {Tacchella}, {{\"U}bler}, {Venturi}, {Williams}, \& {Willmer}}]{2024Natur.633..318C}
{Carniani}, S., {Hainline}, K., {D'Eugenio}, F., {et~al.} 2024, \nat, 633, 318, \dodoi{10.1038/s41586-024-07860-9}

\bibitem[{{Casey} {et~al.}(2023){Casey}, {Kartaltepe}, {Drakos}, {Franco}, {Harish}, {Paquereau}, {Ilbert}, {Rose}, {Cox}, {Nightingale}, {Robertson}, {Silverman}, {Koekemoer}, {Massey}, {McCracken}, {Rhodes}, {Akins}, {Allen}, {Amvrosiadis}, {Arango-Toro}, {Bagley}, {Bongiorno}, {Capak}, {Champagne}, {Chartab}, {Ch{\'a}vez Ortiz}, {Chworowsky}, {Cooke}, {Cooper}, {Darvish}, {Ding}, {Faisst}, {Finkelstein}, {Fujimoto}, {Gentile}, {Gillman}, {Gould}, {Gozaliasl}, {Hayward}, {He}, {Hemmati}, {Hirschmann}, {Jahnke}, {Jin}, {Khostovan}, {Kokorev}, {Lambrides}, {Laigle}, {Larson}, {Leung}, {Liu}, {Liaudat}, {Long}, {Magdis}, {Mahler}, {Mainieri}, {Manning}, {Maraston}, {Martin}, {McCleary}, {McKinney}, {McPartland}, {Mobasher}, {Pattnaik}, {Renzini}, {Rich}, {Sanders}, {Sattari}, {Scognamiglio}, {Scoville}, {Sheth}, {Shuntov}, {Sparre}, {Suzuki}, {Talia}, {Toft}, {Trakhtenbrot}, {Urry}, {Valentino}, {Vanderhoof}, {Vardoulaki}, {Weaver}, {Whitaker}, {Wilkins}, {Yang}, \& {Zavala}}]{2023ApJ...954...31C}
{Casey}, C.~M., {Kartaltepe}, J.~S., {Drakos}, N.~E., {et~al.} 2023, \apj, 954, 31, \dodoi{10.3847/1538-4357/acc2bc}

\bibitem[{{Casey} {et~al.}(2024){Casey}, {Akins}, {Shuntov}, {Ilbert}, {Paquereau}, {Franco}, {Hayward}, {Finkelstein}, {Boylan-Kolchin}, {Robertson}, {Allen}, {Brinch}, {Cooper}, {Ding}, {Drakos}, {Faisst}, {Fujimoto}, {Gillman}, {Harish}, {Hirschmann}, {Jin}, {Kartaltepe}, {Koekemoer}, {Kokorev}, {Liu}, {Long}, {Magdis}, {Maraston}, {Martin}, {McCracken}, {McKinney}, {Mobasher}, {Rhodes}, {Rich}, {Sanders}, {Silverman}, {Toft}, {Vijayan}, {Weaver}, {Wilkins}, {Yang}, \& {Zavala}}]{2024ApJ...965...98C}
{Casey}, C.~M., {Akins}, H.~B., {Shuntov}, M., {et~al.} 2024, \apj, 965, 98, \dodoi{10.3847/1538-4357/ad2075}

\bibitem[{{Castellano} {et~al.}(2022){Castellano}, {Fontana}, {Treu}, {Santini}, {Merlin}, {Leethochawalit}, {Trenti}, {Vanzella}, {Mestric}, {Bonchi}, {Belfiori}, {Nonino}, {Paris}, {Polenta}, {Roberts-Borsani}, {Boyett}, {Brada{\v{c}}}, {Calabr{\`o}}, {Glazebrook}, {Grillo}, {Mascia}, {Mason}, {Mercurio}, {Morishita}, {Nanayakkara}, {Pentericci}, {Rosati}, {Vulcani}, {Wang}, \& {Yang}}]{2022ApJ...938L..15C}
{Castellano}, M., {Fontana}, A., {Treu}, T., {et~al.} 2022, \apjl, 938, L15, \dodoi{10.3847/2041-8213/ac94d0}

\bibitem[{{Castellano} {et~al.}(2023){Castellano}, {Fontana}, {Treu}, {Merlin}, {Santini}, {Bergamini}, {Grillo}, {Rosati}, {Acebron}, {Leethochawalit}, {Paris}, {Bonchi}, {Belfiori}, {Calabr{\`o}}, {Correnti}, {Nonino}, {Polenta}, {Trenti}, {Boyett}, {Brammer}, {Broadhurst}, {Caminha}, {Chen}, {Filippenko}, {Fortuni}, {Glazebrook}, {Mascia}, {Mason}, {Menci}, {Meneghetti}, {Mercurio}, {Metha}, {Morishita}, {Nanayakkara}, {Pentericci}, {Roberts-Borsani}, {Roy}, {Vanzella}, {Vulcani}, {Yang}, \& {Wang}}]{2023ApJ...948L..14C}
---. 2023, \apjl, 948, L14, \dodoi{10.3847/2041-8213/accea5}

\bibitem[{{Castellano} {et~al.}(2024){Castellano}, {Napolitano}, {Fontana}, {Roberts-Borsani}, {Treu}, {Vanzella}, {Zavala}, {Arrabal Haro}, {Calabr{\`o}}, {Llerena}, {Mascia}, {Merlin}, {Paris}, {Pentericci}, {Santini}, {Bakx}, {Bergamini}, {Cupani}, {Dickinson}, {Filippenko}, {Glazebrook}, {Grillo}, {Kelly}, {Malkan}, {Mason}, {Morishita}, {Nanayakkara}, {Rosati}, {Sani}, {Wang}, \& {Yoon}}]{2024ApJ...972..143C}
{Castellano}, M., {Napolitano}, L., {Fontana}, A., {et~al.} 2024, \apj, 972, 143, \dodoi{10.3847/1538-4357/ad5f88}

\bibitem[{{Castellano} {et~al.}(2025){Castellano}, {Fontana}, {Merlin}, {Santini}, {Napolitano}, {Menci}, {Calabr{\`o}}, {Paris}, {Pentericci}, {Zavala}, {Dickinson}, {Finkelstein}, {Treu}, {Amorin}, {Arrabal Haro}, {Bergamini}, {Bisigello}, {Daddi}, {Dayal}, {Dekel}, {Ferrara}, {Fortuni}, {Gandolfi}, {Giavalisco}, {Grillo}, {Guida}, {Hathi}, {Holwerda}, {Koekemoer}, {Kokorev}, {Li}, {Llerena}, {Lucas}, {Mascia}, {Metha}, {Morishita}, {Nanayakkara}, {Pacucci}, {P{\'e}rez-Gonz{\'a}lez}, {Roberts-Borsani}, {Rodighiero}, {Rosati}, {Salazar}, {Schneider}, {Somerville}, {Taylor}, {Trenti}, {Trinca}, {Wang}, {Watson}, {Yang}, \& {Yung}}]{2025arXiv250405893C}
{Castellano}, M., {Fontana}, A., {Merlin}, E., {et~al.} 2025, arXiv e-prints, arXiv:2504.05893, \dodoi{10.48550/arXiv.2504.05893}

\bibitem[{{Chabrier}(2003)}]{2003PASP..115..763C}
{Chabrier}, G. 2003, \pasp, 115, 763, \dodoi{10.1086/376392}

\bibitem[{{Conroy} \& {Gunn}(2010)}]{2010ascl.soft10043C}
{Conroy}, C., \& {Gunn}, J.~E. 2010, {FSPS: Flexible Stellar Population Synthesis}, Astrophysics Source Code Library, record ascl:1010.043.
\newblock \doeprint{1010.043}

\bibitem[{{Conselice} {et~al.}(2024){Conselice}, {Adams}, {Harvey}, {Austin}, {Ferreira}, {Ormerod}, {Duan}, {Trussler}, {Li}, {Juodzbalis}, {Westcott}, {Harris}, {Seeyave}, {Bluck}, {Windhorst}, {Bhatawdekar}, {Coe}, {Cohen}, {Cheng}, {Driver}, {Frye}, {Furtak}, {Grogin}, {Hathi}, {Holwerda}, {Jansen}, {Koekemoer}, {Marshall}, {Nonino}, {Robotham}, {Summers}, {Wilkins}, {Willmer}, {Yan}, \& {Zitrin}}]{2024arXiv240714973C}
{Conselice}, C.~J., {Adams}, N., {Harvey}, T., {et~al.} 2024, arXiv e-prints, arXiv:2407.14973, \dodoi{10.48550/arXiv.2407.14973}

\bibitem[{{Cullen} {et~al.}(2023){Cullen}, {McLure}, {McLeod}, {Dunlop}, {Donnan}, {Carnall}, {Bowler}, {Begley}, {Hamadouche}, \& {Stanton}}]{2023MNRAS.520...14C}
{Cullen}, F., {McLure}, R.~J., {McLeod}, D.~J., {et~al.} 2023, \mnras, 520, 14, \dodoi{10.1093/mnras/stad073}

\bibitem[{{Curti} {et~al.}(2023){Curti}, {D'Eugenio}, {Carniani}, {Maiolino}, {Sandles}, {Witstok}, {Baker}, {Bennett}, {Piotrowska}, {Tacchella}, {Charlot}, {Nakajima}, {Maheson}, {Mannucci}, {Amiri}, {Arribas}, {Belfiore}, {Bonaventura}, {Bunker}, {Chevallard}, {Cresci}, {Curtis-Lake}, {Hayden-Pawson}, {Jones}, {Kumari}, {Laseter}, {Looser}, {Marconi}, {Maseda}, {Scholtz}, {Smit}, {{\"U}bler}, \& {Wallace}}]{2023MNRAS.518..425C}
{Curti}, M., {D'Eugenio}, F., {Carniani}, S., {et~al.} 2023, \mnras, 518, 425, \dodoi{10.1093/mnras/stac2737}

\bibitem[{{Curtis-Lake} {et~al.}(2023){Curtis-Lake}, {Carniani}, {Cameron}, {Charlot}, {Jakobsen}, {Maiolino}, {Bunker}, {Witstok}, {Smit}, {Chevallard}, {Willott}, {Ferruit}, {Arribas}, {Bonaventura}, {Curti}, {D'Eugenio}, {Franx}, {Giardino}, {Looser}, {L{\"u}tzgendorf}, {Maseda}, {Rawle}, {Rix}, {Rodr{\'\i}guez del Pino}, {{\"U}bler}, {Sirianni}, {Dressler}, {Egami}, {Eisenstein}, {Endsley}, {Hainline}, {Hausen}, {Johnson}, {Rieke}, {Robertson}, {Shivaei}, {Stark}, {Tacchella}, {Williams}, {Willmer}, {Bhatawdekar}, {Bowler}, {Boyett}, {Chen}, {de Graaff}, {Helton}, {Hviding}, {Jones}, {Kumari}, {Lyu}, {Nelson}, {Perna}, {Sandles}, {Saxena}, {Suess}, {Sun}, {Topping}, {Wallace}, \& {Whitler}}]{2023NatAs...7..622C}
{Curtis-Lake}, E., {Carniani}, S., {Cameron}, A., {et~al.} 2023, Nature Astronomy, 7, 622, \dodoi{10.1038/s41550-023-01918-w}

\bibitem[{{Dekel} {et~al.}(2023){Dekel}, {Sarkar}, {Birnboim}, {Mandelker}, \& {Li}}]{2023MNRAS.523.3201D}
{Dekel}, A., {Sarkar}, K.~C., {Birnboim}, Y., {Mandelker}, N., \& {Li}, Z. 2023, \mnras, 523, 3201, \dodoi{10.1093/mnras/stad1557}

\bibitem[{{Donnan} {et~al.}(2023){Donnan}, {McLeod}, {Dunlop}, {McLure}, {Carnall}, {Begley}, {Cullen}, {Hamadouche}, {Bowler}, {Magee}, {McCracken}, {Milvang-Jensen}, {Moneti}, \& {Targett}}]{2023MNRAS.518.6011D}
{Donnan}, C.~T., {McLeod}, D.~J., {Dunlop}, J.~S., {et~al.} 2023, \mnras, 518, 6011, \dodoi{10.1093/mnras/stac3472}

\bibitem[{{Donnan} {et~al.}(2024){Donnan}, {McLure}, {Dunlop}, {McLeod}, {Magee}, {Arellano-C{\'o}rdova}, {Barrufet}, {Begley}, {Bowler}, {Carnall}, {Cullen}, {Ellis}, {Fontana}, {Illingworth}, {Grogin}, {Hamadouche}, {Koekemoer}, {Liu}, {Mason}, {Santini}, \& {Stanton}}]{2024MNRAS.533.3222D}
{Donnan}, C.~T., {McLure}, R.~J., {Dunlop}, J.~S., {et~al.} 2024, \mnras, 533, 3222, \dodoi{10.1093/mnras/stae2037}

\bibitem[{{Efstathiou} {et~al.}(1988){Efstathiou}, {Ellis}, \& {Peterson}}]{1988MNRAS.232..431E}
{Efstathiou}, G., {Ellis}, R.~S., \& {Peterson}, B.~A. 1988, \mnras, 232, 431, \dodoi{10.1093/mnras/232.2.431}

\bibitem[{{Eisenstein} {et~al.}(2023){Eisenstein}, {Willott}, {Alberts}, {Arribas}, {Bonaventura}, {Bunker}, {Cameron}, {Carniani}, {Charlot}, {Curtis-Lake}, {D'Eugenio}, {Endsley}, {Ferruit}, {Giardino}, {Hainline}, {Hausen}, {Jakobsen}, {Johnson}, {Maiolino}, {Rieke}, {Rieke}, {Rix}, {Robertson}, {Stark}, {Tacchella}, {Williams}, {Willmer}, {Baker}, {Baum}, {Bhatawdekar}, {Boyett}, {Chen}, {Chevallard}, {Circosta}, {Curti}, {Danhaive}, {DeCoursey}, {de Graaff}, {Dressler}, {Egami}, {Helton}, {Hviding}, {Ji}, {Jones}, {Kumari}, {L{\"u}tzgendorf}, {Laseter}, {Looser}, {Lyu}, {Maseda}, {Nelson}, {Parlanti}, {Perna}, {Pusk{\'a}s}, {Rawle}, {Rodr{\'\i}guez Del Pino}, {Sandles}, {Saxena}, {Scholtz}, {Sharpe}, {Shivaei}, {Silcock}, {Simmonds}, {Skarbinski}, {Smit}, {Stone}, {Suess}, {Sun}, {Tang}, {Topping}, {{\"U}bler}, {Villanueva}, {Wallace}, {Whitler}, {Witstok}, \& {Woodrum}}]{2023arXiv230602465E}
{Eisenstein}, D.~J., {Willott}, C., {Alberts}, S., {et~al.} 2023, arXiv e-prints, arXiv:2306.02465, \dodoi{10.48550/arXiv.2306.02465}

\bibitem[{{Eldridge} {et~al.}(2017){Eldridge}, {Stanway}, {Xiao}, {McClelland}, {Taylor}, {Ng}, {Greis}, \& {Bray}}]{2017PASA...34...58E}
{Eldridge}, J.~J., {Stanway}, E.~R., {Xiao}, L., {et~al.} 2017, \pasa, 34, e058, \dodoi{10.1017/pasa.2017.51}

\bibitem[{{Erb} {et~al.}(2010){Erb}, {Pettini}, {Shapley}, {Steidel}, {Law}, \& {Reddy}}]{2010ApJ...719.1168E}
{Erb}, D.~K., {Pettini}, M., {Shapley}, A.~E., {et~al.} 2010, \apj, 719, 1168, \dodoi{10.1088/0004-637X/719/2/1168}

\bibitem[{{Falc{\'o}n-Barroso} \& {Knapen}(2013)}]{2013seg..book.....F}
{Falc{\'o}n-Barroso}, J., \& {Knapen}, J.~H. 2013, {Secular Evolution of Galaxies} (IAC)

\bibitem[{{Feldmann} {et~al.}(2025){Feldmann}, {Boylan-Kolchin}, {Bullock}, {{\c{C}}atmabacak}, {Faucher-Gigu{\`e}re}, {Hayward}, {Kere{\v{s}}}, {Lazar}, {Liang}, {Moreno}, {Oesch}, {Quataert}, {Shen}, \& {Sun}}]{2025MNRAS.536..988F}
{Feldmann}, R., {Boylan-Kolchin}, M., {Bullock}, J.~S., {et~al.} 2025, \mnras, 536, 988, \dodoi{10.1093/mnras/stae2633}

\bibitem[{{Ferrara}(2024)}]{2024A&A...689A.310F}
{Ferrara}, A. 2024, \aap, 689, A310, \dodoi{10.1051/0004-6361/202450944}

\bibitem[{{Ferrara} {et~al.}(2023){Ferrara}, {Pallottini}, \& {Dayal}}]{2023MNRAS.522.3986F}
{Ferrara}, A., {Pallottini}, A., \& {Dayal}, P. 2023, \mnras, 522, 3986, \dodoi{10.1093/mnras/stad1095}

\bibitem[{{Finkelstein} {et~al.}(2022){Finkelstein}, {Bagley}, {Arrabal Haro}, {Dickinson}, {Ferguson}, {Kartaltepe}, {Papovich}, {Burgarella}, {Kocevski}, {Huertas-Company}, {Iyer}, {Koekemoer}, {Larson}, {P{\'e}rez-Gonz{\'a}lez}, {Rose}, {Tacchella}, {Wilkins}, {Chworowsky}, {Medrano}, {Morales}, {Somerville}, {Yung}, {Fontana}, {Giavalisco}, {Grazian}, {Grogin}, {Kewley}, {Kirkpatrick}, {Kurczynski}, {Lotz}, {Pentericci}, {Pirzkal}, {Ravindranath}, {Ryan}, {Trump}, {Yang}, {Almaini}, {Amor{\'\i}n}, {Annunziatella}, {Backhaus}, {Barro}, {Behroozi}, {Bell}, {Bhatawdekar}, {Bisigello}, {Bromm}, {Buat}, {Buitrago}, {Calabr{\`o}}, {Casey}, {Castellano}, {Ch{\'a}vez Ortiz}, {Ciesla}, {Cleri}, {Cohen}, {Cole}, {Cooke}, {Cooper}, {Cooray}, {Costantin}, {Cox}, {Croton}, {Daddi}, {Dav{\'e}}, {de La Vega}, {Dekel}, {Elbaz}, {Estrada-Carpenter}, {Faber}, {Fern{\'a}ndez}, {Finkelstein}, {Freundlich}, {Fujimoto}, {Garc{\'\i}a-Argum{\'a}nez}, {Gardner}, {Gawiser}, {G{\'o}mez-Guijarro}, {Guo}, {Hamblin}, {Hamilton},
  {Hathi}, {Holwerda}, {Hirschmann}, {Hutchison}, {Jaskot}, {Jha}, {Jogee}, {Juneau}, {Jung}, {Kassin}, {Le Bail}, {Leung}, {Lucas}, {Magnelli}, {Mantha}, {Matharu}, {McGrath}, {McIntosh}, {Merlin}, {Mobasher}, {Newman}, {Nicholls}, {Pandya}, {Rafelski}, {Ronayne}, {Santini}, {Seill{\'e}}, {Shah}, {Shen}, {Simons}, {Snyder}, {Stanway}, {Straughn}, {Teplitz}, {Vanderhoof}, {Vega-Ferrero}, {Wang}, {Weiner}, {Willmer}, {Wuyts}, {Zavala}, \& {Ceers Team}}]{2022ApJ...940L..55F}
{Finkelstein}, S.~L., {Bagley}, M.~B., {Arrabal Haro}, P., {et~al.} 2022, \apjl, 940, L55, \dodoi{10.3847/2041-8213/ac966e}

\bibitem[{{Finkelstein} {et~al.}(2023){Finkelstein}, {Bagley}, {Ferguson}, {Wilkins}, {Kartaltepe}, {Papovich}, {Yung}, {Arrabal Haro}, {Behroozi}, {Dickinson}, {Kocevski}, {Koekemoer}, {Larson}, {Le Bail}, {Morales}, {P{\'e}rez-Gonz{\'a}lez}, {Burgarella}, {Dav{\'e}}, {Hirschmann}, {Somerville}, {Wuyts}, {Bromm}, {Casey}, {Fontana}, {Fujimoto}, {Gardner}, {Giavalisco}, {Grazian}, {Grogin}, {Hathi}, {Hutchison}, {Jha}, {Jogee}, {Kewley}, {Kirkpatrick}, {Long}, {Lotz}, {Pentericci}, {Pierel}, {Pirzkal}, {Ravindranath}, {Ryan}, {Trump}, {Yang}, {Bhatawdekar}, {Bisigello}, {Buat}, {Calabr{\`o}}, {Castellano}, {Cleri}, {Cooper}, {Croton}, {Daddi}, {Dekel}, {Elbaz}, {Franco}, {Gawiser}, {Holwerda}, {Huertas-Company}, {Jaskot}, {Leung}, {Lucas}, {Mobasher}, {Pandya}, {Tacchella}, {Weiner}, \& {Zavala}}]{2023ApJ...946L..13F}
{Finkelstein}, S.~L., {Bagley}, M.~B., {Ferguson}, H.~C., {et~al.} 2023, \apjl, 946, L13, \dodoi{10.3847/2041-8213/acade4}

\bibitem[{{Finkelstein} {et~al.}(2024){Finkelstein}, {Leung}, {Bagley}, {Dickinson}, {Ferguson}, {Papovich}, {Akins}, {Arrabal Haro}, {Dav{\'e}}, {Dekel}, {Kartaltepe}, {Kocevski}, {Koekemoer}, {Pirzkal}, {Somerville}, {Yung}, {Amor{\'\i}n}, {Backhaus}, {Behroozi}, {Bisigello}, {Bromm}, {Casey}, {Ch{\'a}vez Ortiz}, {Cheng}, {Chworowsky}, {Cleri}, {Cooper}, {Davis}, {de la Vega}, {Elbaz}, {Franco}, {Fontana}, {Fujimoto}, {Giavalisco}, {Grogin}, {Holwerda}, {Huertas-Company}, {Hirschmann}, {Iyer}, {Jogee}, {Jung}, {Larson}, {Lucas}, {Mobasher}, {Morales}, {Morley}, {Mukherjee}, {P{\'e}rez-Gonz{\'a}lez}, {Ravindranath}, {Rodighiero}, {Rowland}, {Tacchella}, {Taylor}, {Trump}, \& {Wilkins}}]{2024ApJ...969L...2F}
{Finkelstein}, S.~L., {Leung}, G. C.~K., {Bagley}, M.~B., {et~al.} 2024, \apjl, 969, L2, \dodoi{10.3847/2041-8213/ad4495}

\bibitem[{{Finkelstein} {et~al.}(2025){Finkelstein}, {Bagley}, {Arrabal Haro}, {Dickinson}, {Ferguson}, {Kartaltepe}, {Kocevski}, {Koekemoer}, {Lotz}, {Papovich}, {Perez-Gonzalez}, {Pirzkal}, {Somerville}, {Trump}, {Yang}, {Yung}, {Fontana}, {Grazian}, {Grogin}, {Kewley}, {Kirkpatrick}, {Larson}, {Pentericci}, {Ravindranath}, {Wilkins}, {Almaini}, {Amorin}, {Barro}, {Bhatawdekar}, {Bisigello}, {Brooks}, {Buitrago}, {Calabro}, {Castellano}, {Cheng}, {Cleri}, {Cole}, {Cooper}, {Cooper}, {Costantin}, {Cox}, {Croton}, {Daddi}, {Davis}, {Dekel}, {Elbaz}, {Fernandez}, {Fujimoto}, {Gandolfi}, {Gardner}, {Gawiser}, {Giavalisco}, {Gomez-Guijarro}, {Guo}, {Gupta}, {Hathi}, {Harish}, {Henry}, {Hirschmann}, {Hu}, {Hutchison}, {Iyer}, {Jaskot}, {Jha}, {Jung}, {Kokorev}, {Kurczynski}, {Leung}, {Llerena}, {Long}, {Lucas}, {Lu}, {McGrath}, {McIntosh}, {Merlin}, {Morales}, {Napolitano}, {Pacucci}, {Pandya}, {Rafelski}, {Rodighiero}, {Rose}, {Santini}, {Seille}, {Simons}, {Shen}, {Straughn}, {Tacchella}, {Vanderhoof},
  {Vega-Ferrero}, {Weiner}, {Willmer}, {Zhu}, {Bell}, {Wuyts}, {Holwerda}, {Wang}, {Wang}, \& {Zavala}}]{2025arXiv250104085F}
{Finkelstein}, S.~L., {Bagley}, M.~B., {Arrabal Haro}, P., {et~al.} 2025, arXiv e-prints, arXiv:2501.04085, \dodoi{10.48550/arXiv.2501.04085}

\bibitem[{{Foreman-Mackey} {et~al.}(2014){Foreman-Mackey}, {Sick}, \& {Johnson}}]{2014zndo.....12157F}
{Foreman-Mackey}, D., {Sick}, J., \& {Johnson}, B. 2014, {python-fsps: Python bindings to FSPS (v0.1.1)}, v0.1.1,  Zenodo, \dodoi{10.5281/zenodo.12157}

\bibitem[{{Fruchter} \& {Hook}(2002)}]{2002PASP..114..144F}
{Fruchter}, A.~S., \& {Hook}, R.~N. 2002, \pasp, 114, 144, \dodoi{10.1086/338393}

\bibitem[{{Gaia Collaboration} {et~al.}(2016{\natexlab{a}}){Gaia Collaboration}, {Prusti}, {de Bruijne}, {Brown}, {Vallenari}, {Babusiaux}, {Bailer-Jones}, {Bastian}, {Biermann}, {Evans}, {Eyer}, {Jansen}, {Jordi}, {Klioner}, {Lammers}, {Lindegren}, {Luri}, {Mignard}, {Milligan}, {Panem}, {Poinsignon}, {Pourbaix}, {Randich}, {Sarri}, {Sartoretti}, {Siddiqui}, {Soubiran}, {Valette}, {van Leeuwen}, {Walton}, {Aerts}, {Arenou}, {Cropper}, {Drimmel}, {H{\o}g}, {Katz}, {Lattanzi}, {O'Mullane}, {Grebel}, {Holland}, {Huc}, {Passot}, {Bramante}, {Cacciari}, {Casta{\~n}eda}, {Chaoul}, {Cheek}, {De Angeli}, {Fabricius}, {Guerra}, {Hern{\'a}ndez}, {Jean-Antoine-Piccolo}, {Masana}, {Messineo}, {Mowlavi}, {Nienartowicz}, {Ord{\'o}{\~n}ez-Blanco}, {Panuzzo}, {Portell}, {Richards}, {Riello}, {Seabroke}, {Tanga}, {Th{\'e}venin}, {Torra}, {Els}, {Gracia-Abril}, {Comoretto}, {Garcia-Reinaldos}, {Lock}, {Mercier}, {Altmann}, {Andrae}, {Astraatmadja}, {Bellas-Velidis}, {Benson}, {Berthier}, {Blomme}, {Busso}, {Carry}, {Cellino},
  {Clementini}, {Cowell}, {Creevey}, {Cuypers}, {Davidson}, {De Ridder}, {de Torres}, {Delchambre}, {Dell'Oro}, {Ducourant}, {Fr{\'e}mat}, {Garc{\'\i}a-Torres}, {Gosset}, {Halbwachs}, {Hambly}, {Harrison}, {Hauser}, {Hestroffer}, {Hodgkin}, {Huckle}, {Hutton}, {Jasniewicz}, {Jordan}, {Kontizas}, {Korn}, {Lanzafame}, {Manteiga}, {Moitinho}, {Muinonen}, {Osinde}, {Pancino}, {Pauwels}, {Petit}, {Recio-Blanco}, {Robin}, {Sarro}, {Siopis}, {Smith}, {Smith}, {Sozzetti}, {Thuillot}, {van Reeven}, {Viala}, {Abbas}, {Abreu Aramburu}, {Accart}, {Aguado}, {Allan}, {Allasia}, {Altavilla}, {{\'A}lvarez}, {Alves}, {Anderson}, {Andrei}, {Anglada Varela}, {Antiche}, {Antoja}, {Ant{\'o}n}, {Arcay}, {Atzei}, {Ayache}, {Bach}, {Baker}, {Balaguer-N{\'u}{\~n}ez}, {Barache}, {Barata}, {Barbier}, {Barblan}, {Baroni}, {Barrado y Navascu{\'e}s}, {Barros}, {Barstow}, {Becciani}, {Bellazzini}, {Bellei}, {Bello Garc{\'\i}a}, {Belokurov}, {Bendjoya}, {Berihuete}, {Bianchi}, {Bienaym{\'e}}, {Billebaud}, {Blagorodnova}, {Blanco-Cuaresma},
  {Boch}, {Bombrun}, {Borrachero}, {Bouquillon}, {Bourda}, {Bouy}, {Bragaglia}, {Breddels}, {Brouillet}, {Br{\"u}semeister}, {Bucciarelli}, {Budnik}, {Burgess}, {Burgon}, {Burlacu}, {Busonero}, {Buzzi}, {Caffau}, {Cambras}, {Campbell}, {Cancelliere}, {Cantat-Gaudin}, {Carlucci}, {Carrasco}, {Castellani}, {Charlot}, {Charnas}, {Charvet}, {Chassat}, {Chiavassa}, {Clotet}, {Cocozza}, {Collins}, {Collins}, {Costigan}, {Crifo}, {Cross}, {Crosta}, {Crowley}, {Dafonte}, {Damerdji}, {Dapergolas}, {David}, {David}, {De Cat}, {de Felice}, {de Laverny}, {De Luise}, {De March}, {de Martino}, {de Souza}, {Debosscher}, {del Pozo}, {Delbo}, {Delgado}, {Delgado}, {di Marco}, {Di Matteo}, {Diakite}, {Distefano}, {Dolding}, {Dos Anjos}, {Drazinos}, {Dur{\'a}n}, {Dzigan}, {Ecale}, {Edvardsson}, {Enke}, {Erdmann}, {Escolar}, {Espina}, {Evans}, {Eynard Bontemps}, {Fabre}, {Fabrizio}, {Faigler}, {Falc{\~a}o}, {Farr{\`a}s Casas}, {Faye}, {Federici}, {Fedorets}, {Fern{\'a}ndez-Hern{\'a}ndez}, {Fernique}, {Fienga}, {Figueras},
  {Filippi}, {Findeisen}, {Fonti}, {Fouesneau}, {Fraile}, {Fraser}, {Fuchs}, {Furnell}, {Gai}, {Galleti}, {Galluccio}, {Garabato}, {Garc{\'\i}a-Sedano}, {Gar{\'e}}, {Garofalo}, {Garralda}, {Gavras}, {Gerssen}, {Geyer}, {Gilmore}, {Girona}, {Giuffrida}, {Gomes}, {Gonz{\'a}lez-Marcos}, {Gonz{\'a}lez-N{\'u}{\~n}ez}, {Gonz{\'a}lez-Vidal}, {Granvik}, {Guerrier}, {Guillout}, {Guiraud}, {G{\'u}rpide}, {Guti{\'e}rrez-S{\'a}nchez}, {Guy}, {Haigron}, {Hatzidimitriou}, {Haywood}, {Heiter}, {Helmi}, {Hobbs}, {Hofmann}, {Holl}, {Holland}, {Hunt}, {Hypki}, {Icardi}, {Irwin}, {Jevardat de Fombelle}, {Jofr{\'e}}, {Jonker}, {Jorissen}, {Julbe}, {Karampelas}, {Kochoska}, {Kohley}, {Kolenberg}, {Kontizas}, {Koposov}, {Kordopatis}, {Koubsky}, {Kowalczyk}, {Krone-Martins}, {Kudryashova}, {Kull}, {Bachchan}, {Lacoste-Seris}, {Lanza}, {Lavigne}, {Le Poncin-Lafitte}, {Lebreton}, {Lebzelter}, {Leccia}, {Leclerc}, {Lecoeur-Taibi}, {Lemaitre}, {Lenhardt}, {Leroux}, {Liao}, {Licata}, {Lindstr{\o}m}, {Lister}, {Livanou}, {Lobel},
  {L{\"o}ffler}, {L{\'o}pez}, {Lopez-Lozano}, {Lorenz}, {Loureiro}, {MacDonald}, {Magalh{\~a}es Fernandes}, {Managau}, {Mann}, {Mantelet}, {Marchal}, {Marchant}, {Marconi}, {Marie}, {Marinoni}, {Marrese}, {Marschalk{\'o}}, {Marshall}, {Mart{\'\i}n-Fleitas}, {Martino}, {Mary}, {Matijevi{\v{c}}}, {Mazeh}, {McMillan}, {Messina}, {Mestre}, {Michalik}, {Millar}, {Miranda}, {Molina}, {Molinaro}, {Molinaro}, {Moln{\'a}r}, {Moniez}, {Montegriffo}, {Monteiro}, {Mor}, {Mora}, {Morbidelli}, {Morel}, {Morgenthaler}, {Morley}, {Morris}, {Mulone}, {Muraveva}, {Musella}, {Narbonne}, {Nelemans}, {Nicastro}, {Noval}, {Ord{\'e}novic}, {Ordieres-Mer{\'e}}, {Osborne}, {Pagani}, {Pagano}, {Pailler}, {Palacin}, {Palaversa}, {Parsons}, {Paulsen}, {Pecoraro}, {Pedrosa}, {Pentik{\"a}inen}, {Pereira}, {Pichon}, {Piersimoni}, {Pineau}, {Plachy}, {Plum}, {Poujoulet}, {Pr{\v{s}}a}, {Pulone}, {Ragaini}, {Rago}, {Rambaux}, {Ramos-Lerate}, {Ranalli}, {Rauw}, {Read}, {Regibo}, {Renk}, {Reyl{\'e}}, {Ribeiro}, {Rimoldini}, {Ripepi}, {Riva},
  {Rixon}, {Roelens}, {Romero-G{\'o}mez}, {Rowell}, {Royer}, {Rudolph}, {Ruiz-Dern}, {Sadowski}, {Sagrist{\`a} Sell{\'e}s}, {Sahlmann}, {Salgado}, {Salguero}, {Sarasso}, {Savietto}, {Schnorhk}, {Schultheis}, {Sciacca}, {Segol}, {Segovia}, {Segransan}, {Serpell}, {Shih}, {Smareglia}, {Smart}, {Smith}, {Solano}, {Solitro}, {Sordo}, {Soria Nieto}, {Souchay}, {Spagna}, {Spoto}, {Stampa}, {Steele}, {Steidelm{\"u}ller}, {Stephenson}, {Stoev}, {Suess}, {S{\"u}veges}, {Surdej}, {Szabados}, {Szegedi-Elek}, {Tapiador}, {Taris}, {Tauran}, {Taylor}, {Teixeira}, {Terrett}, {Tingley}, {Trager}, {Turon}, {Ulla}, {Utrilla}, {Valentini}, {van Elteren}, {Van Hemelryck}, {van Leeuwen}, {Varadi}, {Vecchiato}, {Veljanoski}, {Via}, {Vicente}, {Vogt}, {Voss}, {Votruba}, {Voutsinas}, {Walmsley}, {Weiler}, {Weingrill}, {Werner}, {Wevers}, {Whitehead}, {Wyrzykowski}, {Yoldas}, {{\v{Z}}erjal}, {Zucker}, {Zurbach}, {Zwitter}, {Alecu}, {Allen}, {Allende Prieto}, {Amorim}, {Anglada-Escud{\'e}}, {Arsenijevic}, {Azaz}, {Balm}, {Beck},
  {Bernstein}, {Bigot}, {Bijaoui}, {Blasco}, {Bonfigli}, {Bono}, {Boudreault}, {Bressan}, {Brown}, {Brunet}, {Bunclark}, {Buonanno}, {Butkevich}, {Carret}, {Carrion}, {Chemin}, {Ch{\'e}reau}, {Corcione}, {Darmigny}, {de Boer}, {de Teodoro}, {de Zeeuw}, {Delle Luche}, {Domingues}, {Dubath}, {Fodor}, {Fr{\'e}zouls}, {Fries}, {Fustes}, {Fyfe}, {Gallardo}, {Gallegos}, {Gardiol}, {Gebran}, {Gomboc}, {G{\'o}mez}, {Grux}, {Gueguen}, {Heyrovsky}, {Hoar}, {Iannicola}, {Isasi Parache}, {Janotto}, {Joliet}, {Jonckheere}, {Keil}, {Kim}, {Klagyivik}, {Klar}, {Knude}, {Kochukhov}, {Kolka}, {Kos}, {Kutka}, {Lainey}, {LeBouquin}, {Liu}, {Loreggia}, {Makarov}, {Marseille}, {Martayan}, {Martinez-Rubi}, {Massart}, {Meynadier}, {Mignot}, {Munari}, {Nguyen}, {Nordlander}, {Ocvirk}, {O'Flaherty}, {Olias Sanz}, {Ortiz}, {Osorio}, {Oszkiewicz}, {Ouzounis}, {Palmer}, {Park}, {Pasquato}, {Peltzer}, {Peralta}, {P{\'e}turaud}, {Pieniluoma}, {Pigozzi}, {Poels}, {Prat}, {Prod'homme}, {Raison}, {Rebordao}, {Risquez}, {Rocca-Volmerange},
  {Rosen}, {Ruiz-Fuertes}, {Russo}, {Sembay}, {Serraller Vizcaino}, {Short}, {Siebert}, {Silva}, {Sinachopoulos}, {Slezak}, {Soffel}, {Sosnowska}, {Strai{\v{z}}ys}, {ter Linden}, {Terrell}, {Theil}, {Tiede}, {Troisi}, {Tsalmantza}, {Tur}, {Vaccari}, {Vachier}, {Valles}, {Van Hamme}, {Veltz}, {Virtanen}, {Wallut}, {Wichmann}, {Wilkinson}, {Ziaeepour}, \& {Zschocke}}]{2016A&A...595A...1G}
{Gaia Collaboration}, {Prusti}, T., {de Bruijne}, J.~H.~J., {et~al.} 2016{\natexlab{a}}, \aap, 595, A1, \dodoi{10.1051/0004-6361/201629272}

\bibitem[{{Gaia Collaboration} {et~al.}(2016{\natexlab{b}}){Gaia Collaboration}, {Brown}, {Vallenari}, {Prusti}, {de Bruijne}, {Mignard}, {Drimmel}, {Babusiaux}, {Bailer-Jones}, {Bastian}, {Biermann}, {Evans}, {Eyer}, {Jansen}, {Jordi}, {Katz}, {Klioner}, {Lammers}, {Lindegren}, {Luri}, {O'Mullane}, {Panem}, {Pourbaix}, {Randich}, {Sartoretti}, {Siddiqui}, {Soubiran}, {Valette}, {van Leeuwen}, {Walton}, {Aerts}, {Arenou}, {Cropper}, {H{\o}g}, {Lattanzi}, {Grebel}, {Holland}, {Huc}, {Passot}, {Perryman}, {Bramante}, {Cacciari}, {Casta{\~n}eda}, {Chaoul}, {Cheek}, {De Angeli}, {Fabricius}, {Guerra}, {Hern{\'a}ndez}, {Jean-Antoine-Piccolo}, {Masana}, {Messineo}, {Mowlavi}, {Nienartowicz}, {Ord{\'o}{\~n}ez-Blanco}, {Panuzzo}, {Portell}, {Richards}, {Riello}, {Seabroke}, {Tanga}, {Th{\'e}venin}, {Torra}, {Els}, {Gracia-Abril}, {Comoretto}, {Garcia-Reinaldos}, {Lock}, {Mercier}, {Altmann}, {Andrae}, {Astraatmadja}, {Bellas-Velidis}, {Benson}, {Berthier}, {Blomme}, {Busso}, {Carry}, {Cellino}, {Clementini},
  {Cowell}, {Creevey}, {Cuypers}, {Davidson}, {De Ridder}, {de Torres}, {Delchambre}, {Dell'Oro}, {Ducourant}, {Fr{\'e}mat}, {Garc{\'\i}a-Torres}, {Gosset}, {Halbwachs}, {Hambly}, {Harrison}, {Hauser}, {Hestroffer}, {Hodgkin}, {Huckle}, {Hutton}, {Jasniewicz}, {Jordan}, {Kontizas}, {Korn}, {Lanzafame}, {Manteiga}, {Moitinho}, {Muinonen}, {Osinde}, {Pancino}, {Pauwels}, {Petit}, {Recio-Blanco}, {Robin}, {Sarro}, {Siopis}, {Smith}, {Smith}, {Sozzetti}, {Thuillot}, {van Reeven}, {Viala}, {Abbas}, {Abreu Aramburu}, {Accart}, {Aguado}, {Allan}, {Allasia}, {Altavilla}, {{\'A}lvarez}, {Alves}, {Anderson}, {Andrei}, {Anglada Varela}, {Antiche}, {Antoja}, {Ant{\'o}n}, {Arcay}, {Bach}, {Baker}, {Balaguer-N{\'u}{\~n}ez}, {Barache}, {Barata}, {Barbier}, {Barblan}, {Barrado y Navascu{\'e}s}, {Barros}, {Barstow}, {Becciani}, {Bellazzini}, {Bello Garc{\'\i}a}, {Belokurov}, {Bendjoya}, {Berihuete}, {Bianchi}, {Bienaym{\'e}}, {Billebaud}, {Blagorodnova}, {Blanco-Cuaresma}, {Boch}, {Bombrun}, {Borrachero}, {Bouquillon},
  {Bourda}, {Bouy}, {Bragaglia}, {Breddels}, {Brouillet}, {Br{\"u}semeister}, {Bucciarelli}, {Burgess}, {Burgon}, {Burlacu}, {Busonero}, {Buzzi}, {Caffau}, {Cambras}, {Campbell}, {Cancelliere}, {Cantat-Gaudin}, {Carlucci}, {Carrasco}, {Castellani}, {Charlot}, {Charnas}, {Chiavassa}, {Clotet}, {Cocozza}, {Collins}, {Costigan}, {Crifo}, {Cross}, {Crosta}, {Crowley}, {Dafonte}, {Damerdji}, {Dapergolas}, {David}, {David}, {De Cat}, {de Felice}, {de Laverny}, {De Luise}, {De March}, {de Martino}, {de Souza}, {Debosscher}, {del Pozo}, {Delbo}, {Delgado}, {Delgado}, {Di Matteo}, {Diakite}, {Distefano}, {Dolding}, {Dos Anjos}, {Drazinos}, {Duran}, {Dzigan}, {Edvardsson}, {Enke}, {Evans}, {Eynard Bontemps}, {Fabre}, {Fabrizio}, {Faigler}, {Falc{\~a}o}, {Farr{\`a}s Casas}, {Federici}, {Fedorets}, {Fern{\'a}ndez-Hern{\'a}ndez}, {Fernique}, {Fienga}, {Figueras}, {Filippi}, {Findeisen}, {Fonti}, {Fouesneau}, {Fraile}, {Fraser}, {Fuchs}, {Gai}, {Galleti}, {Galluccio}, {Garabato}, {Garc{\'\i}a-Sedano}, {Garofalo},
  {Garralda}, {Gavras}, {Gerssen}, {Geyer}, {Gilmore}, {Girona}, {Giuffrida}, {Gomes}, {Gonz{\'a}lez-Marcos}, {Gonz{\'a}lez-N{\'u}{\~n}ez}, {Gonz{\'a}lez-Vidal}, {Granvik}, {Guerrier}, {Guillout}, {Guiraud}, {G{\'u}rpide}, {Guti{\'e}rrez-S{\'a}nchez}, {Guy}, {Haigron}, {Hatzidimitriou}, {Haywood}, {Heiter}, {Helmi}, {Hobbs}, {Hofmann}, {Holl}, {Holland}, {Hunt}, {Hypki}, {Icardi}, {Irwin}, {Jevardat de Fombelle}, {Jofr{\'e}}, {Jonker}, {Jorissen}, {Julbe}, {Karampelas}, {Kochoska}, {Kohley}, {Kolenberg}, {Kontizas}, {Koposov}, {Kordopatis}, {Koubsky}, {Krone-Martins}, {Kudryashova}, {Kull}, {Bachchan}, {Lacoste-Seris}, {Lanza}, {Lavigne}, {Le Poncin-Lafitte}, {Lebreton}, {Lebzelter}, {Leccia}, {Leclerc}, {Lecoeur-Taibi}, {Lemaitre}, {Lenhardt}, {Leroux}, {Liao}, {Licata}, {Lindstr{\o}m}, {Lister}, {Livanou}, {Lobel}, {L{\"o}ffler}, {L{\'o}pez}, {Lorenz}, {MacDonald}, {Magalh{\~a}es Fernandes}, {Managau}, {Mann}, {Mantelet}, {Marchal}, {Marchant}, {Marconi}, {Marinoni}, {Marrese}, {Marschalk{\'o}}, {Marshall},
  {Mart{\'\i}n-Fleitas}, {Martino}, {Mary}, {Matijevi{\v{c}}}, {Mazeh}, {McMillan}, {Messina}, {Michalik}, {Millar}, {Miranda}, {Molina}, {Molinaro}, {Molinaro}, {Moln{\'a}r}, {Moniez}, {Montegriffo}, {Mor}, {Mora}, {Morbidelli}, {Morel}, {Morgenthaler}, {Morris}, {Mulone}, {Muraveva}, {Musella}, {Narbonne}, {Nelemans}, {Nicastro}, {Noval}, {Ord{\'e}novic}, {Ordieres-Mer{\'e}}, {Osborne}, {Pagani}, {Pagano}, {Pailler}, {Palacin}, {Palaversa}, {Parsons}, {Pecoraro}, {Pedrosa}, {Pentik{\"a}inen}, {Pichon}, {Piersimoni}, {Pineau}, {Plachy}, {Plum}, {Poujoulet}, {Pr{\v{s}}a}, {Pulone}, {Ragaini}, {Rago}, {Rambaux}, {Ramos-Lerate}, {Ranalli}, {Rauw}, {Read}, {Regibo}, {Reyl{\'e}}, {Ribeiro}, {Rimoldini}, {Ripepi}, {Riva}, {Rixon}, {Roelens}, {Romero-G{\'o}mez}, {Rowell}, {Royer}, {Ruiz-Dern}, {Sadowski}, {Sagrist{\`a} Sell{\'e}s}, {Sahlmann}, {Salgado}, {Salguero}, {Sarasso}, {Savietto}, {Schultheis}, {Sciacca}, {Segol}, {Segovia}, {Segransan}, {Shih}, {Smareglia}, {Smart}, {Solano}, {Solitro}, {Sordo}, {Soria
  Nieto}, {Souchay}, {Spagna}, {Spoto}, {Stampa}, {Steele}, {Steidelm{\"u}ller}, {Stephenson}, {Stoev}, {Suess}, {S{\"u}veges}, {Surdej}, {Szabados}, {Szegedi-Elek}, {Tapiador}, {Taris}, {Tauran}, {Taylor}, {Teixeira}, {Terrett}, {Tingley}, {Trager}, {Turon}, {Ulla}, {Utrilla}, {Valentini}, {van Elteren}, {Van Hemelryck}, {van Leeuwen}, {Varadi}, {Vecchiato}, {Veljanoski}, {Via}, {Vicente}, {Vogt}, {Voss}, {Votruba}, {Voutsinas}, {Walmsley}, {Weiler}, {Weingrill}, {Wevers}, {Wyrzykowski}, {Yoldas}, {{\v{Z}}erjal}, {Zucker}, {Zurbach}, {Zwitter}, {Alecu}, {Allen}, {Allende Prieto}, {Amorim}, {Anglada-Escud{\'e}}, {Arsenijevic}, {Azaz}, {Balm}, {Beck}, {Bernstein}, {Bigot}, {Bijaoui}, {Blasco}, {Bonfigli}, {Bono}, {Boudreault}, {Bressan}, {Brown}, {Brunet}, {Bunclark}, {Buonanno}, {Butkevich}, {Carret}, {Carrion}, {Chemin}, {Ch{\'e}reau}, {Corcione}, {Darmigny}, {de Boer}, {de Teodoro}, {de Zeeuw}, {Delle Luche}, {Domingues}, {Dubath}, {Fodor}, {Fr{\'e}zouls}, {Fries}, {Fustes}, {Fyfe}, {Gallardo}, {Gallegos},
  {Gardiol}, {Gebran}, {Gomboc}, {G{\'o}mez}, {Grux}, {Gueguen}, {Heyrovsky}, {Hoar}, {Iannicola}, {Isasi Parache}, {Janotto}, {Joliet}, {Jonckheere}, {Keil}, {Kim}, {Klagyivik}, {Klar}, {Knude}, {Kochukhov}, {Kolka}, {Kos}, {Kutka}, {Lainey}, {LeBouquin}, {Liu}, {Loreggia}, {Makarov}, {Marseille}, {Martayan}, {Martinez-Rubi}, {Massart}, {Meynadier}, {Mignot}, {Munari}, {Nguyen}, {Nordlander}, {Ocvirk}, {O'Flaherty}, {Olias Sanz}, {Ortiz}, {Osorio}, {Oszkiewicz}, {Ouzounis}, {Palmer}, {Park}, {Pasquato}, {Peltzer}, {Peralta}, {P{\'e}turaud}, {Pieniluoma}, {Pigozzi}, {Poels}, {Prat}, {Prod'homme}, {Raison}, {Rebordao}, {Risquez}, {Rocca-Volmerange}, {Rosen}, {Ruiz-Fuertes}, {Russo}, {Sembay}, {Serraller Vizcaino}, {Short}, {Siebert}, {Silva}, {Sinachopoulos}, {Slezak}, {Soffel}, {Sosnowska}, {Strai{\v{z}}ys}, {ter Linden}, {Terrell}, {Theil}, {Tiede}, {Troisi}, {Tsalmantza}, {Tur}, {Vaccari}, {Vachier}, {Valles}, {Van Hamme}, {Veltz}, {Virtanen}, {Wallut}, {Wichmann}, {Wilkinson}, {Ziaeepour}, \&
  {Zschocke}}]{2016A&A...595A...2G}
{Gaia Collaboration}, {Brown}, A.~G.~A., {Vallenari}, A., {et~al.} 2016{\natexlab{b}}, \aap, 595, A2, \dodoi{10.1051/0004-6361/201629512}

\bibitem[{{Gandolfi} {et~al.}(2025){Gandolfi}, {Rodighiero}, {Bisigello}, {Grazian}, {Finkelstein}, {Dickinson}, {Castellano}, {Merlin}, {Calabr{\`o}}, {Papovich}, {Bianchetti}, {Ba{\~n}ados}, {Benotto}, {Buitrago}, {Daddi}, {Girardi}, {Giulietti}, {Hirschmann}, {Holwerda}, {Arrabal Haro}, {Lapi}, {Lucas}, {Lyu}, {Massardi}, {Pacucci}, {P{\'e}rez-Gonz{\'a}lez}, {Ronconi}, {Tarrasse}, {Wilkins}, {Vulcani}, {Yung}, {Zavala}, {Backhaus}, {Bagley}, {Buat}, {Burgarella}, {Kartaltepe}, {Khusanova}, {Kirkpatrick}, {Kocevski}, {Koekemoer}, {Lambrides}, {Pirzkal}, \& {Yang}}]{2025arXiv250202637G}
{Gandolfi}, G., {Rodighiero}, G., {Bisigello}, L., {et~al.} 2025, arXiv e-prints, arXiv:2502.02637, \dodoi{10.48550/arXiv.2502.02637}

\bibitem[{{Gelli} {et~al.}(2024){Gelli}, {Mason}, \& {Hayward}}]{2024ApJ...975..192G}
{Gelli}, V., {Mason}, C., \& {Hayward}, C.~C. 2024, \apj, 975, 192, \dodoi{10.3847/1538-4357/ad7b36}

\bibitem[{{Golini} {et~al.}(2024){Golini}, {Montes}, {Carrasco}, {Rom{\'a}n}, \& {Trujillo}}]{2024A&A...684A..99G}
{Golini}, G., {Montes}, M., {Carrasco}, E.~R., {Rom{\'a}n}, J., \& {Trujillo}, I. 2024, \aap, 684, A99, \dodoi{10.1051/0004-6361/202348300}

\bibitem[{{Hainline} {et~al.}(2024{\natexlab{a}}){Hainline}, {Johnson}, {Robertson}, {Tacchella}, {Helton}, {Sun}, {Eisenstein}, {Simmonds}, {Topping}, {Whitler}, {Willmer}, {Rieke}, {Suess}, {Hviding}, {Cameron}, {Alberts}, {Baker}, {Baum}, {Bhatawdekar}, {Bonaventura}, {Boyett}, {Bunker}, {Carniani}, {Charlot}, {Chevallard}, {Chen}, {Curti}, {Curtis-Lake}, {D'Eugenio}, {Egami}, {Endsley}, {Hausen}, {Ji}, {Looser}, {Lyu}, {Maiolino}, {Nelson}, {Pusk{\'a}s}, {Rawle}, {Sandles}, {Saxena}, {Smit}, {Stark}, {Williams}, {Willott}, \& {Witstok}}]{2024ApJ...964...71H}
{Hainline}, K.~N., {Johnson}, B.~D., {Robertson}, B., {et~al.} 2024{\natexlab{a}}, \apj, 964, 71, \dodoi{10.3847/1538-4357/ad1ee4}

\bibitem[{{Hainline} {et~al.}(2024{\natexlab{b}}){Hainline}, {Helton}, {Johnson}, {Sun}, {Topping}, {Leisenring}, {Baker}, {Eisenstein}, {Hausen}, {Hviding}, {Lyu}, {Robertson}, {Tacchella}, {Williams}, {Willmer}, \& {Roellig}}]{2024ApJ...964...66H}
{Hainline}, K.~N., {Helton}, J.~M., {Johnson}, B.~D., {et~al.} 2024{\natexlab{b}}, \apj, 964, 66, \dodoi{10.3847/1538-4357/ad20d1}

\bibitem[{{Harikane} {et~al.}(2024){Harikane}, {Nakajima}, {Ouchi}, {Umeda}, {Isobe}, {Ono}, {Xu}, \& {Zhang}}]{2024ApJ...960...56H}
{Harikane}, Y., {Nakajima}, K., {Ouchi}, M., {et~al.} 2024, \apj, 960, 56, \dodoi{10.3847/1538-4357/ad0b7e}

\bibitem[{{Harikane} {et~al.}(2022){Harikane}, {Ono}, {Ouchi}, {Liu}, {Sawicki}, {Shibuya}, {Behroozi}, {He}, {Shimasaku}, {Arnouts}, {Coupon}, {Fujimoto}, {Gwyn}, {Huang}, {Inoue}, {Kashikawa}, {Komiyama}, {Matsuoka}, \& {Willott}}]{2022ApJS..259...20H}
{Harikane}, Y., {Ono}, Y., {Ouchi}, M., {et~al.} 2022, \apjs, 259, 20, \dodoi{10.3847/1538-4365/ac3dfc}

\bibitem[{{Harikane} {et~al.}(2023){Harikane}, {Ouchi}, {Oguri}, {Ono}, {Nakajima}, {Isobe}, {Umeda}, {Mawatari}, \& {Zhang}}]{2023ApJS..265....5H}
{Harikane}, Y., {Ouchi}, M., {Oguri}, M., {et~al.} 2023, \apjs, 265, 5, \dodoi{10.3847/1538-4365/acaaa9}

\bibitem[{{Harvey} {et~al.}(2025){Harvey}, {Conselice}, {Adams}, {Austin}, {Juod{\v{z}}balis}, {Trussler}, {Li}, {Ormerod}, {Ferreira}, {Lovell}, {Duan}, {Westcott}, {Harris}, {Bhatawdekar}, {Coe}, {Cohen}, {Caruana}, {Cheng}, {Driver}, {Frye}, {Furtak}, {Grogin}, {Hathi}, {Holwerda}, {Jansen}, {Koekemoer}, {Marshall}, {Nonino}, {Vijayan}, {Wilkins}, {Windhorst}, {Willmer}, {Yan}, \& {Zitrin}}]{2025ApJ...978...89H}
{Harvey}, T., {Conselice}, C.~J., {Adams}, N.~J., {et~al.} 2025, \apj, 978, 89, \dodoi{10.3847/1538-4357/ad8c29}

\bibitem[{{Helton} {et~al.}(2024){Helton}, {Rieke}, {Alberts}, {Wu}, {Eisenstein}, {Hainline}, {Carniani}, {Ji}, {Baker}, {Bhatawdekar}, {Bunker}, {Cargile}, {Charlot}, {Chevallard}, {D'Eugenio}, {Egami}, {Johnson}, {Jones}, {Lyu}, {Maiolino}, {P{\'e}rez-Gonz{\'a}lez}, {Rieke}, {Robertson}, {Saxena}, {Scholtz}, {Shivaei}, {Sun}, {Tacchella}, {Whitler}, {Williams}, {Willmer}, {Willott}, {Witstok}, \& {Zhu}}]{2024arXiv240518462H}
{Helton}, J.~M., {Rieke}, G.~H., {Alberts}, S., {et~al.} 2024, arXiv e-prints, arXiv:2405.18462, \dodoi{10.48550/arXiv.2405.18462}

\bibitem[{{Johnson} {et~al.}(2021){Johnson}, {Leja}, {Conroy}, \& {Speagle}}]{2021ApJS..254...22J}
{Johnson}, B.~D., {Leja}, J., {Conroy}, C., \& {Speagle}, J.~S. 2021, \apjs, 254, 22, \dodoi{10.3847/1538-4365/abef67}

\bibitem[{{Katz} {et~al.}(2024){Katz}, {Cameron}, {Saxena}, {Barrufet}, {Choustikov}, {Cleri}, {de Graaff}, {Ellis}, {Fosbury}, {Heintz}, {Maseda}, {Matthee}, {McConchie}, \& {Oesch}}]{2024arXiv240803189K}
{Katz}, H., {Cameron}, A.~J., {Saxena}, A., {et~al.} 2024, arXiv e-prints, arXiv:2408.03189, \dodoi{10.48550/arXiv.2408.03189}

\bibitem[{{Kennicutt}(1998)}]{1998ARA&A..36..189K}
{Kennicutt}, Jr., R.~C. 1998, \araa, 36, 189, \dodoi{10.1146/annurev.astro.36.1.189}

\bibitem[{{Killi} {et~al.}(2023){Killi}, {Watson}, {Brammer}, {McPartland}, {Antwi-Danso}, {Newshore}, {Coe}, {Allen}, {Fynbo}, {Gould}, {Heintz}, {Rusakov}, \& {Vejlgaard}}]{2023arXiv231203065K}
{Killi}, M., {Watson}, D., {Brammer}, G., {et~al.} 2023, arXiv e-prints, arXiv:2312.03065, \dodoi{10.48550/arXiv.2312.03065}

\bibitem[{{Kokorev} {et~al.}(2024){Kokorev}, {Atek}, {Chisholm}, {Endsley}, {Chemerynska}, {Mu{\~n}oz}, {Furtak}, {Pan}, {Berg}, {Fujimoto}, {Oesch}, {Weibel}, {Adamo}, {Blaizot}, {Bouwens}, {Dessauges-Zavadsky}, {Khullar}, {Korber}, {Goovaerts}, {Jecmen}, {Labb{\'e}}, {Leclercq}, {Marques-Chaves}, {Mason}, {McQuinn}, {Naidu}, {Natarajan}, {Nelson}, {Rosdahl}, {Saldana-Lopez}, {Schaerer}, {Trebitsch}, {Volonteri}, \& {Zitrin}}]{2024arXiv241113640K}
{Kokorev}, V., {Atek}, H., {Chisholm}, J., {et~al.} 2024, arXiv e-prints, arXiv:2411.13640, \dodoi{10.48550/arXiv.2411.13640}

\bibitem[{{Kron}(1980)}]{1980ApJS...43..305K}
{Kron}, R.~G. 1980, \apjs, 43, 305, \dodoi{10.1086/190669}

\bibitem[{{Kroupa}(2001)}]{2001MNRAS.322..231K}
{Kroupa}, P. 2001, \mnras, 322, 231, \dodoi{10.1046/j.1365-8711.2001.04022.x}

\bibitem[{{Larson} {et~al.}(2022){Larson}, {Hutchison}, {Bagley}, {Finkelstein}, {Yung}, {Somerville}, {Hirschmann}, {Brammer}, {Holwerda}, {Papovich}, {Morales}, \& {Wilkins}}]{2022arXiv221110035L}
{Larson}, R.~L., {Hutchison}, T.~A., {Bagley}, M., {et~al.} 2022, arXiv e-prints, arXiv:2211.10035.
\newblock \doarXiv{2211.10035}

\bibitem[{{Lei} {et~al.}(2025){Lei}, {Wang}, {Yuan}, {Wang}, {Groenewegen}, \& {Fan}}]{2025ApJ...980..249L}
{Lei}, L., {Wang}, Y.-Y., {Yuan}, G.-W., {et~al.} 2025, \apj, 980, 249, \dodoi{10.3847/1538-4357/ada93b}

\bibitem[{{Leung} {et~al.}(2023){Leung}, {Bagley}, {Finkelstein}, {Ferguson}, {Koekemoer}, {P{\'e}rez-Gonz{\'a}lez}, {Morales}, {Kocevski}, {Yang}, {Somerville}, {Wilkins}, {Yung}, {Fujimoto}, {Larson}, {Papovich}, {Pirzkal}, {Berg}, {Lotz}, {Castellano}, {Ch{\'a}vez Ortiz}, {Cheng}, {Dickinson}, {Giavalisco}, {Hathi}, {Hutchison}, {Jung}, {Kartaltepe}, {Natarajan}, \& {Rothberg}}]{2023ApJ...954L..46L}
{Leung}, G. C.~K., {Bagley}, M.~B., {Finkelstein}, S.~L., {et~al.} 2023, \apjl, 954, L46, \dodoi{10.3847/2041-8213/acf365}

\bibitem[{{Li} {et~al.}(2024){Li}, {Dekel}, {Sarkar}, {Aung}, {Giavalisco}, {Mandelker}, \& {Tacchella}}]{2024A&A...690A.108L}
{Li}, Z., {Dekel}, A., {Sarkar}, K.~C., {et~al.} 2024, \aap, 690, A108, \dodoi{10.1051/0004-6361/202348727}

\bibitem[{{Lu} {et~al.}(2025){Lu}, {Frenk}, {Bose}, {Lacey}, {Cole}, {Baugh}, \& {Helly}}]{2025MNRAS.536.1018L}
{Lu}, S., {Frenk}, C.~S., {Bose}, S., {et~al.} 2025, \mnras, 536, 1018, \dodoi{10.1093/mnras/stae2646}

\bibitem[{{Madau} \& {Dickinson}(2014)}]{2014ARA&A..52..415M}
{Madau}, P., \& {Dickinson}, M. 2014, \araa, 52, 415, \dodoi{10.1146/annurev-astro-081811-125615}

\bibitem[{{Maiolino} {et~al.}(2024){Maiolino}, {Scholtz}, {Witstok}, {Carniani}, {D'Eugenio}, {de Graaff}, {{\"U}bler}, {Tacchella}, {Curtis-Lake}, {Arribas}, {Bunker}, {Charlot}, {Chevallard}, {Curti}, {Looser}, {Maseda}, {Rawle}, {Rodr{\'\i}guez del Pino}, {Willott}, {Egami}, {Eisenstein}, {Hainline}, {Robertson}, {Williams}, {Willmer}, {Baker}, {Boyett}, {DeCoursey}, {Fabian}, {Helton}, {Ji}, {Jones}, {Kumari}, {Laporte}, {Nelson}, {Perna}, {Sandles}, {Shivaei}, \& {Sun}}]{2024Natur.627...59M}
{Maiolino}, R., {Scholtz}, J., {Witstok}, J., {et~al.} 2024, \nat, 627, 59, \dodoi{10.1038/s41586-024-07052-5}

\bibitem[{{Mason} {et~al.}(2023){Mason}, {Trenti}, \& {Treu}}]{2023MNRAS.521..497M}
{Mason}, C.~A., {Trenti}, M., \& {Treu}, T. 2023, \mnras, 521, 497, \dodoi{10.1093/mnras/stad035}

\bibitem[{{Matteri} {et~al.}(2025){Matteri}, {Ferrara}, \& {Pallottini}}]{2025arXiv250318850M}
{Matteri}, A., {Ferrara}, A., \& {Pallottini}, A. 2025, arXiv e-prints, arXiv:2503.18850, \dodoi{10.48550/arXiv.2503.18850}

\bibitem[{{McLeod} {et~al.}(2024){McLeod}, {Donnan}, {McLure}, {Dunlop}, {Magee}, {Begley}, {Carnall}, {Cullen}, {Ellis}, {Hamadouche}, \& {Stanton}}]{2024MNRAS.527.5004M}
{McLeod}, D.~J., {Donnan}, C.~T., {McLure}, R.~J., {et~al.} 2024, \mnras, 527, 5004, \dodoi{10.1093/mnras/stad3471}

\bibitem[{{Menon} {et~al.}(2024){Menon}, {Lancaster}, {Burkhart}, {Somerville}, {Dekel}, \& {Krumholz}}]{Menon2024}
{Menon}, S.~H., {Lancaster}, L., {Burkhart}, B., {et~al.} 2024, \apjl, 967, L28, \dodoi{10.3847/2041-8213/ad462d}

\bibitem[{{Merlin} {et~al.}(2024){Merlin}, {Santini}, {Paris}, {Castellano}, {Fontana}, {Treu}, {Finkelstein}, {Dunlop}, {Arrabal Haro}, {Bagley}, {Boyett}, {Calabr{\`o}}, {Correnti}, {Davis}, {Dickinson}, {Donnan}, {Ferguson}, {Fortuni}, {Giavalisco}, {Glazebrook}, {Grazian}, {Grogin}, {Hathi}, {Hirschmann}, {Kartaltepe}, {Kewley}, {Kirkpatrick}, {Kocevski}, {Koekemoer}, {Leung}, {Lotz}, {Lucas}, {Magee}, {Marchesini}, {Mascia}, {McLeod}, {McLure}, {Nanayakkara}, {Napolitano}, {Nonino}, {Papovich}, {Pentericci}, {P{\'e}rez-Gonz{\'a}lez}, {Pirzkal}, {Ravindranath}, {Roberts-Borsani}, {Somerville}, {Trenti}, {Trump}, {Vulcani}, {Wang}, {Watson}, {Wilkins}, {Yang}, \& {Yung}}]{2024A&A...691A.240M}
{Merlin}, E., {Santini}, P., {Paris}, D., {et~al.} 2024, \aap, 691, A240, \dodoi{10.1051/0004-6361/202451409}

\bibitem[{{Morales} {et~al.}(2024){Morales}, {Finkelstein}, {Leung}, {Bagley}, {Cleri}, {Dave}, {Dickinson}, {Ferguson}, {Hathi}, {Jones}, {Koekemoer}, {Papovich}, {P{\'e}rez-Gonz{\'a}lez}, {Pirzkal}, {Smith}, {Wilkins}, \& {Yung}}]{2024ApJ...964L..24M}
{Morales}, A.~M., {Finkelstein}, S.~L., {Leung}, G. C.~K., {et~al.} 2024, \apjl, 964, L24, \dodoi{10.3847/2041-8213/ad2de4}

\bibitem[{{Morishita} {et~al.}(2024){Morishita}, {Mason}, {Kreilgaard}, {Trenti}, {Treu}, {Vulcani}, {Zhang}, {Abdurro'uf}, {Alavi}, {Atek}, {Bahe}, {Bradac}, {Bradley}, {Bunker}, {Coe}, {Colbert}, {Gelli}, {Hayes}, {Jones}, {Kodama}, {Leethochawalit}, {Liu}, {Malkan}, {Mehta}, {Metha}, {Newman}, {Rafelski}, {Roberts-Borsani}, {Rutkowski}, {Scarlata}, {Stiavelli}, {Sutanto}, {Takahashi}, {Teplitz}, \& {Wang}}]{2024arXiv241204211M}
{Morishita}, T., {Mason}, C.~A., {Kreilgaard}, K.~C., {et~al.} 2024, arXiv e-prints, arXiv:2412.04211, \dodoi{10.48550/arXiv.2412.04211}

\bibitem[{{Muzzin} {et~al.}(2013){Muzzin}, {Marchesini}, {Stefanon}, {Franx}, {Milvang-Jensen}, {Dunlop}, {Fynbo}, {Brammer}, {Labb{\'e}}, \& {van Dokkum}}]{2013ApJS..206....8M}
{Muzzin}, A., {Marchesini}, D., {Stefanon}, M., {et~al.} 2013, \apjs, 206, 8, \dodoi{10.1088/0067-0049/206/1/8}

\bibitem[{{Naidu} {et~al.}(2022){Naidu}, {Oesch}, {Setton}, {Matthee}, {Conroy}, {Johnson}, {Weaver}, {Bouwens}, {Brammer}, {Dayal}, {Illingworth}, {Barrufet}, {Belli}, {Bezanson}, {Bose}, {Heintz}, {Leja}, {Leonova}, {Marques-Chaves}, {Stefanon}, {Toft}, {van der Wel}, {van Dokkum}, {Weibel}, \& {Whitaker}}]{2022arXiv220802794N}
{Naidu}, R.~P., {Oesch}, P.~A., {Setton}, D.~J., {et~al.} 2022, arXiv e-prints, arXiv:2208.02794, \dodoi{10.48550/arXiv.2208.02794}

\bibitem[{{Naidu} {et~al.}(2025){Naidu}, {Oesch}, {Brammer}, {Weibel}, {Li}, {Matthee}, {Chisholm}, {Pollock}, {Heintz}, {Johnson}, {Shen}, {Hviding}, {Leja}, {Tacchella}, {Ganguly}, {Witten}, {Atek}, {Belli}, {Bose}, {Bouwens}, {Dayal}, {Decarli}, {de Graaff}, {Fudamoto}, {Giovinazzo}, {Greene}, {Illingworth}, {Inoue}, {Kane}, {Labbe}, {Leonova}, {Marques-Chaves}, {Meyer}, {Nelson}, {Roberts-Borsani}, {Schaerer}, {Simcoe}, {Stefanon}, {Sugahara}, {Toft}, {van der Wel}, {van Dokkum}, {Walter}, {Watson}, {Weaver}, \& {Whitaker}}]{2025arXiv250511263N}
{Naidu}, R.~P., {Oesch}, P.~A., {Brammer}, G., {et~al.} 2025, arXiv e-prints, arXiv:2505.11263.
\newblock \doarXiv{2505.11263}

\bibitem[{{Nakajima} {et~al.}(2023){Nakajima}, {Ouchi}, {Isobe}, {Harikane}, {Zhang}, {Ono}, {Umeda}, \& {Oguri}}]{2023ApJS..269...33N}
{Nakajima}, K., {Ouchi}, M., {Isobe}, Y., {et~al.} 2023, \apjs, 269, 33, \dodoi{10.3847/1538-4365/acd556}

\bibitem[{{Narayanan} {et~al.}(2025){Narayanan}, {Stark}, {Finkelstein}, {Torrey}, {Li}, {Cullen}, {Topping}, {Marinacci}, {Sales}, {Shen}, \& {Vogelsberger}}]{2025ApJ...982....7N}
{Narayanan}, D., {Stark}, D.~P., {Finkelstein}, S.~L., {et~al.} 2025, \apj, 982, 7, \dodoi{10.3847/1538-4357/adb41c}

\bibitem[{{Oesch} {et~al.}(2018){Oesch}, {Bouwens}, {Illingworth}, {Labb{\'e}}, \& {Stefanon}}]{2018ApJ...855..105O}
{Oesch}, P.~A., {Bouwens}, R.~J., {Illingworth}, G.~D., {Labb{\'e}}, I., \& {Stefanon}, M. 2018, \apj, 855, 105, \dodoi{10.3847/1538-4357/aab03f}

\bibitem[{{Oke} \& {Gunn}(1983)}]{1983ApJ...266..713O}
{Oke}, J.~B., \& {Gunn}, J.~E. 1983, \apj, 266, 713, \dodoi{10.1086/160817}

\bibitem[{{{\"O}stlin} {et~al.}(2024){{\"O}stlin}, {P{\'e}rez-Gonz{\'a}lez}, {Melinder}, {Gillman}, {Iani}, {Costantin}, {Boogaard}, {Rinaldi}, {Colina}, {N{\o}rgaard-Nielsen}, {Dicken}, {Greve}, {Wright}, {Alonso-Herrero}, {Alvarez-Marquez}, {Annunziatella}, {Bik}, {Bosman}, {Caputi}, {Crespo Gomez}, {Eckart}, {Garcia-Marin}, {Hjorth}, {Ilbert}, {Jermann}, {Kendrew}, {Labiano}, {Langeroodi}, {Le Fevre}, {Libralato}, {Meyer}, {Moutard}, {Peissker}, {Pye}, {Tikkanen}, {Topinka}, {Walter}, {Ward}, {van der Werf}, {van Dishoeck}, {Henning}, {Lagage}, {Ray}, \& {Vandenbussche}}]{2024arXiv241119686O}
{{\"O}stlin}, G., {P{\'e}rez-Gonz{\'a}lez}, P.~G., {Melinder}, J., {et~al.} 2024, arXiv e-prints, arXiv:2411.19686, \dodoi{10.48550/arXiv.2411.19686}

\bibitem[{{Pasha} \& {Miller}(2023)}]{2023JOSS....8.5703P}
{Pasha}, I., \& {Miller}, T.~B. 2023, The Journal of Open Source Software, 8, 5703, \dodoi{10.21105/joss.05703}

\bibitem[{{P{\'e}rez-Gonz{\'a}lez} {et~al.}(2003){P{\'e}rez-Gonz{\'a}lez}, {Gil de Paz}, {Zamorano}, {Gallego}, {Alonso-Herrero}, \& {Arag{\'o}n-Salamanca}}]{2003MNRAS.338..508P}
{P{\'e}rez-Gonz{\'a}lez}, P.~G., {Gil de Paz}, A., {Zamorano}, J., {et~al.} 2003, \mnras, 338, 508, \dodoi{10.1046/j.1365-8711.2003.06077.x}

\bibitem[{{P{\'e}rez-Gonz{\'a}lez} {et~al.}(2008){P{\'e}rez-Gonz{\'a}lez}, {Rieke}, {Villar}, {Barro}, {Blaylock}, {Egami}, {Gallego}, {Gil de Paz}, {Pascual}, {Zamorano}, \& {Donley}}]{2008ApJ...675..234P}
{P{\'e}rez-Gonz{\'a}lez}, P.~G., {Rieke}, G.~H., {Villar}, V., {et~al.} 2008, \apj, 675, 234, \dodoi{10.1086/523690}

\bibitem[{{P{\'e}rez-Gonz{\'a}lez} {et~al.}(2023{\natexlab{a}}){P{\'e}rez-Gonz{\'a}lez}, {Costantin}, {Langeroodi}, {Rinaldi}, {Annunziatella}, {Ilbert}, {Colina}, {N{\o}rgaard-Nielsen}, {Greve}, {{\"O}stlin}, {Wright}, {Alonso-Herrero}, {{\'A}lvarez-M{\'a}rquez}, {Caputi}, {Eckart}, {Le F{\`e}vre}, {Labiano}, {Garc{\'\i}a-Mar{\'\i}n}, {Hjorth}, {Kendrew}, {Pye}, {Tikkanen}, {van der Werf}, {Walter}, {Ward}, {Bik}, {Boogaard}, {Bosman}, {G{\'o}mez}, {Gillman}, {Iani}, {Jermann}, {Melinder}, {Meyer}, {Moutard}, {van Dishoek}, {Henning}, {Lagage}, {Guedel}, {Peissker}, {Ray}, {Vandenbussche}, {Garc{\'\i}a-Argum{\'a}nez}, \& {Mar{\'\i}a M{\'e}rida}}]{2023ApJ...951L...1P}
{P{\'e}rez-Gonz{\'a}lez}, P.~G., {Costantin}, L., {Langeroodi}, D., {et~al.} 2023{\natexlab{a}}, \apjl, 951, L1, \dodoi{10.3847/2041-8213/acd9d0}

\bibitem[{{P{\'e}rez-Gonz{\'a}lez} {et~al.}(2023{\natexlab{b}}){P{\'e}rez-Gonz{\'a}lez}, {Barro}, {Annunziatella}, {Costantin}, {Garc{\'\i}a-Argum{\'a}nez}, {McGrath}, {M{\'e}rida}, {Zavala}, {Arrabal Haro}, {Bagley}, {Backhaus}, {Behroozi}, {Bell}, {Bisigello}, {Buat}, {Calabr{\`o}}, {Casey}, {Cleri}, {Coogan}, {Cooper}, {Cooray}, {Dekel}, {Dickinson}, {Elbaz}, {Ferguson}, {Finkelstein}, {Fontana}, {Franco}, {Gardner}, {Giavalisco}, {G{\'o}mez-Guijarro}, {Grazian}, {Grogin}, {Guo}, {Huertas-Company}, {Jogee}, {Kartaltepe}, {Kewley}, {Kirkpatrick}, {Kocevski}, {Koekemoer}, {Long}, {Lotz}, {Lucas}, {Papovich}, {Pirzkal}, {Ravindranath}, {Somerville}, {Tacchella}, {Trump}, {Wang}, {Wilkins}, {Wuyts}, {Yang}, \& {Yung}}]{2023ApJ...946L..16P}
{P{\'e}rez-Gonz{\'a}lez}, P.~G., {Barro}, G., {Annunziatella}, M., {et~al.} 2023{\natexlab{b}}, \apjl, 946, L16, \dodoi{10.3847/2041-8213/acb3a5}

\bibitem[{{P{\'e}rez-Gonz{\'a}lez} {et~al.}(2024){P{\'e}rez-Gonz{\'a}lez}, {Barro}, {Rieke}, {Lyu}, {Rieke}, {Alberts}, {Williams}, {Hainline}, {Sun}, {Pusk{\'a}s}, {Annunziatella}, {Baker}, {Bunker}, {Egami}, {Ji}, {Johnson}, {Robertson}, {Rodr{\'\i}guez Del Pino}, {Rujopakarn}, {Shivaei}, {Tacchella}, {Willmer}, \& {Willott}}]{2024ApJ...968....4P}
{P{\'e}rez-Gonz{\'a}lez}, P.~G., {Barro}, G., {Rieke}, G.~H., {et~al.} 2024, \apj, 968, 4, \dodoi{10.3847/1538-4357/ad38bb}

\bibitem[{{Rieke} {et~al.}(2023){Rieke}, {Robertson}, {Tacchella}, {Hainline}, {Johnson}, {Hausen}, {Ji}, {Willmer}, {Eisenstein}, {Pusk{\'a}s}, {Alberts}, {Arribas}, {Baker}, {Baum}, {Bhatawdekar}, {Bonaventura}, {Boyett}, {Bunker}, {Cameron}, {Carniani}, {Charlot}, {Chevallard}, {Chen}, {Curti}, {Curtis-Lake}, {Danhaive}, {DeCoursey}, {Dressler}, {Egami}, {Endsley}, {Helton}, {Hviding}, {Kumari}, {Looser}, {Lyu}, {Maiolino}, {Maseda}, {Nelson}, {Rieke}, {Rix}, {Sandles}, {Saxena}, {Sharpe}, {Shivaei}, {Skarbinski}, {Smit}, {Stark}, {Stone}, {Suess}, {Sun}, {Topping}, {{\"U}bler}, {Villanueva}, {Wallace}, {Williams}, {Willott}, {Whitler}, {Witstok}, \& {Woodrum}}]{2023ApJS..269...16R}
{Rieke}, M.~J., {Robertson}, B., {Tacchella}, S., {et~al.} 2023, \apjs, 269, 16, \dodoi{10.3847/1538-4365/acf44d}

\bibitem[{{Rigby} {et~al.}(2023){Rigby}, {Perrin}, {McElwain}, {Kimble}, {Friedman}, {Lallo}, {Doyon}, {Feinberg}, {Ferruit}, {Glasse}, {Rieke}, {Rieke}, {Wright}, {Willott}, {Colon}, {Milam}, {Neff}, {Stark}, {Valenti}, {Abell}, {Abney}, {Abul-Huda}, {Acton}, {Adams}, {Adler}, {Aguilar}, {Ahmed}, {Albert}, {Alberts}, {Aldridge}, {Allen}, {Altenburg}, {{\'A}lvarez-M{\'a}rquez}, {Alves de Oliveira}, {Andersen}, {Anderson}, {Anderson}, {Argyriou}, {Armstrong}, {Arribas}, {Artigau}, {Arvai}, {Atkinson}, {Bacon}, {Bair}, {Banks}, {Barrientes}, {Barringer}, {Bartosik}, {Bast}, {Baudoz}, {Beatty}, {Bechtold}, {Beck}, {Bergeron}, {Bergkoetter}, {Bhatawdekar}, {Birkmann}, {Blazek}, {Blome}, {Boccaletti}, {B{\"o}ker}, {Boia}, {Bonaventura}, {Bond}, {Bosley}, {Boucarut}, {Bourque}, {Bouwman}, {Bower}, {Bowers}, {Boyer}, {Bradley}, {Brady}, {Braun}, {Breda}, {Bresnahan}, {Bright}, {Britt}, {Bromenschenkel}, {Brooks}, {Brooks}, {Brown}, {Brown}, {Brown}, {Bunker}, {Burger}, {Bushouse}, {Cale}, {Cameron}, {Cameron},
  {Canipe}, {Caplinger}, {Caputo}, {Cara}, {Carey}, {Carniani}, {Carrasquilla}, {Carruthers}, {Case}, {Catherine}, {Chance}, {Chapman}, {Charlot}, {Charlow}, {Chayer}, {Chen}, {Cherinka}, {Chichester}, {Chilton}, {Chonis}, {Clampin}, {Clark}, {Clark}, {Coe}, {Coleman}, {Comber}, {Comeau}, {Connolly}, {Cooper}, {Cooper}, {Coppock}, {Correnti}, {Cossou}, {Coulais}, {Coyle}, {Cracraft}, {Curti}, {Cuturic}, {Davis}, {Davis}, {Dean}, {DeLisa}, {deMeester}, {Dencheva}, {Dencheva}, {DePasquale}, {Deschenes}, {Hunor Detre}, {Diaz}, {Dicken}, {DiFelice}, {Dillman}, {Dixon}, {Doggett}, {Donaldson}, {Douglas}, {DuPrie}, {Dupuis}, {Durning}, {Easmin}, {Eck}, {Edeani}, {Egami}, {Ehrenwinkler}, {Eisenhamer}, {Eisenhower}, {Elie}, {Elliott}, {Elliott}, {Ellis}, {Engesser}, {Espinoza}, {Etienne}, {Etxaluze}, {Falini}, {Feeney}, {Ferry}, {Filippazzo}, {Fincham}, {Fix}, {Flagey}, {Florian}, {Flynn}, {Fontanella}, {Ford}, {Forshay}, {Fox}, {Franz}, {Fu}, {Fullerton}, {Galkin}, {Galyer}, {Garc{\'\i}a Mar{\'\i}n}, {Gardner},
  {Gardner}, {Garland}, {Garrett}, {Gasman}, {Gaspar}, {Gaudreau}, {Gauthier}, {Geers}, {Geithner}, {Gennaro}, {Giardino}, {Girard}, {Giuliano}, {Glassmire}, \& {Glauser}}]{2023PASP..135d8001R}
{Rigby}, J., {Perrin}, M., {McElwain}, M., {et~al.} 2023, \pasp, 135, 048001, \dodoi{10.1088/1538-3873/acb293}

\bibitem[{{Rix} {et~al.}(2004){Rix}, {Barden}, {Beckwith}, {Bell}, {Borch}, {Caldwell}, {H{\"a}ussler}, {Jahnke}, {Jogee}, {McIntosh}, {Meisenheimer}, {Peng}, {Sanchez}, {Somerville}, {Wisotzki}, \& {Wolf}}]{2004ApJS..152..163R}
{Rix}, H.-W., {Barden}, M., {Beckwith}, S. V.~W., {et~al.} 2004, \apjs, 152, 163, \dodoi{10.1086/420885}

\bibitem[{{Robertson} {et~al.}(2024){Robertson}, {Johnson}, {Tacchella}, {Eisenstein}, {Hainline}, {Arribas}, {Baker}, {Bunker}, {Carniani}, {Cargile}, {Carreira}, {Charlot}, {Chevallard}, {Curti}, {Curtis-Lake}, {D'Eugenio}, {Egami}, {Hausen}, {Helton}, {Jakobsen}, {Ji}, {Jones}, {Maiolino}, {Maseda}, {Nelson}, {P{\'e}rez-Gonz{\'a}lez}, {Pusk{\'a}s}, {Rieke}, {Smit}, {Sun}, {{\"U}bler}, {Whitler}, {Williams}, {Willmer}, {Willott}, \& {Witstok}}]{2024ApJ...970...31R}
{Robertson}, B., {Johnson}, B.~D., {Tacchella}, S., {et~al.} 2024, \apj, 970, 31, \dodoi{10.3847/1538-4357/ad463d}

\bibitem[{{Robertson} {et~al.}(2023){Robertson}, {Tacchella}, {Johnson}, {Hainline}, {Whitler}, {Eisenstein}, {Endsley}, {Rieke}, {Stark}, {Alberts}, {Dressler}, {Egami}, {Hausen}, {Rieke}, {Shivaei}, {Williams}, {Willmer}, {Arribas}, {Bonaventura}, {Bunker}, {Cameron}, {Carniani}, {Charlot}, {Chevallard}, {Curti}, {Curtis-Lake}, {D'Eugenio}, {Jakobsen}, {Looser}, {L{\"u}tzgendorf}, {Maiolino}, {Maseda}, {Rawle}, {Rix}, {Smit}, {{\"U}bler}, {Willott}, {Witstok}, {Baum}, {Bhatawdekar}, {Boyett}, {Chen}, {de Graaff}, {Florian}, {Helton}, {Hviding}, {Ji}, {Kumari}, {Lyu}, {Nelson}, {Sandles}, {Saxena}, {Suess}, {Sun}, {Topping}, \& {Wallace}}]{2023NatAs...7..611R}
{Robertson}, B.~E., {Tacchella}, S., {Johnson}, B.~D., {et~al.} 2023, Nature Astronomy, 7, 611, \dodoi{10.1038/s41550-023-01921-1}

\bibitem[{{Sanders} {et~al.}(2024){Sanders}, {Shapley}, {Topping}, {Reddy}, \& {Brammer}}]{2024ApJ...962...24S}
{Sanders}, R.~L., {Shapley}, A.~E., {Topping}, M.~W., {Reddy}, N.~A., \& {Brammer}, G.~B. 2024, \apj, 962, 24, \dodoi{10.3847/1538-4357/ad15fc}

\bibitem[{{Schechter}(1976)}]{1976ApJ...203..297S}
{Schechter}, P. 1976, \apj, 203, 297, \dodoi{10.1086/154079}

\bibitem[{{Sersic}(1968)}]{1968adga.book.....S}
{Sersic}, J.~L. 1968, {Atlas de Galaxias Australes} (Observatorio Astron\'omico, C\'ordoba, Argentina)

\bibitem[{{Shen} {et~al.}(2023){Shen}, {Vogelsberger}, {Boylan-Kolchin}, {Tacchella}, \& {Kannan}}]{2023MNRAS.525.3254S}
{Shen}, X., {Vogelsberger}, M., {Boylan-Kolchin}, M., {Tacchella}, S., \& {Kannan}, R. 2023, \mnras, 525, 3254, \dodoi{10.1093/mnras/stad2508}

\bibitem[{{Shibuya} {et~al.}(2015){Shibuya}, {Ouchi}, \& {Harikane}}]{2015ApJS..219...15S}
{Shibuya}, T., {Ouchi}, M., \& {Harikane}, Y. 2015, \apjs, 219, 15, \dodoi{10.1088/0067-0049/219/2/15}

\bibitem[{{Silk} {et~al.}(2024){Silk}, {Begelman}, {Norman}, {Nusser}, \& {Wyse}}]{2024ApJ...961L..39S}
{Silk}, J., {Begelman}, M.~C., {Norman}, C., {Nusser}, A., \& {Wyse}, R. F.~G. 2024, \apjl, 961, L39, \dodoi{10.3847/2041-8213/ad1bf0}

\bibitem[{{Somerville} {et~al.}(2008){Somerville}, {Hopkins}, {Cox}, {Robertson}, \& {Hernquist}}]{Somerville2008}
{Somerville}, R.~S., {Hopkins}, P.~F., {Cox}, T.~J., {Robertson}, B.~E., \& {Hernquist}, L. 2008, \mnras, 391, 481, \dodoi{10.1111/j.1365-2966.2008.13805.x}

\bibitem[{{Somerville} {et~al.}(2025){Somerville}, {Yung}, {Lancaster}, {Menon}, {Sommovigo}, \& {Finkelstein}}]{2025arXiv250505442S}
{Somerville}, R.~S., {Yung}, L.~Y.~A., {Lancaster}, L., {et~al.} 2025, arXiv e-prints, arXiv:2505.05442, \dodoi{10.48550/arXiv.2505.05442}

\bibitem[{{Speagle}(2020)}]{2020MNRAS.493.3132S}
{Speagle}, J.~S. 2020, \mnras, 493, 3132, \dodoi{10.1093/mnras/staa278}

\bibitem[{{Stiavelli} {et~al.}(2025){Stiavelli}, {Morishita}, {Chiaberge}, {Leethochawalit}, {Norman}, {Ricotti}, {Roberts-Borsani}, {Treu}, {Vanzella}, {Wyse}, {Zhang}, \& {Boyett}}]{2025ApJ...981..136S}
{Stiavelli}, M., {Morishita}, T., {Chiaberge}, M., {et~al.} 2025, \apj, 981, 136, \dodoi{10.3847/1538-4357/adb5f3}

\bibitem[{{Theios} {et~al.}(2019){Theios}, {Steidel}, {Strom}, {Rudie}, {Trainor}, \& {Reddy}}]{2019ApJ...871..128T}
{Theios}, R.~L., {Steidel}, C.~C., {Strom}, A.~L., {et~al.} 2019, \apj, 871, 128, \dodoi{10.3847/1538-4357/aaf386}

\bibitem[{{Tinker} {et~al.}(2008){Tinker}, {Kravtsov}, {Klypin}, {Abazajian}, {Warren}, {Yepes}, {Gottl{\"o}ber}, \& {Holz}}]{2008ApJ...688..709T}
{Tinker}, J., {Kravtsov}, A.~V., {Klypin}, A., {et~al.} 2008, \apj, 688, 709, \dodoi{10.1086/591439}

\bibitem[{{Trapp} \& {Furlanetto}(2020)}]{2020MNRAS.499.2401T}
{Trapp}, A.~C., \& {Furlanetto}, S.~R. 2020, \mnras, 499, 2401, \dodoi{10.1093/mnras/staa2828}

\bibitem[{{Trenti} \& {Stiavelli}(2008)}]{2008ApJ...676..767T}
{Trenti}, M., \& {Stiavelli}, M. 2008, \apj, 676, 767, \dodoi{10.1086/528674}

\bibitem[{{Treu} {et~al.}(2022){Treu}, {Roberts-Borsani}, {Bradac}, {Brammer}, {Fontana}, {Henry}, {Mason}, {Morishita}, {Pentericci}, {Wang}, {Acebron}, {Bagley}, {Bergamini}, {Belfiori}, {Bonchi}, {Boyett}, {Boutsia}, {Calabr{\'o}}, {Caminha}, {Castellano}, {Dressler}, {Glazebrook}, {Grillo}, {Jacobs}, {Jones}, {Kelly}, {Leethochawalit}, {Malkan}, {Marchesini}, {Mascia}, {Mercurio}, {Merlin}, {Nanayakkara}, {Nonino}, {Paris}, {Poggianti}, {Rosati}, {Santini}, {Scarlata}, {Shipley}, {Strait}, {Trenti}, {Tubthong}, {Vanzella}, {Vulcani}, \& {Yang}}]{2022ApJ...935..110T}
{Treu}, T., {Roberts-Borsani}, G., {Bradac}, M., {et~al.} 2022, \apj, 935, 110, \dodoi{10.3847/1538-4357/ac8158}

\bibitem[{{Trujillo-Gomez} {et~al.}(2011){Trujillo-Gomez}, {Klypin}, {Primack}, \& {Romanowsky}}]{2011ApJ...742...16T}
{Trujillo-Gomez}, S., {Klypin}, A., {Primack}, J., \& {Romanowsky}, A.~J. 2011, \apj, 742, 16, \dodoi{10.1088/0004-637X/742/1/16}

\bibitem[{{Trump} {et~al.}(2023){Trump}, {Arrabal Haro}, {Simons}, {Backhaus}, {Amor{\'\i}n}, {Dickinson}, {Fern{\'a}ndez}, {Papovich}, {Nicholls}, {Kewley}, {Brunker}, {Salzer}, {Wilkins}, {Almaini}, {Bagley}, {Berg}, {Bhatawdekar}, {Bisigello}, {Buat}, {Burgarella}, {Calabr{\`o}}, {Casey}, {Ciesla}, {Cleri}, {Cole}, {Cooper}, {Cooray}, {Costantin}, {Croton}, {Ferguson}, {Finkelstein}, {Fujimoto}, {Gardner}, {Gawiser}, {Giavalisco}, {Grazian}, {Grogin}, {Hathi}, {Hirschmann}, {Holwerda}, {Huertas-Company}, {Hutchison}, {Jogee}, {Juneau}, {Jung}, {Kartaltepe}, {Kirkpatrick}, {Kocevski}, {Koekemoer}, {Lotz}, {Lucas}, {Magnelli}, {Matharu}, {P{\'e}rez-Gonz{\'a}lez}, {Pirzkal}, {Rafelski}, {Rose}, {Seill{\'e}}, {Somerville}, {Straughn}, {Tacchella}, {Vanderhoof}, {Weiner}, {Wuyts}, {Yung}, \& {Zavala}}]{2023ApJ...945...35T}
{Trump}, J.~R., {Arrabal Haro}, P., {Simons}, R.~C., {et~al.} 2023, \apj, 945, 35, \dodoi{10.3847/1538-4357/acba8a}

\bibitem[{{Ucci} {et~al.}(2021){Ucci}, {Dayal}, {Hutter}, {Yepes}, {Gottl{\"o}ber}, {Legrand}, {Pentericci}, {Castellano}, \& {Choudhury}}]{2021MNRAS.506..202U}
{Ucci}, G., {Dayal}, P., {Hutter}, A., {et~al.} 2021, \mnras, 506, 202, \dodoi{10.1093/mnras/stab1229}

\bibitem[{{Weibel} {et~al.}(2025){Weibel}, {Oesch}, {Williams}, {Jespersen}, {Shuntov}, {Whitaker}, {Atek}, {Bezanson}, {Brammer}, {Chemerynska}, {Cloonan}, {Dayal}, {Furtak}, {Hutter}, {Ji}, {Maseda}, \& {Xiao}}]{2025arXiv250706292W}
{Weibel}, A., {Oesch}, P.~A., {Williams}, C.~C., {et~al.} 2025, arXiv e-prints, arXiv:2507.06292, \dodoi{10.48550/arXiv.2507.06292}

\bibitem[{{Whitaker} {et~al.}(2019){Whitaker}, {Ashas}, {Illingworth}, {Magee}, {Leja}, {Oesch}, {van Dokkum}, {Mowla}, {Bouwens}, {Franx}, {Holden}, {Labb{\'e}}, {Rafelski}, {Teplitz}, \& {Gonzalez}}]{2019ApJS..244...16W}
{Whitaker}, K.~E., {Ashas}, M., {Illingworth}, G., {et~al.} 2019, \apjs, 244, 16, \dodoi{10.3847/1538-4365/ab3853}

\bibitem[{{Whitler} {et~al.}(2025){Whitler}, {Stark}, {Topping}, {Robertson}, {Rieke}, {Hainline}, {Endsley}, {Chen}, {Baker}, {Bhatawdekar}, {Bunker}, {Carniani}, {Charlot}, {Chevallard}, {Curtis-Lake}, {Egami}, {Eisenstein}, {Helton}, {Ji}, {Johnson}, {P{\'e}rez-Gonz{\'a}lez}, {Rinaldi}, {Tacchella}, {Williams}, {Willmer}, {Willott}, \& {Witstok}}]{2025arXiv250100984W}
{Whitler}, L., {Stark}, D.~P., {Topping}, M.~W., {et~al.} 2025, arXiv e-prints, arXiv:2501.00984, \dodoi{10.48550/arXiv.2501.00984}

\bibitem[{{Willmer}(2018)}]{2018ApJS..236...47W}
{Willmer}, C. N.~A. 2018, \apjs, 236, 47, \dodoi{10.3847/1538-4365/aabfdf}

\bibitem[{{Willott} {et~al.}(2024){Willott}, {Desprez}, {Asada}, {Sarrouh}, {Abraham}, {Brada{\v{c}}}, {Brammer}, {Estrada-Carpenter}, {Iyer}, {Martis}, {Matharu}, {Mowla}, {Muzzin}, {Noirot}, {Sawicki}, {Strait}, {Rihtar{\v{s}}i{\v{c}}}, \& {Withers}}]{2024ApJ...966...74W}
{Willott}, C.~J., {Desprez}, G., {Asada}, Y., {et~al.} 2024, \apj, 966, 74, \dodoi{10.3847/1538-4357/ad35bc}

\bibitem[{{Witstok} {et~al.}(2025){Witstok}, {Jakobsen}, {Maiolino}, {Helton}, {Johnson}, {Robertson}, {Tacchella}, {Cameron}, {Smit}, {Bunker}, {Saxena}, {Sun}, {Alberts}, {Arribas}, {Baker}, {Bhatawdekar}, {Boyett}, {Cargile}, {Carniani}, {Charlot}, {Chevallard}, {Curti}, {Curtis-Lake}, {D'Eugenio}, {Eisenstein}, {Hainline}, {Jones}, {Kumari}, {Maseda}, {P{\'e}rez-Gonz{\'a}lez}, {Rinaldi}, {Scholtz}, {{\"U}bler}, {Williams}, {Willmer}, {Willott}, \& {Zhu}}]{2025Natur.639..897W}
{Witstok}, J., {Jakobsen}, P., {Maiolino}, R., {et~al.} 2025, \nat, 639, 897, \dodoi{10.1038/s41586-025-08779-5}

\bibitem[{{Yan} {et~al.}(2023{\natexlab{a}}){Yan}, {Ma}, {Ling}, {Cheng}, \& {Huang}}]{2023ApJ...942L...9Y}
{Yan}, H., {Ma}, Z., {Ling}, C., {Cheng}, C., \& {Huang}, J.-S. 2023{\natexlab{a}}, \apjl, 942, L9, \dodoi{10.3847/2041-8213/aca80c}

\bibitem[{{Yan} {et~al.}(2023{\natexlab{b}}){Yan}, {Sun}, {Ma}, \& {Ling}}]{2023arXiv231115121Y}
{Yan}, H., {Sun}, B., {Ma}, Z., \& {Ling}, C. 2023{\natexlab{b}}, arXiv e-prints, arXiv:2311.15121, \dodoi{10.48550/arXiv.2311.15121}

\bibitem[{{Yan} {et~al.}(2023{\natexlab{c}}){Yan}, {Cohen}, {Windhorst}, {Jansen}, {Ma}, {Beacom}, {Ling}, {Cheng}, {Huang}, {Grogin}, {Willner}, {Yun}, {Hammel}, {Milam}, {Conselice}, {Driver}, {Frye}, {Marshall}, {Koekemoer}, {Willmer}, {Robotham}, {D'Silva}, {Summers}, {Lim}, {Harrington}, {Ferreira}, {Diego}, {Pirzkal}, {Wilkins}, {Wang}, {Hathi}, {Zitrin}, {Bhatawdekar}, {Adams}, {Furtak}, {Maksym}, {Rutkowski}, \& {Fazio}}]{2023ApJ...942L...8Y}
{Yan}, H., {Cohen}, S.~H., {Windhorst}, R.~A., {et~al.} 2023{\natexlab{c}}, \apjl, 942, L8, \dodoi{10.3847/2041-8213/aca974}

\bibitem[{{Yung} {et~al.}(2024{\natexlab{a}}){Yung}, {Somerville}, {Finkelstein}, {Wilkins}, \& {Gardner}}]{2024MNRAS.527.5929Y}
{Yung}, L.~Y.~A., {Somerville}, R.~S., {Finkelstein}, S.~L., {Wilkins}, S.~M., \& {Gardner}, J.~P. 2024{\natexlab{a}}, \mnras, 527, 5929, \dodoi{10.1093/mnras/stad3484}

\bibitem[{{Yung} {et~al.}(2025){Yung}, {Somerville}, \& {Iyer}}]{2025arXiv250418618Y}
{Yung}, L.~Y.~A., {Somerville}, R.~S., \& {Iyer}, K.~G. 2025, arXiv e-prints, arXiv:2504.18618, \dodoi{10.48550/arXiv.2504.18618}

\bibitem[{{Yung} {et~al.}(2024{\natexlab{b}}){Yung}, {Somerville}, {Nguyen}, {Behroozi}, {Modi}, \& {Gardner}}]{2024MNRAS.530.4868Y}
{Yung}, L.~Y.~A., {Somerville}, R.~S., {Nguyen}, T., {et~al.} 2024{\natexlab{b}}, \mnras, 530, 4868, \dodoi{10.1093/mnras/stae1188}

\bibitem[{{Zackrisson} {et~al.}(2011){Zackrisson}, {Rydberg}, {Schaerer}, {{\"O}stlin}, \& {Tuli}}]{2011ApJ...740...13Z}
{Zackrisson}, E., {Rydberg}, C.-E., {Schaerer}, D., {{\"O}stlin}, G., \& {Tuli}, M. 2011, \apj, 740, 13, \dodoi{10.1088/0004-637X/740/1/13}

\bibitem[{{Zavala} {et~al.}(2022){Zavala}, {Buat}, {Casey}, {Burgarella}, {Finkelstein}, {Bagley}, {Ciesla}, {Daddi}, {Dickinson}, {Ferguson}, {Franco}, {Jim'enez-Andrade}, {Kartaltepe}, {Koekemoer}, {Le Bail}, {Murphy}, {Papovich}, {Tacchella}, {Wilkins}, {Aretxaga}, {Behroozi}, {Champagne}, {Fontana}, {Giavalisco}, {Grazian}, {Grogin}, {Kewley}, {Kocevski}, {Kirkpatrick}, {Lotz}, {Pentericci}, {Perez-Gonzalez}, {Pirzkal}, {Ravindranath}, {Somerville}, {Trump}, {Yang}, {Yung}, {Almaini}, {Amorin}, {Annunziatella}, {Arrabal Haro}, {Backhaus}, {Barro}, {Bell}, {Bhatawdekar}, {Bisigello}, {Buitrago}, {Calabro}, {Castellano}, {Chavez Ortiz}, {Chworowsky}, {Cleri}, {Cohen}, {Cole}, {Cooke}, {Cooper}, {Cooray}, {Costantin}, {Cox}, {Croton}, {Dave}, {de la Vega}, {Dekel}, {Elbaz}, {Estrada-Carpenter}, {Fern{\'a}ndez}, {Finkelstein}, {Freundlich}, {Fujimoto}, {Garc{\'\i}a-Argum{\'a}nez}, {Gardner}, {Gawiser}, {G{\'o}mez-Guijarro}, {Guo}, {Hamilton}, {Hathi}, {Holwerda}, {Hirschmann}, {Huertas-Company}, {Hutchison},
  {Iyer}, {Jaskot}, {Jha}, {Jogee}, {Juneau}, {Jung}, {Kassin}, {Kurczynski}, {Larson}, {Leung}, {Long}, {Lucas}, {Magnelli}, {Mantha}, {Matharu}, {McGrath}, {McIntosh}, {Medrano}, {Merlin}, {Mobasher}, {Morales}, {Newman}, {Nicholls}, {Pandya}, {Rafelski}, {Ronayne}, {Rose}, {Ryan}, {Santini}, {Seill{\'e}}, {Shah}, {Shen}, {Simons}, {Snyder}, {Stanway}, {Straughn}, {Teplitz}, {Vanderhoof}, {Vega-Ferrero}, {Wang}, {Weiner}, {Willmer}, \& {Wuyts}}]{2022arXiv220801816Z}
{Zavala}, J.~A., {Buat}, V., {Casey}, C.~M., {et~al.} 2022, arXiv e-prints, arXiv:2208.01816, \dodoi{10.48550/arXiv.2208.01816}

\bibitem[{{Zavala} {et~al.}(2023){Zavala}, {Buat}, {Casey}, {Finkelstein}, {Burgarella}, {Bagley}, {Ciesla}, {Daddi}, {Dickinson}, {Ferguson}, {Franco}, {Jim{\'e}nez-Andrade}, {Kartaltepe}, {Koekemoer}, {Le Bail}, {Murphy}, {Papovich}, {Tacchella}, {Wilkins}, {Aretxaga}, {Behroozi}, {Champagne}, {Fontana}, {Giavalisco}, {Grazian}, {Grogin}, {Kewley}, {Kocevski}, {Kirkpatrick}, {Lotz}, {Pentericci}, {P{\'e}rez-Gonz{\'a}lez}, {Pirzkal}, {Ravindranath}, {Somerville}, {Trump}, {Yang}, {Yung}, {Almaini}, {Amor{\'\i}n}, {Annunziatella}, {Arrabal Haro}, {Backhaus}, {Barro}, {Bell}, {Bhatawdekar}, {Bisigello}, {Buitrago}, {Calabr{\`o}}, {Castellano}, {Ch{\'a}vez Ortiz}, {Chworowsky}, {Cleri}, {Cohen}, {Cole}, {Cooke}, {Cooper}, {Cooray}, {Costantin}, {Cox}, {Croton}, {Dav{\'e}}, {de La Vega}, {Dekel}, {Elbaz}, {Estrada-Carpenter}, {Fern{\'a}ndez}, {Finkelstein}, {Freundlich}, {Fujimoto}, {Garc{\'\i}a-Argum{\'a}nez}, {Gardner}, {Gawiser}, {G{\'o}mez-Guijarro}, {Guo}, {Hamilton}, {Hathi}, {Holwerda}, {Hirschmann},
  {Huertas-Company}, {Hutchison}, {Iyer}, {Jaskot}, {Jha}, {Jogee}, {Juneau}, {Jung}, {Kassin}, {Kurczynski}, {Larson}, {Leung}, {Long}, {Lucas}, {Magnelli}, {Mantha}, {Matharu}, {McGrath}, {McIntosh}, {Medrano}, {Merlin}, {Mobasher}, {Morales}, {Newman}, {Nicholls}, {Pandya}, {Rafelski}, {Ronayne}, {Rose}, {Ryan}, {Santini}, {Seill{\'e}}, {Shah}, {Shen}, {Simons}, {Snyder}, {Stanway}, {Straughn}, {Teplitz}, {Vanderhoof}, {Vega-Ferrero}, {Wang}, {Weiner}, {Willmer}, {Wuyts}, \& {Ceers Team}}]{2023ApJ...943L...9Z}
---. 2023, \apjl, 943, L9, \dodoi{10.3847/2041-8213/acacfe}

\bibitem[{{Zavala} {et~al.}(2025){Zavala}, {Castellano}, {Akins}, {Bakx}, {Burgarella}, {Casey}, {Ch{\'a}vez Ortiz}, {Dickinson}, {Finkelstein}, {Mitsuhashi}, {Nakajima}, {P{\'e}rez-Gonz{\'a}lez}, {Arrabal Haro}, {Bergamini}, {Buat}, {Backhaus}, {Calabr{\`o}}, {Cleri}, {Fern{\'a}ndez-Arenas}, {Fontana}, {Franco}, {Grillo}, {Giavalisco}, {Grogin}, {Hathi}, {Hirschmann}, {Ikeda}, {Jung}, {Kartaltepe}, {Koekemoer}, {Larson}, {McKinney}, {Papovich}, {Rosati}, {Saito}, {Santini}, {Terlevich}, {Terlevich}, {Treu}, \& {Yung}}]{2025NatAs...9..155Z}
{Zavala}, J.~A., {Castellano}, M., {Akins}, H.~B., {et~al.} 2025, Nature Astronomy, 9, 155, \dodoi{10.1038/s41550-024-02397-3}

\end{thebibliography}
\bibliographystyle{aasjournal}

\appendix
\section{Noise estimation on JWST/NIRCam images}
\label{app:noise}

The selection of high-redshift galaxy candidates carried out in this paper depends strongly on the determination of photometric uncertainties, which affect the estimation of photometric redshift probabilities through the zPDF. Our method to identify $z>16$ galaxies considers that the candidates must present a strong break and null flux bluewards of it, so the zPDF favors the $z>10$ solution over the low redshift alternative. Having this in mind, in this Appendix we present a detailed analysis of the noise properties of the imaging data used in this work, also comparing it with other similar works in the literature.

\begin{figure}[ht!]
\centering
\includegraphics[clip, trim=1.4cm 0.7cm 2.0cm 2.0cm,scale=0.5]{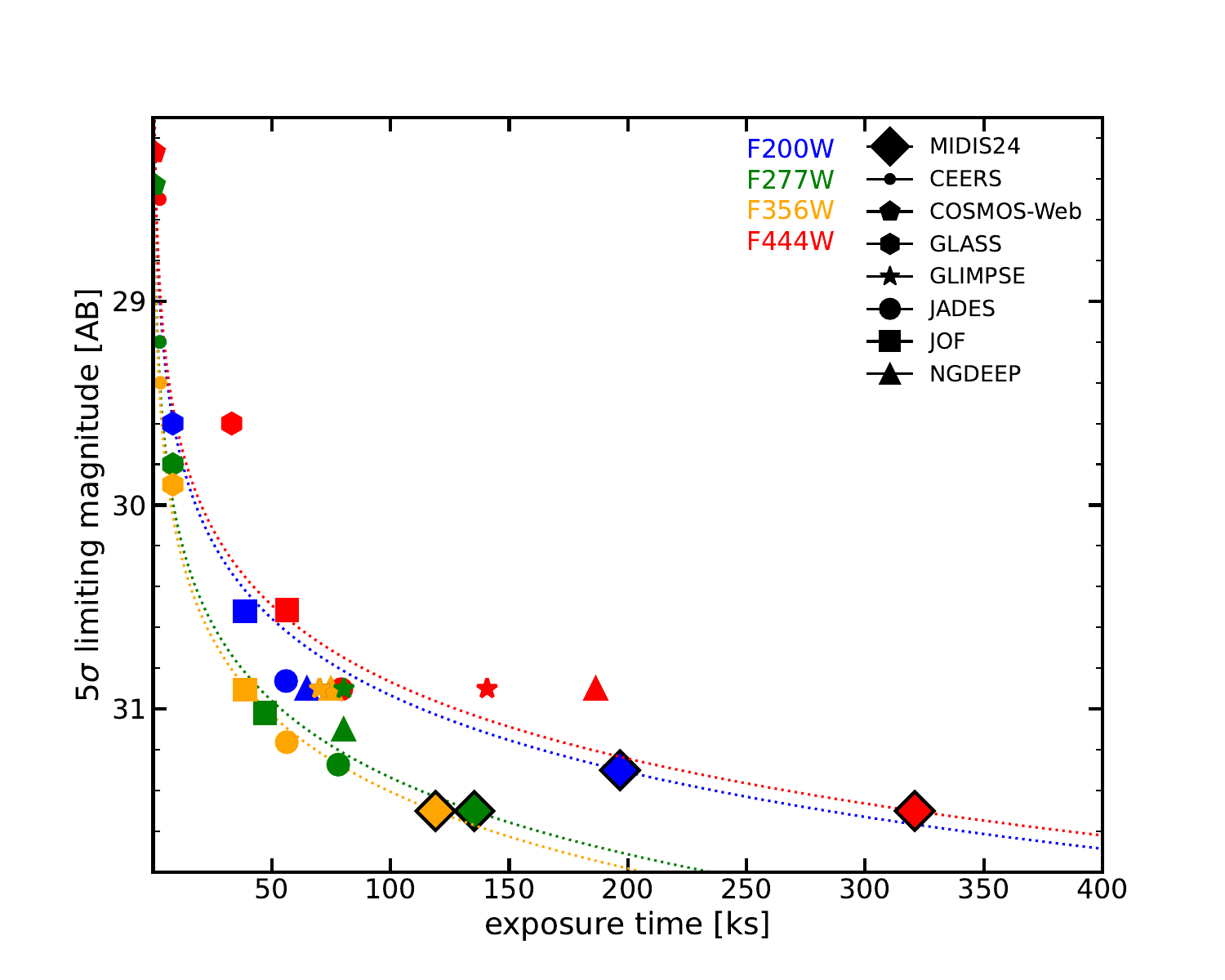}
\caption{\label{fig:expdepth}Depths ($5\sigma$ limiting magnitudes for point-like sources) of the most relevant JWST/NIRCam surveys which have been used in the literature to search for high redshift galaxy candidates. The depths have been taken from the references given below and homogenized to $5\sigma$ point-source limiting magnitudes measured in a diameter $d=0.2\arcsec$ circular aperture (corrected for PSF effects). The points are color-coded according to the NIRCam band. For clarity, we only show the most relevant bands for our work, which probe F200W and F277W dropouts. The dotted lines show the expected behavior of the limiting magnitude with respect to exposure time, normalized to our measurements for the deepest region covered by MIDIS and NGDEEP. Apart from MIDIS24 (depth achieved by MIDIS and NGDEEP for the 2024 MIDIS-RED observations, quoted in this paper and \citealt{2023ApJ...951L...1P}), the surveys we show are: CEERS \citep{2025arXiv250104085F}, COSMOS-Web \citep{2023ApJ...954...31C}, GLASS \citep{2022ApJ...935..110T}, GLIMPSE \citep{2024arXiv241113640K}, JADES \citep{2023arXiv230602465E}, JOF \citep{2024ApJ...970...31R}, and NGDEEP \citep{2024ApJ...965L...6B}. See also \citet{2024A&A...691A.240M} for a compilation of depths.
}
\end{figure}

Our method to measure photometric uncertainties was developed to account for the noise correlation effects introduced by the drizzling method used in the mosaicking of the data \citep{2002PASP..114..144F}. Indeed, the reduction of the data includes a resampling of the original observations to account for the dithering offsets and a stacking of all exposures employing a pixel size which is smaller than the original, more specifically, roughly half the size for the LW filters, 3\% smaller size for SW. Our method gathers a statistical sample of pixels from the local background (after masking objects) to calculate the median value and the rms. The background box around each source was set to $6\arcsec\times6\arcsec$ to ensure not being affected by gradients and be able to obtain enough pixels for the statistical analysis of the background. In order to avoid the effect of signal correlation, the selected pixels must be non-contiguous, with a minimum distance between them of 3~pixels, given that the drizzle method roughly divides the original pixels in half (for LW). Using larger minimum distances have a negligible effect but force to use larger background boxes, since the number of eligible pixels decreases and the statistical calculation is affected. The number of pixels used for the statistical analysis of the background is equal to the number of pixels in the photometric aperture, thus forming a virtual aperture of non-contiguous pixels. Several of these artificial apertures (at least 10, a larger number does not affect the results significantly) are compiled to get a more robust statistical measurement. This is similar to using small circular apertures but we remark that our apertures are virtual since we construct them with randomly-selected non-contiguous (and, consequently, independent) pixels.

In \citet{2023ApJ...951L...1P}, we compared our technique to estimate photometric errors with another method using randomly distributed circular apertures (as in \citealt{2023ApJS..265....5H}). With the latter, we found smaller rms values compared to our method based on randomly-selected non-contiguous pixels: 60\% and 3\% smaller rms values for the LW and SW channels. We repeated the exercise for the actual data and reductions used in this paper, and we obtained an average ratio of 1.3 between ours and the random-aperture methods (our method providing larger rms values, i.e., shallower limiting magnitudes), roughly independent of wavelength.

The depths quoted in this paper are compared in Figure~\ref{fig:expdepth} to calculations found in the literature for other surveys (see caption for references), after homogenization to the same aperture. We note that the specific detector setups and different backgrounds (e.g., zodiacal or intra-cluster light in the case of lensing fields) for each field should also have a noticiable effect on the limiting magnitudes, i.e., depths do not have to scale with exposure time perfectly. The same applies to the reference files (darks and flats), whose differences are not taken into account in this comparison.

The quoted depths in this paper broadly agree with the expected behavior against exposure time (depth proportional to $\sqrt{t_\mathrm{exp}}$). In fact, we obtain shallower than expected data compared to some surveys with shorter integration times, which we interpret as a consequence of  combining the MIDIS and NGDEEP data, which present different exposure times and detector setups (see \citealt{2023ApJ...954L..46L} for a discussion about the shallower than expected depths of NGDEEP Epoch 1). We note that the optimization of the combination of data coming from different surveys is far beyond the scope of this paper.

Apart from the depth estimations found in the survey presentation papers for different programs, we compare here with some other works using the same datasets shown in Figure~\ref{fig:expdepth} (and most of the times, different reductions of the same dataset). For example, \citet{2023ApJ...952L...7A} quote depths for NGDEEP epoch 1 data of 29.5-29.8 (see also \citealt{2024arXiv240714973C} and \citealt{2025ApJ...978...89H}) for F200W through F444W, using a $d=0.32\arcsec$ diameter aperture; taking into account that the exposure time was doubled by NGDEEP and scaling to a $d=0.2\arcsec$ diameter aperture, depths should be around 1~mag deeper, similar to the values used in the figure (which come from \citealt{2024ApJ...965L...6B}).  \citet{2024A&A...691A.240M} quote independent depths for most of the surveys presented in Figure~\ref{fig:expdepth}, obtaining similar results; noticeably, they quote a depth for the NGDEEP F200W band of 30.1~mag (after correcting to the use of Epoch 1$+$2 data), which is 0.6~mag shallower than the value provided by \citet{2024ApJ...965L...6B}, but more consistent with the curve shown in our figure. \citet{2024MNRAS.527.5004M} provide also consistent depths for CEERS, once accounting for a 0.6~mag scaling to match the aperture size (0.35\arcsec\, diameter in that paper, 0.2\arcsec\, in Figure~\ref{fig:expdepth}). 

\begin{figure*}[ht!]
\centering
\includegraphics[clip, trim=1.4cm 0.7cm 2.0cm 2.0cm,scale=0.5]{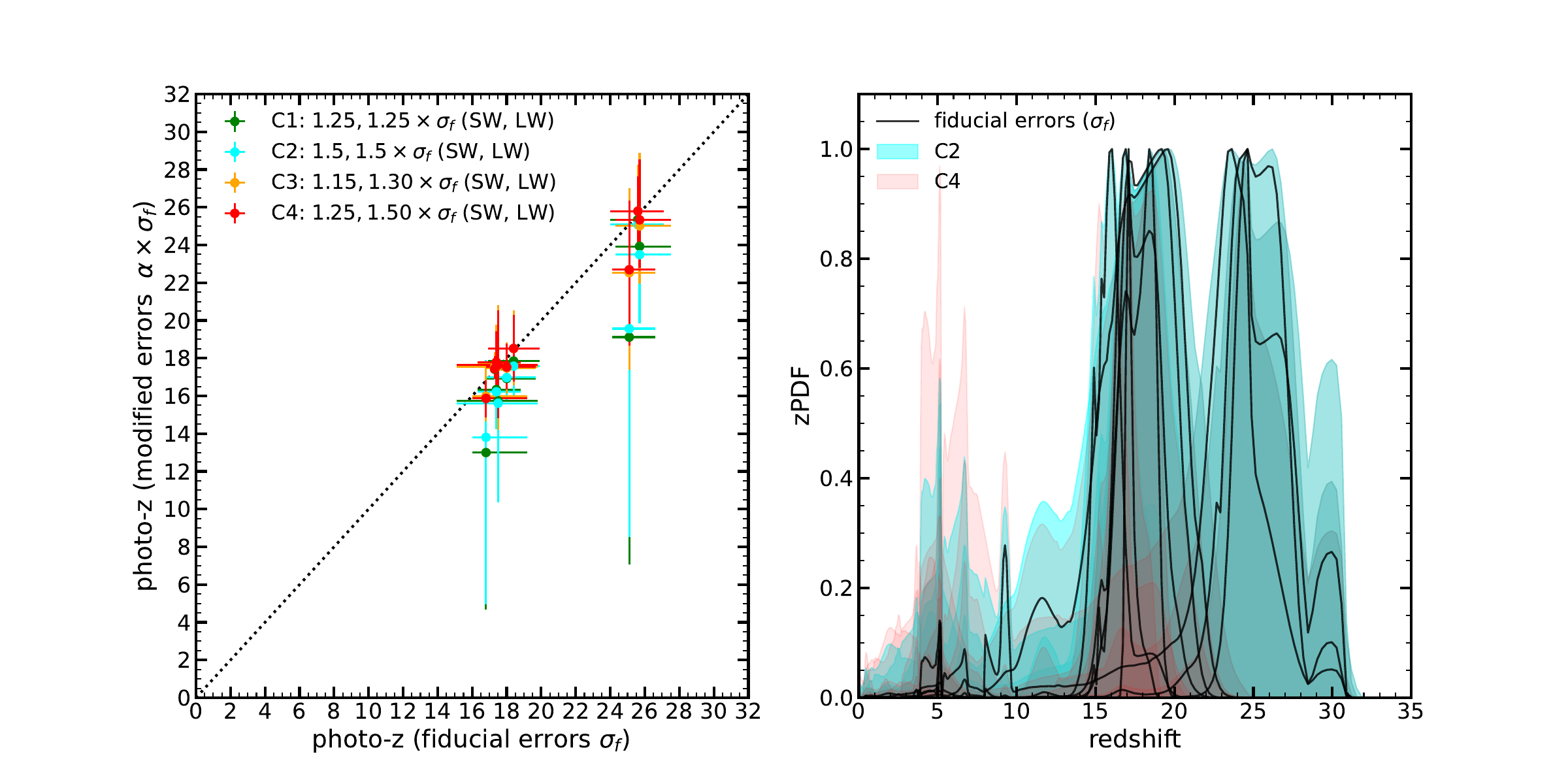}
\caption{\label{fig:error_in_z}Evaluation of the effect of underestimations of the flux uncertainties in the determination of photometric redshifts for our $z>16$ galaxy candidates. On the left, we show the comparison of redshifts calculated with our fiducial flux errors (x-axis) and the estimates obtained after multiplying all errors by 1.25 (case 1, C1, green), 1.5 (C2, cyan), and multiplying the SW errors by 1.15 and the LW errors by 1.3 (C3, orange) or SW errors by 1.25 and the LW errors by 1.5 (C4, red). On the right, we show the corresponding redshift probability distribution functions. For clarity, in this right panel we only show the fiducial zPDFs (black) and the results for C1 (cyan) and C4 (red).
}
\end{figure*}

Figure~\ref{fig:expdepth} demonstrates that our depth estimations, which are directly related to the flux uncertainty calculations, are completely consistent with measurements for other surveys performed independently by other groups. Nevertheless, we explore in Figure~\ref{fig:error_in_z} the effect on the photometric redshift calculations of underestimates of photometric uncertainties. We analyze how a constant for all bands underestimation of our fiducial photometric uncertainties ($\sigma_f$) would affect the redshift determination. We show results for 25\% and 50\% larger uncertainties (case 1 and 2, C1 and C2). We also explore the effects on the redshift estimation in the case that our uncertainties for the SW data would be underestimated by 15\%, while the LW errors would be 30\% larger (case C3), and underestimations of 25\% for SW and 50\% for LW (C4). On the left panels we show that the different estimations of the most probable photometric redshifts would be consistent within uncertainties, with these increasing as photometric errors become larger. On the right panel, we show the effect on the whole (normalized) zPDF for the extreme cases (C2 and C4, the others are not shown for clarity). The overall effect is a larger probability of the low redshift peak (at $z\sim4-7$) and a wider high redshift peak, extending significantly to lower redshift values, which explain the larger error bars shown on the left panel. The most affected sources would be  MIDIS-z17-2,  MIDIS-z17-6, and MIDIS-z25-2, which would fail to pass the $\Delta \chi^2$ threshold for C2, and for underestimations of the uncertainties of 40\% or larger, which are not supported by the analysis performed in Figure~\ref{fig:expdepth}. 

We conclude that our photometric uncertainties are robust, and their underestimation is not supported by any of our tests. Even in the extreme case that they could be underestimated by 50\%, the $z\sim17$ and $z\sim25$ solutions would still dominate, although with lower significance.

\section{High redshift galaxy candidates}
\label{appA}

\begin{figure*}[ht!]
\centering
\includegraphics[clip, trim=1.5cm 5.0cm 2.0cm 1.5cm,scale=0.45]{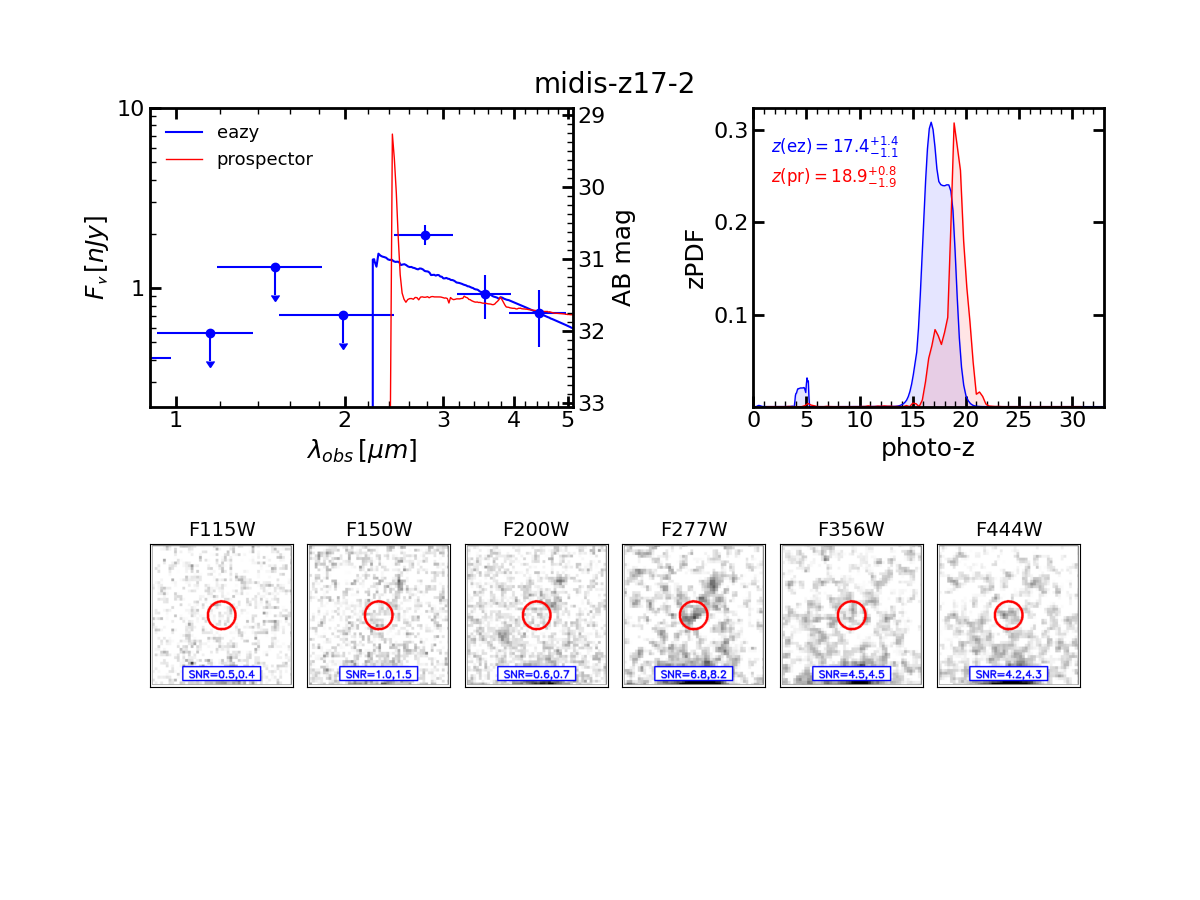}
\caption{\label{fig:z17_2}SED plot and photometric redshift probability distribution function for the $z\sim17$ galaxy candidate midis-z17-2. The figure shows the same information depicted for an example galaxy in the main text (Figure~\ref{fig:sed_examples}).}
\end{figure*}

\begin{figure*}[ht!]
\centering
\includegraphics[clip, trim=1.5cm 5.0cm 2.0cm 1.5cm,scale=0.45]{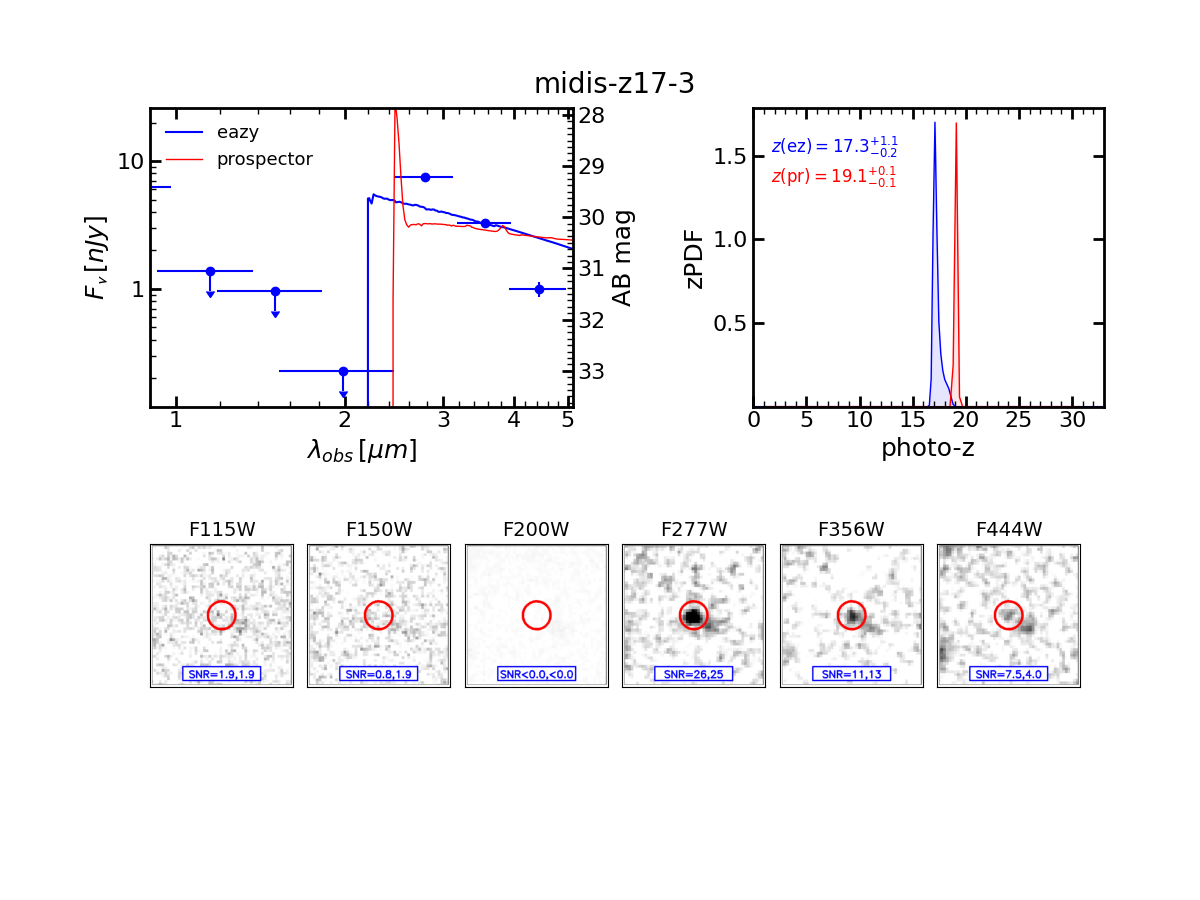}
\caption{\label{fig:z17_3}Same as Figure~\ref{fig:z17_2} for the  $z\sim17$ galaxy candidate midis-z17-3.}
\end{figure*}

\begin{figure*}[ht!]
\centering
\includegraphics[clip, trim=1.5cm 5.0cm 2.0cm 1.5cm,scale=0.45]{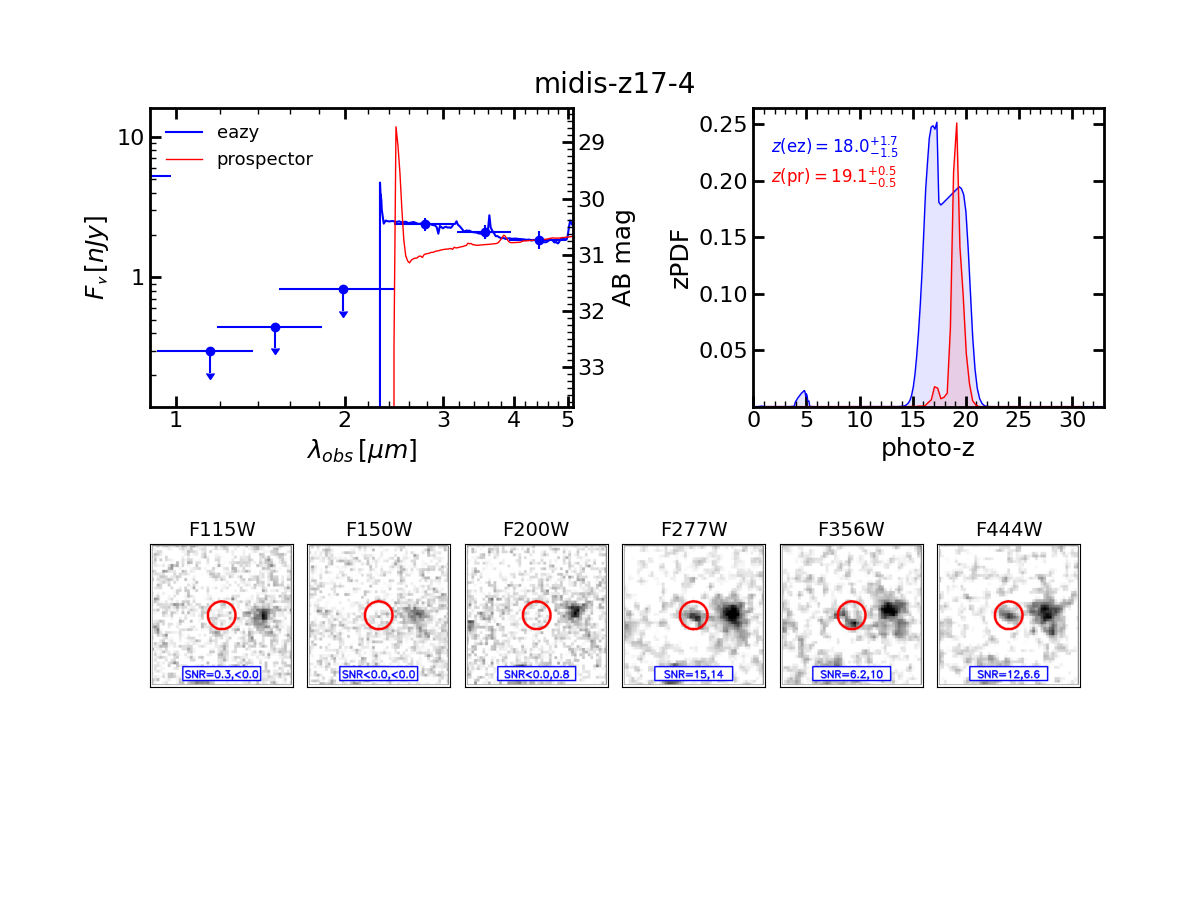}
\caption{\label{fig:z17_4}Same as Figure~\ref{fig:z17_2} for the  $z\sim17$ galaxy candidate midis-z17-4.}
\end{figure*}

\begin{figure*}[ht!]
\centering
\includegraphics[clip, trim=1.5cm 5.0cm 2.0cm 1.5cm,scale=0.45]{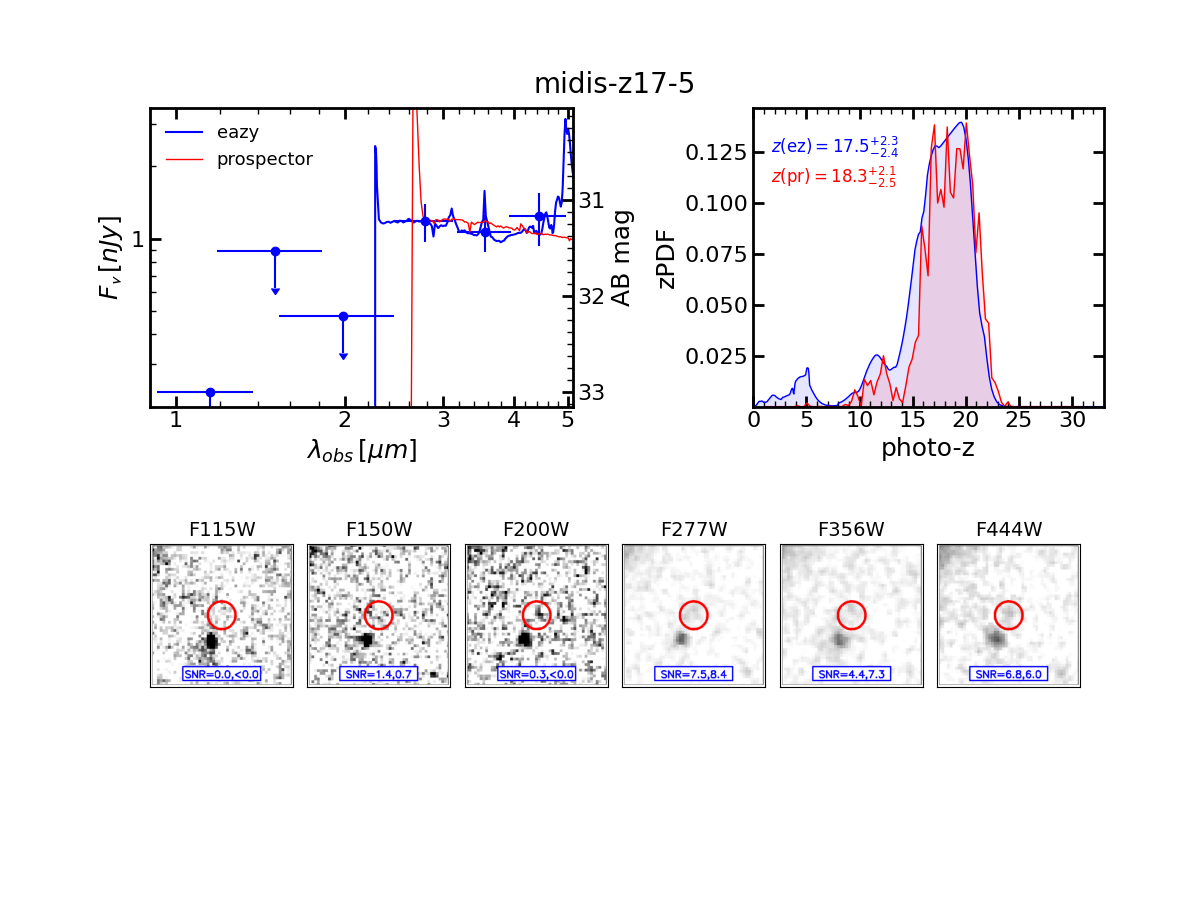}
\caption{\label{fig:z17_5}Same as Figure~\ref{fig:z17_2} for the $z\sim17$ galaxy candidate midis-z17-5.}
\end{figure*}

\begin{figure*}[ht!]
\centering
\includegraphics[clip, trim=1.5cm 5.0cm 2.0cm 1.5cm,scale=0.45]{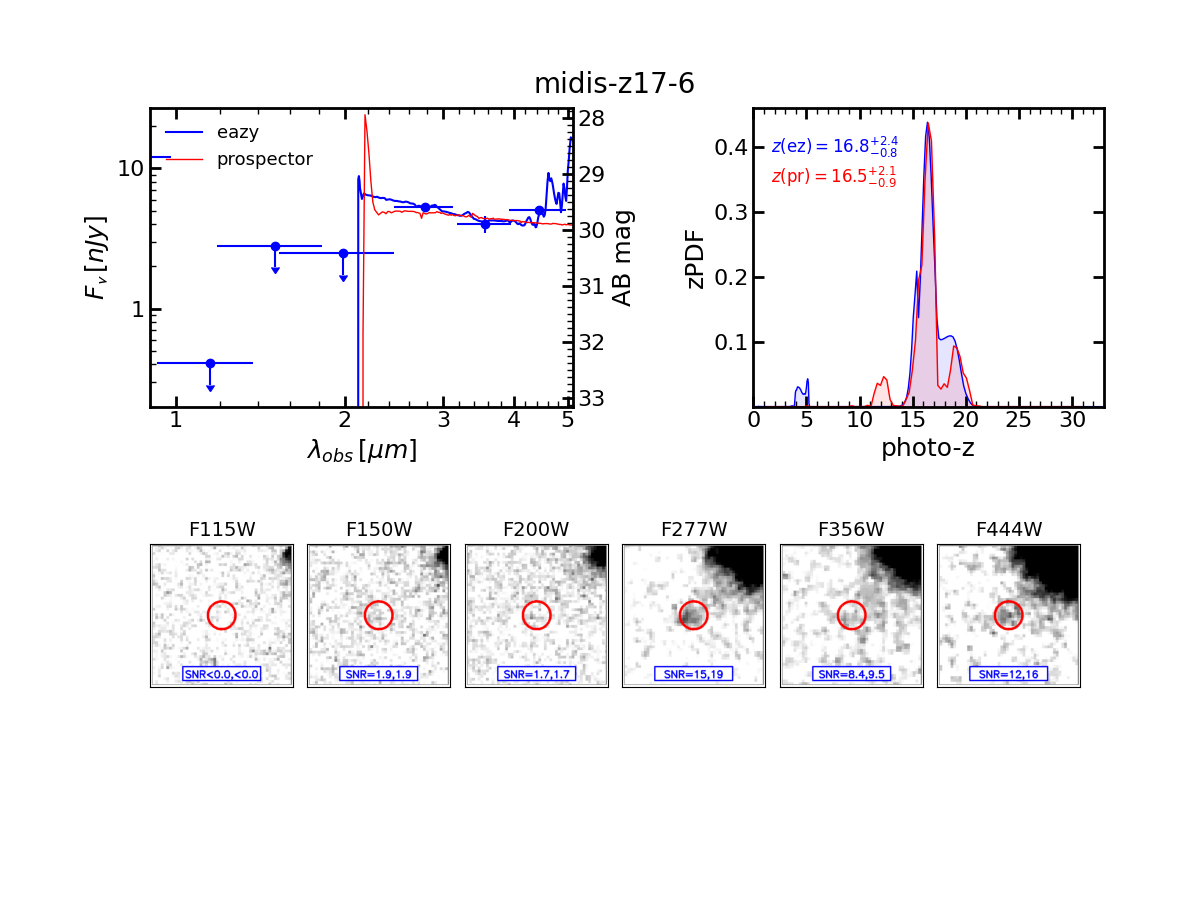}
\caption{\label{fig:z17_6}Same as Figure~\ref{fig:z17_2} for the $z\sim17$ galaxy candidate midis-z17-6.}
\end{figure*}

\begin{figure*}[ht!]
\centering
\includegraphics[clip, trim=1.5cm 5.0cm 2.0cm 1.5cm,scale=0.45]{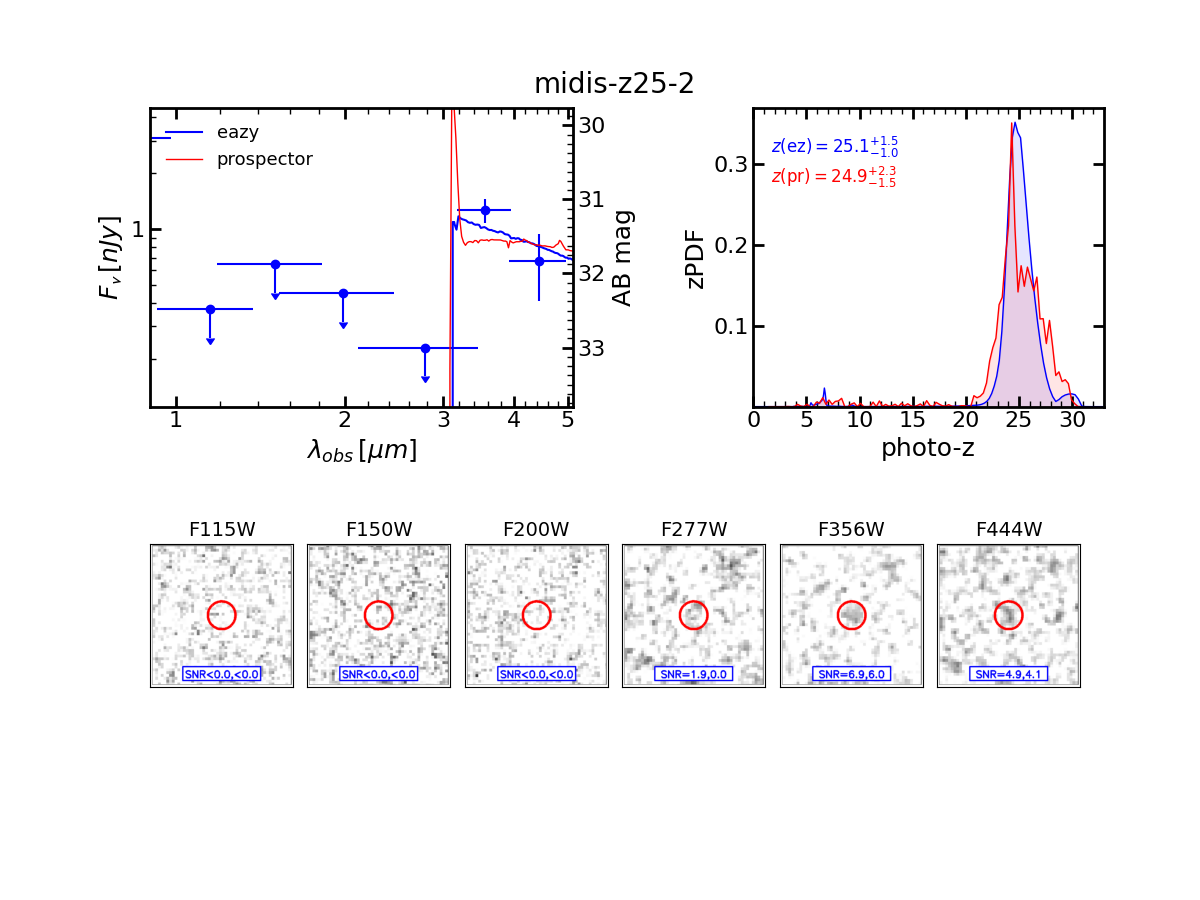}
\caption{\label{fig:z24_2}SED plot and photometric redshift probability distribution function for the $z\sim25$ galaxy candidate midis-z25-2. The figure shows the same information depicted for an example galaxy in the main text (Figure~\ref{fig:sed_examples}).}
\end{figure*}

\begin{figure*}[ht!]
\centering
\includegraphics[clip, trim=1.5cm 5.0cm 2.0cm 1.5cm,scale=0.45]{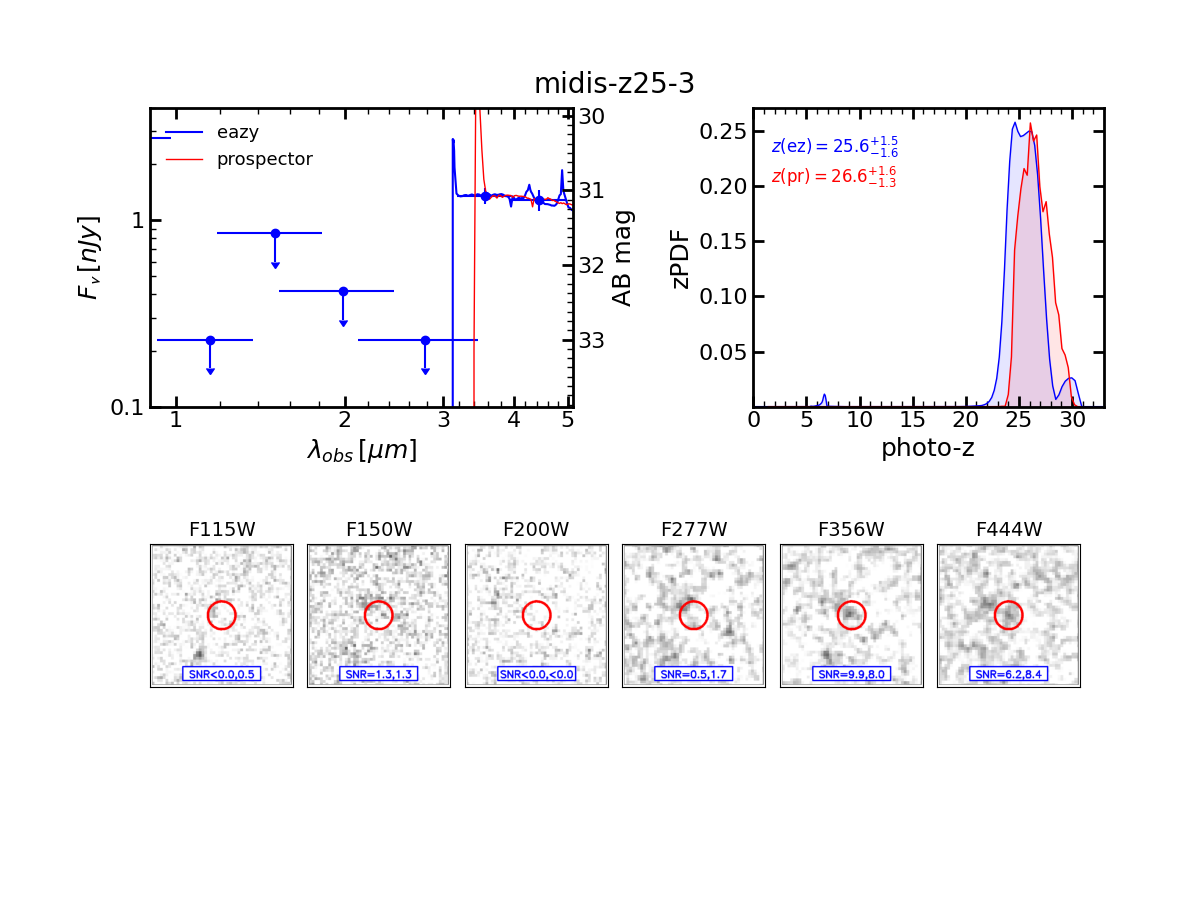}
\caption{\label{fig:z24_3}Same as Figure~\ref{fig:z24_2} for the $z\sim25$ galaxy candidate midis-z25-3.}
\end{figure*}

In the following figures, we show the SED fits, zPDFs, and postage stamps of the galaxy candidates at $z\sim17$ (Figures~\ref{fig:z17_2}, \ref{fig:z17_3}, \ref{fig:z17_4}, \ref{fig:z17_5}, and \ref{fig:z17_6}) and at $z\sim25$  (Figures~\ref{fig:z24_2} and \ref{fig:z24_3}). We also comment here on the validity of the detections of specific sources.

The work presented in this paper relies on the selection and analysis of very faint sources at the edge of the JWST/NIRCam capabilities. Consequently, it is of utmost importance to discuss the solidness of the identification of each source, i.e., the robustness of their non-spurious nature. The analysis also relies heavily on the validity of the photometric uncertainties, which was discussed in Appendix~\ref{app:noise}, as well as on the strength of the break and undetection of the sources in all bands bluewards of the allegedly identified Lyman break. We note, however, that some flux could be present bluewards of the Ly-$\alpha$ line, for example because the red wing of the dropout band partially covers the jump.

To address these points, we have made the following calculations. We have measured how often random aggregations of noise pixels can lead to low significance false detections. Consider a cluster of '$p$' ($>9$) connected pixels summing up to a combined significance of 'n'-$\sigma$ . For $n=1$ there are many such false sources, but they decrease rapidly with increasing $n$. For instance, for $n=3$, one expects ~5400 false sources per square arcminute, which decreases to 0.12 for $n=5$. If we in addition require a false detection at the same position (within 0.06\arcsec) in a second filter at $n>3$, the expected number is $3\times10^{-6}$ per arcmin$^{2}$.
If we have underestimated the error, so that, say $5\sigma$ is really just $4\sigma$, having two coincident $4\sigma$ false sources, only $10^{-5}$ are expected, and even for 3+3 sigma, we only expect 0.02/arcmin$^2$.

With this in mind, we note that all our candidates except one have average SNR in the detection bands, as well as individual bands, above 5, so the spurious nature of any of our candidates has a very low probability. The exception is midis-z25-2, we will comment on it  below.

We also note that although some pixels in the F200W band are gray/black in the postage stamps, their signal is quite low, consistent with random fluctuations of the background, and compatible with the high-z redshift solution. We quantify this statement below for each source.

After these general considerations, we comment briefly about the detection and analysis of each source in our sample.

The source midis-z17-1, whose SED and postage stamps were shown in Figure~\ref{fig:sed_examples}, presents some signal in the F200W. This is also the case for midis-z17-5, and midis-z25-1, and even midis-z17-4, midis-z17-6, but in the latter two cases, the bright pixels are offseted from the centroid of the source.

For midis-z17-1, the bright pixel in the F200W image presents a signal which is $2.2\sigma$ above the background, the adjacent pixels have signals between $0.3\sigma$ and $1.2\sigma$. This means that the FWHM of that signal peak is around or even smaller than two pixels, while the PSF FWHM for F200W is 0.06\arcsec, i.e., 2 pixels. Centering an aperture in that pixel, and with a diameter of d=0.084\arcsec, which encloses 50\% of the flux, we perform a forced measurement in the original F200W image (i.e., not convolved to the PSF of F444W) that maximizes the signal. We get a magnitude of $32.2\pm0.8$ (i.e., $1.3\sigma$), i.e., a very low-significance measurement that cannot be used to claim a detection and is consistent with our quoted magnitude lower limit based on a d=0.2\arcsec\, aperture. The magnitude we measure for this source is $34.1\pm2.7$ for d=0.2\arcsec\, and the flux is negative for d=0.3\arcsec\, and d=0.4\arcsec. We also stress that fluxes were measured in stacked SW data, and all candidates present fluxes with $<2$ significance, our limit to claim a non-negligible flux bluewards of the Lyman-$\alpha$ line.

Perhaps more importantly, we also note that, depending on the exact redshift of the source, some flux in the dropout band is not inconsistent  with the possibility that we are observing the Lyman-$\alpha$ break. In this particular source midis-z17-1, if we force the F200W flux to the measurement given above for a very small aperture ($32.2\pm0.8$~mag; we note that this aperture is extremely small for the standard sizes used for JWST data in this kind of study, see Section~\ref{sec:selection}), we obtain a dominant photometric redshift solution of $z=16.5\pm2.0$ (smaller value and larger error compared to our fiducial calculation).

A similar evaluation is obtained for midis-z17-5, the brightest dark pixel seen in the F200W postage stamp is $0.8\sigma$ above the local background, the adjacent pixels are between 0.8 and 10 times dimmer, except one pixel which has a negative signal. The photometry in F200W gives $32.2\pm1.0$ for d=0.084\arcsec. The $1\sigma$ lower limit for this specific source is 31.6 mag. For the results in the paper, we had $33.8\pm4.5$ for  d=0.2\arcsec\, and negative fluxes for larger apertures. The candidate is then legit based on our selection criteria, and the photometric redshift is not significantly affected even if using the F200W flux quoted above.

For midis-z25-1, the F200W image presents a pixel $1.7\sigma$ above the background, separated 0.06\arcsec\, from the center of the source as measured in F356W. A small aperture with d=0.084\arcsec centered on that pixel gives $31.7\pm0.7$~mag, all larger apertures give negative fluxes. We measured magnitude $32.8\pm0.9$ for d=0.2\arcsec\, and negative fluxes for d=0.3\arcsec\, and 0.4\arcsec for our fiducial apertures centered on the source centroid calculated in F356W. Our quoted magnitude lower limit for this specific source in F200W is 33.0 mag.

It is relevant to discuss the non-spurious nature of midis-z25-2. The significance of the measurement for this source in the F444W band is below the nominal value of 5 to identify detections. However, the SNR of the F356W is higher than 6 for d=0.2\arcsec\, and d=0.3\arcsec. Consequently, the average detection significance is above our limit of 5. Taking the latter, we calculate a probability of a spurious detection in the d=0.2\arcsec\, of a millionth. If our uncertainties were underestimated by 50\%, we could expect around 10 sources of this kind in our data. However, these numbers decrease to 0.2 sources in our data for the significance measured in the d=0.3\arcsec\, aperture, even after allowing a 50\% underestimation of the errors. Given also that we measure some flux in the F277W band at the $1.9\sigma$ level, we kept this source in our sample.

We note that some similar pixel blobs with positive signal are observed in the postage stamps of midis-z25-2 in the three redder bands. They were not selected as real objects for the following reasons. In particular, in order to understand the characteristics of our data based on those postage stamps, we discuss the measurements of one blob to the NW (below and to the right of the "W" in the filter name), another one to the W (just south of the first one), and another to the SE (in the corner of the stamp). The NW-blob is 10-20\% dimmer than our candidate in F356W for the apertures mentioned above, and 60-90\% dimmer for F444W, so their SNR is considerably smaller than 5. The flux in F277W for this blob is 4 times brighter than our candidate in F277W, which probably means that the source is real, but is well below our average detection limit of $SNR>5$ and not compatible with our detection and selection criteria. The W-blob seems like a spurious source and it cannot be compared to the $z\sim25$ candidate given that it is 75\% dimmer than our source in the d=0.2\arcsec\, aperture for F444W and presents a negative flux in the d=0.3\arcsec\, aperture, as well as very small, almost all negative, fluxes in all other bands. For the SE-blob, the flux in the d=0.3\arcsec\, aperture is 70\% dimmer than our candidate, very similar for the d=0.2\arcsec\, aperture. We conclude that this blob is also quite different from our selected sources.

Similarly, two blobs are detected in the neighborhood of the source midis-z25-3 with similar appearance in the PNG files shown in Figure~\ref{fig:z24_3}, one to the NE (very close to the red circle shown in the stamp) and one to the SE  (just above the "SNR" text). Those blobs are identified as real sources, given the detection in F115W in the case of the SE-blob, and the SNR=4-5 fluxes measured in F277W, F356W and F444W for the NE-blob. But their properties are not compatible with dropout sources.

\section{Other high redshift galaxy candidates}
\label{appB}

In the following Table~\ref{tab:others}, we list sources that were in the border of one of our selection criteria and were finally excluded from the luminosity function and subsequent calculations. We list their coordinates, the inferred approximate photometric redshifts, and the reason to discard them.

\begin{deluxetable}{lcccl}[h]
\caption{\label{tab:others}Other galaxy candidates at $14<z<25$.}
\centerwidetable 
\tabletypesize{\scriptsize}
\tablehead{\colhead{MIDIS ID} & \colhead{RA (J2000)} & \colhead{DEC (J2000)} &  \colhead{redshift} & \colhead{Reason to discard}\\
& [degrees] & [degrees] & &}
\startdata
\hline
\hline
midisred0078692 & 53.31222420 & $-$27.88696470 & $z\sim17$ & $r=0.1\arcsec$ aperture gives $z\sim5$\\
midisred0007023 & 53.29276170 & $-$27.87913810 & $z\sim25$ & $r=0.2\arcsec$ aperture gives $z\sim8$\\
midisred0008188 & 53.25742490 & $-$27.87752330 & $z\sim17$ & $r=0.1\arcsec$ gives $SNR=2$ for SW images stacks\\
midisred0081800 & 53.27634280 & $-$27.87423430 & $z\sim17$ & $r=0.1\arcsec$ gives $SNR=2$ for SW images stacks\\
midisred0022529 & 53.30681440 & $-$27.86221840 & $z\sim22$ & $r=0.1\arcsec$ aperture gives $z\sim16$\\
midisred0045193 & 53.25118440 & $-$27.82049580 & $z\sim16$ & $r=0.1\arcsec$ aperture gives $z\sim5$\\
midisred0075426 & 53.26640900 & $-$27.81979570 & $z\sim25$ & in the limit of the $SNR=5$ cut\\
midisred0075755 & 53.25900040 & $-$27.81663240 & $z\sim25$ & in the limit of the $SNR=5$ cut\\
midisred0054262 & 53.24777980 & $-$27.79423420 & $z\sim14$ & $r=0.2\arcsec$ aperture gives $z\sim5$, nearby bright source \\
\hline
\enddata
\end{deluxetable}

\end{document}